\documentclass{iopart}

\pdfoutput=1
\usepackage{iopams} 
\usepackage{graphicx,amssymb}
\expandafter\let\csname equation*\endcsname\relax
\expandafter\let\csname endequation*\endcsname\relax
\usepackage{amsmath}
\usepackage{cite} 
\usepackage{url}

\usepackage{float} 

\usepackage{hyperref} 
\usepackage[varg]{txfonts}
\usepackage[T1]{fontenc}

\usepackage{algorithm}
\usepackage{algpseudocode}
\usepackage{longtable}
\usepackage{booktabs}
\usepackage{siunitx}

\usepackage{empheq,xcolor}
\definecolor{myblue}{rgb}{.8, .8, 1} 
\newcommand*\myblueboxL[1]{\colorbox{myblue}{\hspace{0.7em}#1\hspace{0.7em}}}
\newcommand*\mybluebox[1]{\colorbox{myblue}{\hspace{0.em}#1\hspace{0.em}}}

\DeclareMathAlphabet{\mathcalstd}{OMS}{cmsy}{m}{n}
\DeclareMathAlphabet{\mathpzc}{OT1}{pzc}{m}{it}

\DeclareMathOperator{\diag}{diag}
\DeclareMathOperator{\rank}{rank}

\DeclareMathOperator{\argmax}{argmax}
\DeclareMathOperator{\len}{len}

\newcommand{\bigO}{\mathcal O}

\sloppypar
\eqnobysec 

\begin{document}



\title[Frequency domain reduced order models for GWs from aligned-spin compact binaries]{Frequency domain reduced order models for gravitational waves from aligned-spin compact binaries}

\author{Michael P\"urrer}

\address{School of Physics and Astronomy, Cardiff University, Queens Building, CF24 3AA, Cardiff, United Kingdom}

\ead{Michael.Puerrer@astro.cf.ac.uk}

\begin{abstract}
Black-hole binary coalescences are one of the most promising sources for the first detection of gravitational waves. Fast and accurate theoretical models of the gravitational radiation emitted from these coalescences are highly important for the detection and extraction of physical parameters. Spinning effective-one-body (EOB) models for binaries with aligned spins have been shown to be highly faithful, but are slow to generate and thus have not yet been used for parameter estimation studies. I provide a frequency-domain singular value decomposition (SVD)-based surrogate reduced order model that is thousands of times faster for typical system masses and has a faithfulness mismatch of better than $\sim 0.1\%$ with the original SEOBNRv1 model for advanced LIGO detectors. This model enables parameter estimation studies up to signal-to-noise ratios (SNRs) of 20 and even up to SNR 50 for total masses below $50 M_\odot$. This article discusses various choices for approximations and interpolation over the parameter space that can be made for reduced order models of spinning compact binaries, provides a detailed discussion of errors arising in the construction and assesses the fidelity of such models.
\end{abstract}



\section{Introduction} 
\label{sec:introduction}

The LIGO~\cite{Abbott:2007kv} and VIRGO~\cite{Accadia2012} detectors along with GEO600~\cite{Grote:2010zz} span a world-wide network of ground based gravitational wave detectors using laser interferometry. Currently LIGO and VIRGO are being upgraded into advanced detectors which will increase the detector sensitivities by a factor of 10 and event rates by a factor of 1000.
The advanced LIGO (aLIGO) and Virgo (AdV) detectors~\cite{Harry:2010zz,aVIRGO,Abadie:2010cf} are due to come online in 2015-16 and expected to reach their design sensitivity around 2018~\cite{Aasi:2013wya} when they will very likely make routine detections. The advanced detectors are expected to provide a wealth of astrophysical information (see, e.g.,~\cite{Sathyaprakash:2009xs}) and help establish the field of gravitational-wave astronomy.
Furthermore, an underground cryogenic detector in Japan, KAGRA~\cite{Somiya:2011np}, is under construction.

The coalescence of compact binary systems is one of the most promising sources of gravitational waves (GWs) for these detectors. The theoretical knowledge of the gravitational waveforms emitted by these sources is essential for signal detection and the determination of astrophysical parameters.
The standard technique for observing compact binary coalescences involves matched filtering detector data against a set of ``template'' waveforms which approximate potential astrophysical signals~\cite{Allen:2005fk}.
Furthermore, parameter estimation (PE) algorithms can provide estimates of the parameters of compact binary systems observed with the detection algorithms. The goal of Bayesian PE methods is to estimate the posterior probability density function of the parameters that describe the signal and to evaluate the evidence of different waveform models.

To this end the post-Newtonian (PN) expansion~\cite{Blanchet:2013haa} in $v/c$ of Einstein's equations provides a powerful formalism to describe these waveforms during the inspiral, when the orbital speed is much smaller than the speed of light. Binary neutron stars (BNS) are expected to accumulate the majority of the signal-to-noise ratio (SNR) in the inspiral regime and are therefore well described by high-order PN waveforms. 3.5PN order templates have been shown~\cite{Buonanno:2009zt,Brown:2012nn} to be sufficient for searches of non-spinning compact binaries of total mass up to $12 M_\odot$. For parameter estimation higher accuracy is required.

The PN expansion breaks down in the plunge, merger and ringdown phases of compact binary coalescence. As the total mass of the binary increases the detectors become more sensitive to these strong field effects that manifest themselves at higher frequency. To describe the waveforms radiated from BH binaries in these late stages of evolution numerical solutions of the full nonlinear Einstein equations are required~\cite{Centrella:2010mx}. Complete inspiral-merger-ringdown (IMR) models are created from an amalgam of PN and numerical relativity (NR) waveforms. In practice, both approaches are necessary since generating NR waveforms of more than several tens of orbits is too computationally expensive.

Two broad classes of IMR models have been proposed for BH binaries with spins aligned with the orbital angular momentum. Phenomenological models~\cite{Ajith:2011,Santamaria:2010yb} are constructed in the frequency domain using fits to PN-NR hybrid waveforms. This technique was used to provide models that can be evaluated very fast and are therefore suitable also for the demanding methods required for parameter estimation. In contrast, \emph{effective-one-body} (EOB) models~\cite{Buonanno:1998gg,Buonanno:2000ef,Damour:2000we,Damour:2001tu,Damour:2008gu} use an analytical approach that combines PN expansion, re-summation techniques and perturbation theory with additional calibration of certain model parameters against NR waveforms. EOB waveforms have been shown to be very accurate (e.g. for the non-spinning EOBNRv2 model~\cite{Pan:2011gk} uncertainties in PE are dominated by statistical rather than systematic error~\cite{Littenberg:2012uj}), but are much slower to generate than the aforementioned phenomenological models since a complicated system of ordinary differential equations has to be solved over a long time interval with small time steps.

The Bayesian methods employed in PE studies require the generation of $\sim 10^6 - 10^7$ model waveforms to probe the 11 (for aligned spin binaries) dimensional parameter space~\cite{NINJA2-det-pe::2014tra}.
The more complex EOB models, such as the spinning EOB model SEOBNRv1~\cite{Taracchini:2012ig} on which we focus in this paper, are therefore too expensive for direct use in PE calculations~\cite{NINJA2-det-pe::2014tra}.
For example, to generate an equal-mass SEOBNRv1 waveform for a $20 M_\odot$ binary that fills the aLIGO band down to $\sim 10 \,$ Hz takes about 40 seconds on a current workstation and thus a PE study assuming no parallelism may take about ten years. In this paper we present a surrogate model of SEOBNRv1 that reduces the computational cost for this example by a factor of about three thousand, making it possible to generate the waveforms for such a study on the order of a day.

Even the generation of template banks for detecting GW signals from compact binaries becomes very expensive for aligned-spin systems. Stochastic template bank placement techniques~\cite{Harry:2009ea,Manca:2009xw} produce banks with a size on the order of $\sim 10^6$ templates~\cite{Ajith:2012mn,Brown:2012qf} but require the generation of further waveforms during its construction as the bank becomes denser. In contrast to PE a template bank needs to be built only once, but the cost at low total masses may again be prohibitive for complex models such as SEOBNRv1. Let us note, though, that the nonspinning EOBNR model~\cite{Pan:2011gk} has been used in PE studies and template bank creation.

\emph{Reduced order modeling} (ROM) techniques can be used to build surrogate models that provide fast and accurate compressed approximations of a selected source waveform model. Several such methods have been proposed in the literature.
\emph{Singular value decomposition} (SVD)-based methods have been used~\cite{Smith:2012du,Cannon:2011rj,Cannon:2011xk,Cannon:2010qh} to interpolate time-domain inspiral waveforms. 
Cannon et al~\cite{Cannon:2012gq} calibrate a reduced basis of non-spinning PhenomB waveforms against a set of NR waveforms to obtain a model with improved accuracy.
SVD-based methods have also been used to provide efficient representations of template banks~\cite{Cannon:2012:0004-637X-748-2-136,PhysRevD.89.024003}.
Another class of methods are greedy \emph{reduced basis} (RB) methods which are usually combined with the \emph{empirical interpolation method}~\cite{Barrault:2004}. These techniques have also been applied to GW waveforms~\cite{Field:2013cfa, Field:2012if, Field:2011mf, Blackman:2014maa} and BH ringdown~\cite{Caudill:2011kv}.
Herrmann et al~\cite{Field:2012if} find that the number of RB elements needed to represent aligned-spin TaylorF2~\cite{Sathyaprakash:1991mt,Cutler:1994ys,Droz:1999qx,Buonanno:2009zt,Ajith:2012az}\footnote{TaylorF2 is a closed-form frequency domain PN approximant obtained via the stationary phase approximation.} inspiral waveforms is not much larger than for the non-spinning case. Blackman et al~\cite{Blackman:2014maa} show that precessing PN waveforms can be represented to high accuracy in a surprisingly compact basis. Neither~\cite{Field:2012if} nor~\cite{Blackman:2014maa} build an actual model.
Field et al~\cite{Field:2013cfa} build test models of the non-spinning EOBNR~\cite{Pan:2011gk} waveform family over small mass-ratio intervals, from 1 to 2 and 9 to 10.

Further techniques have recently been proposed to help speed up parameter estimation calculations. One approach is to directly interpolate the likelihood function~\cite{Smith:2013zya} with the help of the SVD; a different method speeds up likelihood evaluations by defining special reduced order quadrature rules~\cite{Canizares:2013ywa,Canizares:2014fya,Antil:2013}.
SVD-interpolated waveforms have been shown to satisfy the stringent waveform-model accuracy criteria imposed by parameter-estimation requirements~\cite{Smith:2012du}.
So far, reduced order models with application for searches and PE have been restricted to non-spinning waveforms and to small subregions of the mass space.

In this study I propose a method for building reduced order models covering a multi-dimensional intrinsic parameter space (mass-ratio and spins) for the first time. The emphasis is on a practical approach with a view towards data analysis requirements for the advanced detector era.
To achieve this I make the following choices. The model approximates SEOBNRv1 waveforms in the frequency domain.
This allows for the fast computation of waveforms and their \emph{match}, which is defined as a frequency domain inner product weighted by the detector noise.
A sparse set of frequency points is introduced that yields a highly accurate representation of waveforms from the BNS regime up to hundreds of solar masses.
I compute reduced bases with the SVD separately for the amplitude and phase of the frequency domain waveform starting from a set of input waveforms on a regular parameter space grid. This separate treatment of amplitude and phase allows for higher compression in the model.

I first build a \emph{single-spin} model over a very wide range of the parameter space, up to mass-ratio $1:100$, which is way beyond the calibration range of the source model, and later a double-spin model up to $1:10$. Single-spin models add one additional parameter compared to non-spinning models and are thus the simplest models describing spinning binaries and provide an ideal testbed for reduced order modeling techniques. The single-spin model only uses equal-spin $\chi_1 = \chi_2$ waveforms and can be used as an \emph{effective single-spin} model~\cite{Purrer:2013ojf} similar to the approach used in phenomenological models for aligned-spin binaries~\cite{Ajith:2011,Santamaria:2010yb,Ajith:2011ec}. The dimensionless spin parameters are defined as $\chi_i = S_i/m_i^2$, where $S_i$ and $m_i$ are the spin and mass of companion (black hole or neutron star) $i$.

For the sake of simplicity, all of the modeling techniques used will be discussed for the single-spin model, but can be extended to the double-spin model in a straightforward manner. I will show that the single-spin model can help identify problems in the source model beyond its calibration range. Thus, reduced order models can be used as a tool that can also be applied to assess waveforms and identify regions in the parameter space where the model should be improved.
The fidelity of the reduced order models is analyzed in terms of the \emph{faithfulness mismatch} against the original SEOBNRv1 model, i.e. I only optimize over time and phase shifts, but not over model parameters.
When speedup and mismatch studies are carried out, the reduced order models are compared against FFTs of SEOBNRv1 waveforms without hybridization.

The outline of this paper is as follows.
I will summarize the SEOBNRv1 waveform model in~\Sref{sec:SEOBNRv1} and give details on how the frequency domain input waveforms are generated in~\Sref{sec:templates}.
I will describe how the SVD can be used to build a reduced basis, compute projection coefficients against a set of input waveforms and give a formal description on how to assemble the surrogate waveform model in~\Sref{sec:build_model}. I will then discuss different choices for interpolating the projection coefficients over the (intrinsic) parameter space in~\Sref{sec:int_par_space} and techniques for generating a suitable set of sparse frequency grid points in~\Sref{sec:sparse_frequency_points}. In~\Sref{sec:errors} I will compare errors arising from approximations introduced in building the model. Model accuracy will be studied for a single-spin model in~\Sref{sec:results-SS} and double-spin model in~\Sref{sec:double-spin-model}. At the end I will compare SVD and greedy basis methods, summarize the performance and storage requirements and address future extensions of the technique in~\Sref{sec:discussion}.


\section{Aligned-spin SVD-based reduced order models in a nutshell}
\label{sec:exsummary}

This section provides an executive summary of the speedup provided by the reduced order models discussed in this paper. It gives an overview of how the models are constructed pointing the reader to section where the various ingredients are discussed in details. At this point the reader is not expected to understand the details of each step. Finally I summarize the fidelity of the single- and double-spin models.

\subsection{The need for fast surrogate models} 
\label{sub:the_need_for_fast_surrogate_models}

\begin{figure}[htbp]
  \centering
    \includegraphics[width=0.47\textwidth]{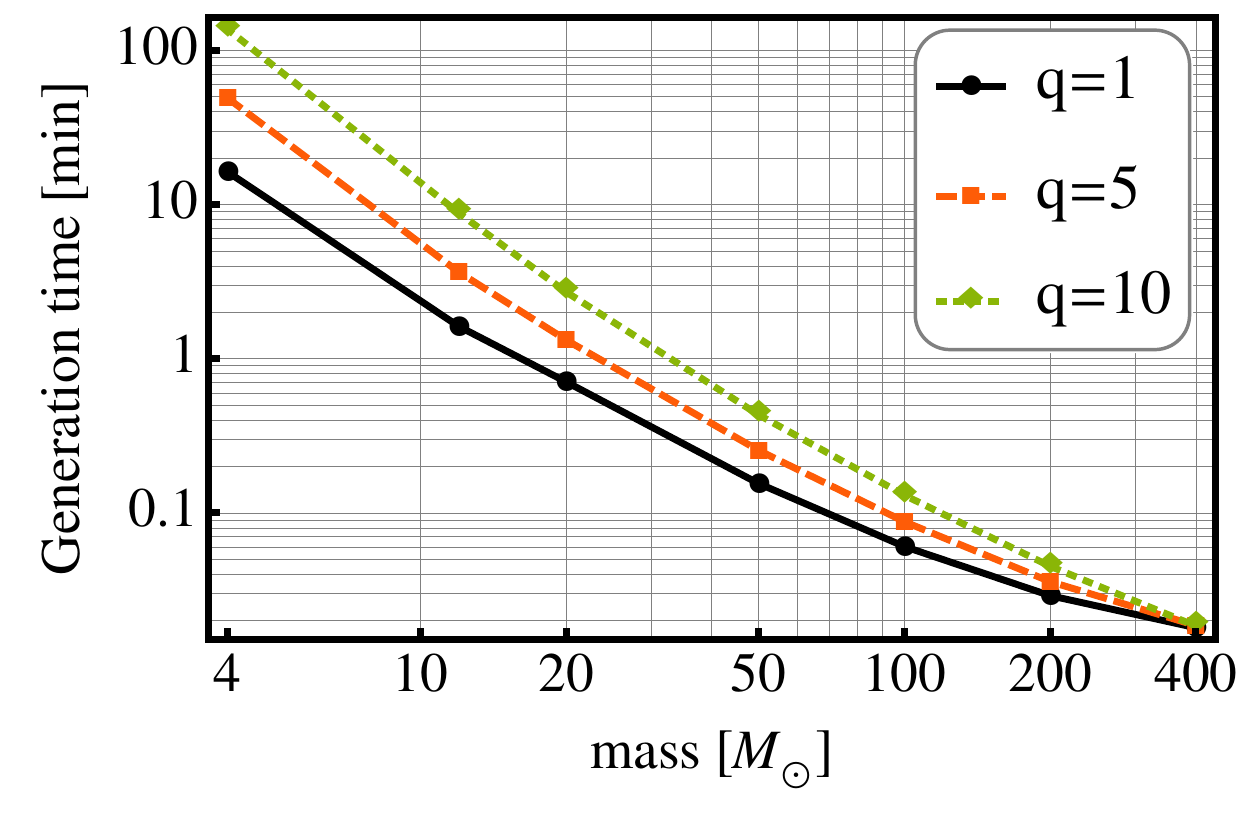}
		\includegraphics[width=0.47\textwidth]{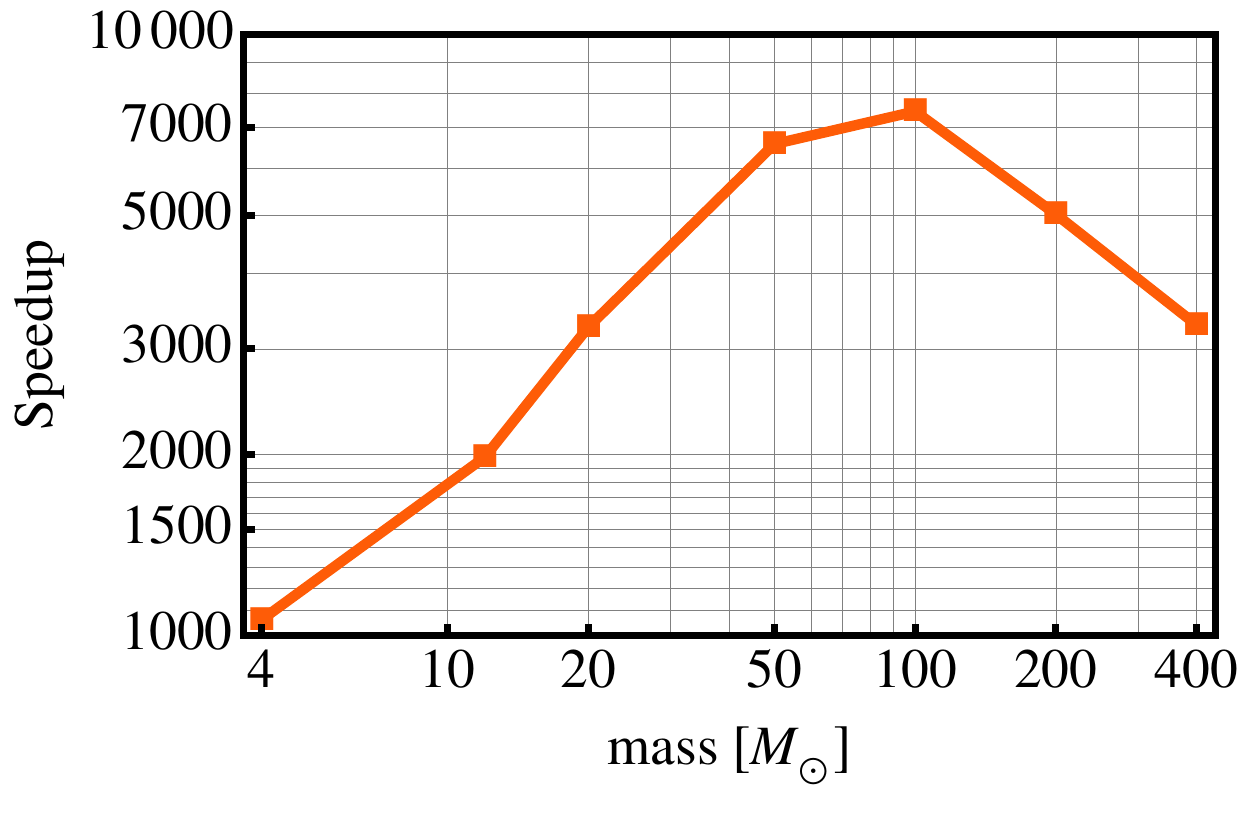}
  \caption{
	The time required for generating a frequency domain SEOBNRv1 waveform as a function of total mass is shown for mass-ratios $q=1,5,10$ and $\chi_i=0$ (left panel). The sampling frequency is $16384 \text{Hz}$ with a LAL starting frequency of $f_\text{Min} = 8 \text{Hz}$. The frequency domain waveforms fill the detector band down to $\sim 11\text{Hz}$.
	In the right panel we display the speedup of a sparse ($\sim 150$ grid points) reduced order model relative to the generation of equal-mass SEOBNRv1 waveforms. 
	In order to fully represent a the FFT of a TD waveform of length $T(M)$ as a function of total mass at the above sample rate $f_\text{max} T (M)$ frequency points are required. This is the number of evaluations used to compute the speedup assuming $f_\text{max} = 0.15/M$.
	The plots show the time for non-spinning configurations; the cost for spinning waveforms is comparable. 
	The waveforms were generated on a $3.07 \text{GHz}$ Intel Xeon using LAL compiled with gcc.
}
  \label{fig:Wf_generation_time_and_speedup}
\end{figure}

I have already pointed out in~\Sref{sec:introduction} that a fast surrogate model of SEOBNRv1 would be extremely useful for PE studies and for the construction of template banks. Let us take a closer look at the time needed for generating SEOBNRv1 waveforms at various total masses and contrast this with the increase in speed obtained from the reduced order models presented in this work.
As the left panel of~\Fref{fig:Wf_generation_time_and_speedup} shows the generation time for SEOBNRv1 waveforms as implemented in the LSC Algorithms Library~\cite{LAL-web} (LAL) becomes very high for low total mass, from tens to hundreds of minutes if the full IMR waveform is needed. The cost also rises with mass-ratio and with sampling frequency.
The evaluation speed for the surrogate models presented in this paper per frequency point is independent of the system mass, but depends on the dimensionality of the parameter space, the number of sparse frequency gridpoints used in the construction of the model (as defined in~\Sref{sec:sparse_frequency_points}) and the dimension of the basis matrices. 
The speedup of the ROM vs SEOBNRv1 is summarized in the right panel of~\Fref{fig:Wf_generation_time_and_speedup} for equal-mass waveforms. The plot shows timings for a C implementation of the ROM; the Mathematica code is about a factor of $30$ slower.

To fully represent the FFT of a time-domain waveform of length $T$ (with power only in the positive frequencies), $T / (2\Delta t) = f_\text{max} T$ frequency points are required and $\Delta t$ is inversely proportional to the sampling frequency. This number of points rises to the order of millions at low total masses and the dominating cost for such large numbers of model evaluations is spline interpolation in frequency for the models discussed in this paper. For this use case a speedup of a factor of 1000 or more can be obtained in the ROM compared to SEOBNRv1. For applications in searches and parameter estimation involving detector data the full number of frequency points according to the used sample rate is needed.
However, when comparing reduced order models against SEOBNRv1 in this paper it turns out that a frequency spacing of $\sim 0.5 \text{Hz}$ for the aLIGO frequency band (corresponding to about 10000 frequency evaluations) already yields sufficiently accurate matches\footnote{In the computation of the match (see~\Eref{eq:match}) one maximizes over time-shifts between two waveforms to minimize the dephasing. If the peak of the correlation function is well enough resolved the match will be accurate. Zero-padding the match integrand is also very beneficial.}.

This massive reduction in the computational cost makes possible detailed parameter estimation studies with spinning EOB models and will be a very valuable tool in the advent of the advanced detector era.
Assuming a cost of about $15$ minutes per SEOBNRv1 waveform in the BNS region and a PE study requiring $10^7$ (serial) waveform evaluations, the total time of the analysis would be around 280 years. Using a C implementation of the reduced order model such a study could be completed in several months using the full set of frequency points.
Further speedup is possible through the use of reduced order quadratures for fast likelihood evaluations~\cite{Canizares:2013ywa,Canizares:2014fya,Antil:2013}.

\subsection{A blueprint for building aligned-spin reduced order models} 
\label{sub:a_blueprint_for_building_aligned_spin_reduced_order_models}

In this section I list all required steps that I use for building reduced order models of SEOBNRv1 as discussed in this study. I simply list the choices I make; details, justifications and alternative methods are discussed in the sections I refer to.

\begin{enumerate}
	\item Generate a set of $n$ input waveforms that cover the multi-dimensional parameter space domain of interest, as densely as desired. 
	\begin{enumerate}
		\item Choose starting frequency, sampling rate and total mass so that the fast Fourier Transforms (FFTs) of the time-domain (TD) waveforms cover a suitable  frequency range (see~\Sref{sec:templates}).
		\item Generate waveforms on a regular grid 
		over the multi-dimensional parameter space (see~\Eref{eq:regular_parameter_space_grid}).
		\item Normalize the amplitudes and subtract a linear frequency fit from the phases.
	\end{enumerate}
	\item Define frequency grids separately for amplitudes and phases.
	\begin{enumerate}
		\item Find $m$ ($\sim 100 - 200$) sparse amplitude and frequency points (see~\Sref{sec:sparse_frequency_points}).
		\item The preferred prescription to generate the frequency points keeps the error of cubic spline interpolation $I_f[\cdot]$  approximately constant in frequency (see~\Sref{sec:const_spline_int_error_points}).
		\item Interpolate input amplitudes and phases onto the sparse frequency grids.
	\end{enumerate}
	\item Compute \emph{reduced bases} for the amplitudes and phases with the SVD (see~\Sref{sec:build_model}).
	\begin{enumerate}
		\item Pack the input amplitudes and phases into the columns of matrices $\mathcal{T}_\mathcal{A}, \mathcal{T}_\Phi \in \mathbb{R}^{m \times n}$.
		\item Take the SVD (see~\Eref{eq:SVD})
		of the input amplitude and phase matrices. 
		\item Obtain orthonormal amplitude $\mathcal{B}_\mathcal{A}$ and phase bases $\mathcal{B}_\Phi$ and a list of decreasing \emph{singular values} $\sigma_{\mathcal{A},i}, \sigma_{\Phi,i}$ indicating the relative importance of the SVD-modes. 
	\end{enumerate}
	\item Interpolation over the parameter space.
	\begin{enumerate}
		\item Calculate projection coefficients $\mathcal{M}$ (see~\Eref{eq:projection_coefficients_M}) of all input waveforms in terms of the reduced bases.
		\item Unpack (see~\Eref{eq:projection_coefficients_M_partitioned}) and interpolate the projection coefficients over the parameter space using tensor product spline interpolation  $I_\otimes[\mathcal{M}]$ (see~\Sref{sec:int_par_space}). 
		\item Interpolate the amplitude normalization factors over the parameter space (see~\Eref{eq:amplitude_normalization_prefactor}). 
	\end{enumerate}
	\item Assemble the frequency domain surrogate model (see~\Eref{eq:model_function})
\begin{empheq}[box=\myblueboxL]{equation*}
	\mkern-28mu \tilde h_\text{m}(q,\chi; M; Mf) := A_0(q,\chi,M) \,
		I_f\left[ \mathcal{B}_\mathcal{A} \cdot I_\otimes[\mathcal{M}_\mathcal{A}](q,\chi) \right](Mf) \,
		\exp\left\{ i \, I_f\left[ \mathcal{B}_\Phi \cdot I_\otimes[\mathcal{M}_\Phi](q,\chi) \right](Mf) \right\}
\end{empheq}
\end{enumerate}


\subsection{Summary of main results} 
\label{sub:summary_of_main_results}

The fidelity of the single- and double-spin models I build in this work in terms of the \emph{faithfulness} mismatch against the SEOBNRv1 model using the ``zero-detuned high-power'' aLIGO noisecurve~\cite{T0900288} is summarized as follows:
\begin{itemize}
	\item \textbf{Single-spin model} ($q \in [1,100], \, \chi \in [-1,0.6]$) for $M \geq 1.35 (1+q) M_\odot$ and $f_\text{min}=15 \text{Hz}$:\\
	$\text{Mismatch} < 0.2\%$ at $m_1 = 1.4 M_\odot$\\
	$\text{Mismatch} < 0.3\%$ at $M \geq 100 M_\odot$ except for isolated configurations far outside the calibration range $q\leq 6$ of SEOBNRv1.
	See a contour plot of the mismatch~\Fref{fig:Faithfulness-sparse-ROM-model-contours} over the parameter space and a histogram~\Fref{fig:Faithfulness-sparse-ROM-model-histogram-CSE} in~\Sref{sec:results-SS} for details.
	Amplitude and phase errors are shown in~\Fref{fig:Phase-error-sparse-ROM-model-histogram-CSE}.
	A time-domain comparison is given in~\Fref{fig:TD-comparison}.
	\item \textbf{Double-spin model} ($q \in [1,10], \, \chi_1,\chi_2 \in [-1,0.6]$) for $M \geq 12 M_\odot$ and $f_\text{min}=11 \text{Hz}$:\\
	$\text{Mismatch} < 0.1\%$  for $M > 100 M_\odot$ \\
	$\text{Mismatch} < 0.01\%$ for $M \lesssim 50 M_\odot$\\
	The total faithfulness mismatch is shown in~\Fref{fig:Faithfulness-full-DS-histogram} of~\Sref{sec:double-spin-model}.
	~\Fref{fig:Faithfulness-modelwfs-DS-histogram} and ~\Fref{fig:Faithfulness-modelwfs-CSE-DS-histogram} explore individual error sources that arise in building reduced order models.
	The faithfulness mismatch in the reduced order model is much smaller than the $0.5\%$ uncertainty of SEOBNRv1 (see~\Sref{sec:SEOBNRv1}).
\end{itemize}



\section{The SEOBNRv1 waveform model}
\label{sec:SEOBNRv1}

The \emph{effective-one-body} (EOB) formalism as introduced by Buonanno and Damour~\cite{Buonanno:1998gg,Buonanno:2000ef,Damour:2000we,Damour:2001tu,Damour:2008gu} is an analytical approach that combines PN and perturbation theory with re-summation techniques in order to go beyond the inspiral and also accurately model the plunge, merger and ringdown signal.
The EOB approach incorporates nonperturbative and strong-field effects that are lost when the dynamics and the waveforms are Taylor-expanded as PN series. 
The conservative dynamics are mapped to the motion for a test particle in an effective background metric, a Schwarzschild or Kerr metric deformed by the symmetric mass-ratio.
Several predictions of the EOB approach, notably the simplicity of the merger signal for non-spinning~\cite{Buonanno:2000ef} and spinning, precessing black holes~\cite{Buonanno:2005xu}, have been 
confirmed by the results of NR simulations.  
Over the years EOB waveforms have been improved~\cite{Damour:2009kr,Pan:2011gk} and calibrated to progressively more accurate NR waveforms and have led to the development of aligned-spin EOB models~\cite{Taracchini:2012ig,Taracchini:2013rva}.

The SEOBNRv1 model~\cite{Taracchini:2012ig} is an effective-one-body (EOB) model of the dominant $(l,m)=(2,2)$ gravitational-wave mode for BH binaries with non-precessing spins. The model is tunable for arbitrary mass-ratio and aligned spins and has been calibrated to five non-spinning  numerical relativity waveforms of mass-ratios $q=1,2,3,4,6$\footnote{We define the mass-ratio as $q = m_1/m_2 \geq 1$.} and two equal-mass spinning waveforms with equal spins $\chi_i \simeq \pm 0.44$.
To compute fits of non-quasicircular coefficients (NQC) over the parameter space, equal-mass highly spinning NR waveforms $\chi_i \simeq -0.95, +0.97$ and test-particle limit results from Teukolsky waveforms were used in addition to the five non-spinning waveforms. 
The calibrated model has overlaps larger than $0.997$ with each of the seven NR waveforms for total masses between $20 M_\odot$ and $200 M_\odot$. These overlaps are computed with the ``zero-detuned high-power'' aLIGO noisecurve~\cite{T0900288} while maximizing only over initial phase and time. Overlaps against an equal-mass highly anti-aligned spinning NR waveform with $\chi_i \simeq -0.95$ which was not used for calibration are larger than $0.995$ for the same range of total masses as above. The dephasing grows up to about 2 rads during the ringdown, while the relative amplitude difference grows up to about $40\%$~\cite{Taracchini:2012ig}.
For highly spinning aligned-spin waveforms the EOB $(2,2)$ mode peaks too early in the orbital evolution where non-quasicircular orbital effects are still negligible which causes the iterative computation of non-quasicircular coefficients to diverge. 
Beyond the EOB ISCO the quasicircular inspiral is followed by the plunge where NQC corrections become important and the dynamics become increasingly relativistic.
Compared to its non-spinning location the EOB ISCO moves to smaller radial separation for aligned spins and larger radial separation for anti-aligned spins. Therefore, the EOB model is expected to perform better for anti-aligned configurations.
The model requires improvement beyond $\chi_i \gtrsim 0.7$. In fact, the LAL~\cite{LAL-web} implementation of SEOBNRv1 is restricted to spins $\chi_i \leq 0.6$.
These shortcomings are addressed in the recent followup model~\cite{Taracchini:2013rva}.

The SEOBNRv1 model has been shown to be quite accurate in the NRAR project~\cite{Hinder:2013oqa} where it compared very well against a non-spinning NR waveform at mass-ratio 10 with a mismatch below $0.4\%$.
The model is not without problems, however, especially if the mass-ratio or the spins are unequal.
Configurations 4 ($q=4, \chi_1 = 0.6, \chi_2 = 0.4$) and 8 ($q=2, \chi_1=\chi_2=0.6$) of the NRAR study showed an unfaithfulness larger than $3\%$, especially at high total mass when the merger is prominent in the detector band. In turned out that for these cases the amplitudes and frequencies of the time-domain SEOBNRv1 waveforms have artificial oscillations around merger that are significant in the frequency domain and lead to the increased mismatch. This is due to limited NR data available for the fits of the NQC over the parameter space.

In this work I go beyond the calibration range and build a single spin reduced order model up to $q=100$ using equal-spin $\chi_1=\chi_2$ waveforms and a double spin model up to $q=10$ while covering the full available spin range $-1 \leq \chi_i \leq 0.6$ for both models. It is then not surprising that one may find unexpected behavior or artifacts in certain regions of the parameter space beyond $q=10$. As discussed in~\Sref{sec:discussion} the variation of the model waveforms on mass-ratio and spin is not found to be uniformly smooth over the parameter space.
This is not intended as a critique of the model, but as a demonstration of the capabilities of reduced order modeling methods and their efficacy in pointing out regions where source models should be improved. Thus, ROM should become a very useful tool to assess and review the quality of waveform models.

The followup model~\cite{Taracchini:2013rva} to SEOBNRv1 has been calibrated to 8 non-spinning and 30 spinning nonprecessing NR waveforms up to mass-ratio $8$ produced by the SXS collaboration~\cite{SXS_web} with mismatches smaller than $1\%$ against the calibration waveforms. It removes the restriction on high aligned spins and strengthens the fits of the NQC due to a much larger number of NR waveforms. At the time of writing the LAL code of this follow-up model was still under development. Therefore, the ROM models constructed in this work are based on the SEOBNRv1 model~\cite{Taracchini:2012ig}.


\section{Production of frequency domain waveforms}
\label{sec:templates}

This section discusses the choice of starting waveforms and the production and storage of frequency domain waveforms from the time-domain SEOBNRv1 model on a set of common frequency grid points. I follow standard practice, but for the sake of completeness give details.

\begin{figure}[htbp]
	\centering
		\includegraphics[width=0.8\textwidth]{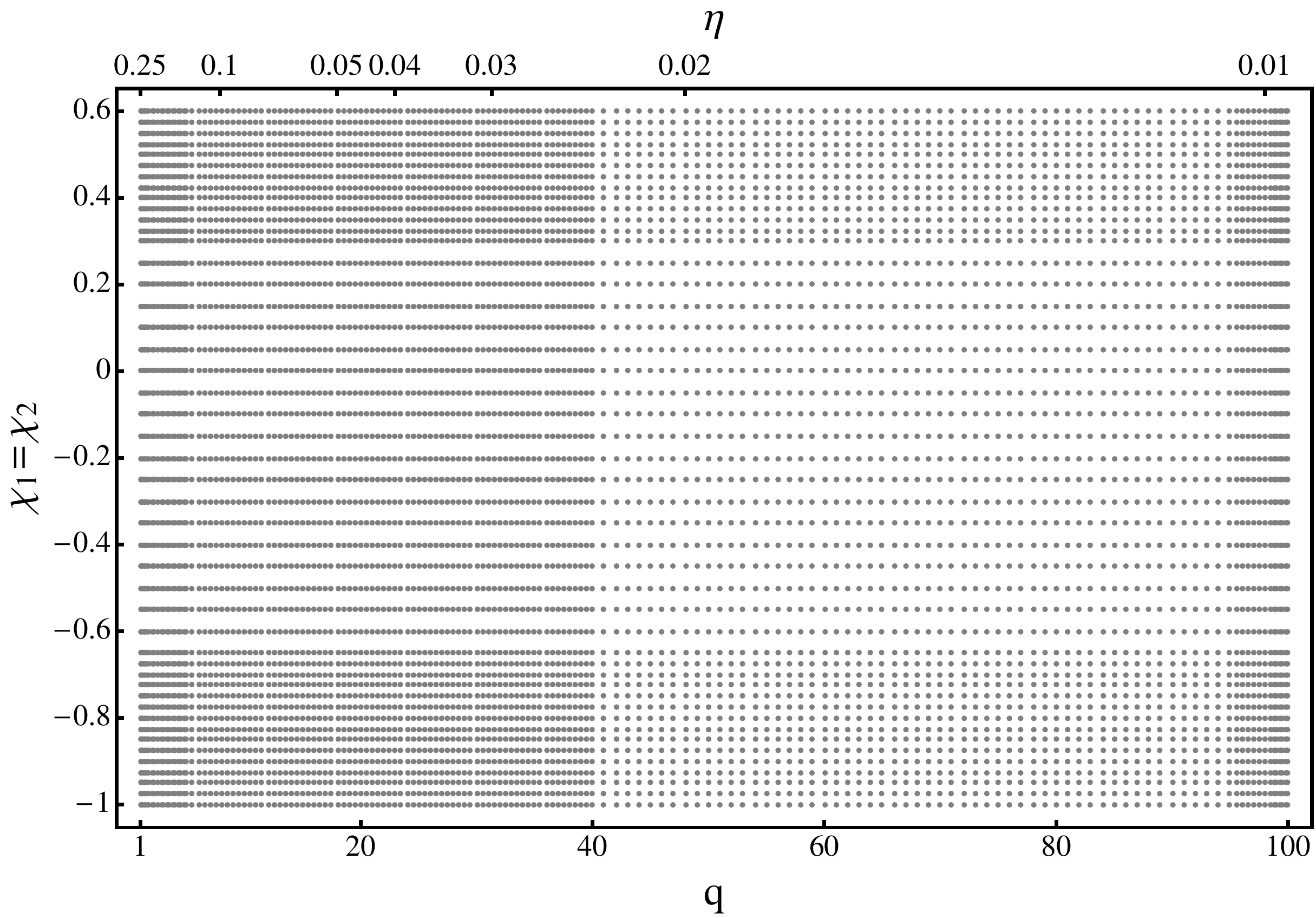}
	\caption{The $157 \times 47 = 7379$ SEOBNRv1 starting waveforms used to construct the single-spin model.}
	\label{fig:plots_Templates_aligned_q1_q100}
\end{figure}

I generate a set of (frequency domain) waveforms $\mathcal{W}$ (separated into their amplitudes $\mathcal{A}$ and phases $\Phi$). These waveforms are to be generated at locations lying in a regular parameter space grid 
\begin{equation}
	\label{eq:regular_parameter_space_grid}
	\Lambda := \mathcal{Q} \times \mathcal{X},
\end{equation}
where $\mathcal{Q}$ and $\mathcal{X}$ are one-dimensional sets covering the desired parameter range.
For the equal-spin case I choose $\mathcal{Q}$ to cover the interval $[1,100]$ in mass-ratio and $\mathcal{X}$ to cover the available extent in spins $[-1,0.6]$. The spacing in the mass-ratio and the spin is chosen in an uneven way to concentrate more points near the boundaries of the domain and at lower mass-ratios. \Fref{fig:plots_Templates_aligned_q1_q100} shows how the parameter space was covered.

Next, I discuss how the frequency domain waveforms were calculated. The time-domain SEOBNRv1 waveforms $h = h_+ - i h_\times$ are sampled at a rate of $f_s = 16384 Hz$. In geometric units  ($G=c=1$) this yields a time spacing of 
\begin{equation}
	\Delta t/M = \frac{1}{M f_s}.
\end{equation}
The total mass $M$ is expressed in solar masses and $M_\odot[s] = G/c^3 M_\odot[kg] \approx 4.93 \times 10^{-6} s$ is the geometrized solar mass in seconds.
At $M=4M_\odot$ we get $\Delta t \sim 3M$ which corresponds to a Nyquist frequency of $f_\text{Ny} = 1/(2\Delta t) \sim 0.17 / M$.

At this time resolution, an equal-mass waveform long enough for a $2 \times 1.35 M_\odot$ BNS system down to $f_\text{low} = 15 Hz$ obtained from the LAL SEOBNRv1 code is given on about $N \sim 15 \times 10^6$ points and is about $T = (N-1) \Delta t \sim 45 \times 10^6 M$ long in time.~\footnote{
I compute waveforms at $4M_\odot$ (for equal-mass), but lower the starting frequency to $8 \,$ Hz to obtain waveforms of sufficient length for BNS configurations, while keeping the full merger ringdown part.) 
}
The spacing in the frequency domain is
\begin{equation}
	\Delta f = \frac{2 f_\text{Ny}}{N}.
\end{equation}
Combining this with the definition of the Nyquist frequency $f_\text{Ny} = 1/(2\Delta t)$ and the relation $\Delta t = T / (N-1)$ we have
\begin{equation}
	\Delta f = \frac{N-1}{N} \frac{1}{T} \sim 1/T.
\end{equation}
With the above example, $\Delta f \sim 2 \times 10^{-8} / M$. 

We taper the waveform with a Planck window~\cite{McKechan:2010kp} at both ends and zero-pad it by doubling the length of the waveform. Then we perform the FFT of the time-domain waveform and obtain $N$ data points at positive frequency. After separating the frequency domain waveform into its amplitude and phase part and unwrapping the phase we only need to keep a fraction of the number of points. We save about 500000 points to data files. 
To make the computation of the input waveforms more efficient, one only needs to make them long enough such that the mass of the smaller companion is at least that of a chosen physical cutoff, say $1.35 M_\odot$. However, all waveforms need to be represented on the same grid for the SVD to make sense. Therefore, I hybridize the waveforms with TaylorF2 to make them of sufficient length in frequency. This is described in~\ref{sec:hybrids}.

The discrete amplitude and phase of each waveform in $\mathcal{W}$ are treated separately. All waveforms are interpolated onto a frequency grid $\mathcal{G}$. A simple choice is an equidistant grid with $m = 10000$ gridpoints in the frequency interval $Mf \in [0.0001, 0.14]$. This is motivated by the starting frequency in geometric units for a BNS system at $10 \text{Hz}$ lower cutoff is $Mf = 10 \text{Hz} \times (2\times 1.35) M_\odot[s] \sim 0.00013$.
The interpolation error associated with this choice and the generation of more efficient sparse frequency grids are discussed in~\Sref{sec:sparse_frequency_points}.
The amplitudes are normalized using the norm induced by the discrete inner product 
$\left( \tilde h_1, \tilde h_2 \right) := \text{Re} \sum_i (\tilde h_1)_i (\tilde h_2)_i^* / m$.
The waveforms are generated at a distance of $1\,$ Mpc with ``face on'' inclination.
In~\Sref{sec:build_model} I will explain how the correct amplitude prefactor is restored.
I subtract a linear fit from each of the hybrid phases in order to approximately factor out two degrees of freedom that have no physical significance. These correspond to time and phase shifts one maximizes over in the computation of the match between two waveforms.


\section{Sparse frequency representation of waveforms}
\label{sec:sparse_frequency_points}

The goal of this section is to find a sparse set of frequency points on which one can represent the amplitude and phase functions of a frequency domain waveform to sufficient accuracy by using interpolation methods. I an algorithm based on the form of the spline interpolation error that can be used to obtain such a set of points and how it performs in terms of a parameter maximized inner product between (normalized) waveforms, the \emph{overlap} or \emph{match}, against the starting waveforms given on a fine equispaced grid.

A natural way of comparing two waveforms $h_1$ and $h_2$ is the scalar product
\begin{equation}
	\langle h_1,h_2 \rangle = 4 \text{Re} \int_{f_\text{min}}^{f_\text{max}} \frac{\tilde h_1(f) \tilde h_2(f)^*}{S_n(f)} df,
\end{equation}
where $S_n(f)$ is the one-sided power-spectral density of the detector noise.
In this paper I use the expected design sensitivity of the Advanced LIGO detector~\cite{Abbott:2007kv,Shoemaker:aLIGO,2010CQGra..27h4006H}. The anticipated optimum sensitivity is given by the ``zero-detuned high-power'' noise curve~\cite{T0900288}. We use a linear interpolation of this expected PSD and choose $f_{\rm min} = 15$\,Hz, and $f_{\rm max} = 8$\,kHz for the single spin model. For the double spin model the waveforms are generated for $12 M_\odot$ and $f_{\rm min} = 11$\,Hz.
$\tilde h(f) = \mathcal{F}[h(t)]$ denotes the Fourier transform of a time-domain waveform $h(t)$.
The scalar product induces a norm $\lVert h \rVert = \sqrt{\langle h, h \rangle}$.
We maximize over time and phase shifts between the two waveforms and define the \emph{overlap} (or \emph{match}) as
\begin{equation}
	\label{eq:match}
	\mathcal{O}(h_1, h_2) = \frac{4}{\lVert h_1 \rVert \, \lVert h_2 \rVert} \max_{t_0} 
	\bigg\lvert \mathcal{F}^{-1} \left[\frac{\tilde h_1(f) \tilde h_2(f)^*}{S_n(f)} \right](t_0) \bigg\rvert,
\end{equation}
where $\mathcal{F}^{-1}[g(f)] = \int_{-\infty}^\infty g(f) e^{-2\pi i f t} df$ is the inverse Fourier transform. We often quote the \emph{mismatch} $1 - O(h_1, h_2)$ instead.

Since the amplitude and the phase are typically fairly simple, smooth functions it should be possible to represent them as an interpolant on a sparse frequency grid. A priori it is not clear how to choose such a grid in an optimal way so that the errors in the computation of the overlap introduced by this approximation are below a desired threshold. We expect the interpolation error in the waveform phase to be the dominant source of error due to the way it appears in the overlap integral
\begin{equation}
	\tilde h \; \mathcal{I}[\tilde h]^* = A(f) \mathcal{I}[A](f) \; \exp\left(i (\phi(f) - \mathcal{I}[\phi](f))\right).
\end{equation}
Accumulating a significant dephasing over many GW cycles will cause the match to degrade more than a comparable level of error in the amplitude will. 

Ideally it would be best to directly generate a grid based on minimizing the error in the overlap, i.e. the mismatch between the waveform on the equispaced grid and on the sparse grid. However, in practice pointwise error estimates are needed to find a grid without making assumptions on how the gridspacing should vary as a function of frequency. Starting iteratively from two endpoints, it is possible to compute the overlap of the spline interpolation of a current grid plus a new grid point taken from the base equispaced grid with the waveform on the full equispaced grid and chose to add the grid point that leads to the highest overlap. However, the overlap as a function of the phase error tends to threshold quickly and such a method turned out not to be well conditioned.

In~\Sref{sec:const_spline_int_error_points} I discuss a method that chooses the gridspacing as a function of frequency such that the local spline interpolation error stays roughly constant. I interpolate amplitudes and phases using cubic splines over the set of chosen frequency points and start with a grid that only consists of the start and end frequency on which the waveforms are given.
An alternative greedy method with the idea to successively add grid points at which the relative or absolute errors are highest is outlined in~\Sref{sec:greedy_frequency_points}. In practise I found the spline error method described in~\Sref{sec:const_spline_int_error_points} to be more accurate and easier to use and I therefore chose it for the models described in this study.

The goal is to generate the smallest set of frequency points required for a given accuracy of the model. In practice one to two hundred points were sufficient for the models considered here. This leads to a substantial compression which makes the final model quite compact and also speeds up the generation of the model significantly since the rank of the basis matrices is then only a few hundred and the computation of the SVD of sparsely sampled waveform matrices takes on the order of a second which makes any performance advantage of the greedy basis algorithm over the SVD a moot point for this application.


\subsection{Constant spline interpolation error (CSE) points}
\label{sec:const_spline_int_error_points}

The idea for this method is guided by the form of the spline interpolation error.
The error of a cubic spline interpolant $I_4[g](x)$ of a function $g(x) \in C^4[a,b]$ with derivatives given at the endpoints on a set of nodes $x_i$ with the largest gridspacing $\Delta :=\max_i \lvert x_{i+1} - x_i \rvert$ is~\cite{DeuflhardHohmannTAM43}
\begin{equation}
	\label{eq:cubic_spline_error}
	\lVert g - I_4[g] \rVert_\infty \leq \frac{5}{384} \Delta^4 \lVert g^{(4)} \rVert_\infty,
\end{equation}
where $g^{(4)}$ is the fourth derivative of $g(x)$. 
This bound makes it clear that in order to find a set of ``good'' nodes the functional form $g(x)$ to be approximated needs to be taken into account. In practice I use cubic splines with a ``not-a-knot'' condition~\cite{deBoor} at the endpoints rather than specifying the actual derivatives. This is the recommended condition for closing the system of equations if only the function values are known (i.e. in the absence of derivative information). The interpolation error of the ``not-a-knot'' spline is of the same order as the error given in~\Eref{eq:cubic_spline_error} and the precise error constant is not important in the following.

Before we think about generating sparse grids let us consider the error associated with the representation of the input waveforms on an equidistant grid. As described in~\Sref{sec:templates} the equal-spin input waveforms $\mathcal{W}$ we proposed to interpolate the amplitudes and phases onto a 10000 point equally-spaced frequency grid
\begin{equation}
	\begin{split}
		G = \{f_\text{min} + i \, \Delta f\}, \quad \text{where} \quad i = 0, \dots, m-1, \quad m=10000\\
		\Delta f = \frac{f_\text{max} - f_\text{min}}{m-1}, \quad f_\text{min} = 0.0001/M, \quad f_\text{max} = 0.14/M.	
	\end{split}
\end{equation}
Using this grid, the maximum relative error in the amplitude is about $0.05\%$ and the absolute error in the phase is less than $0.01 \text{rad}$ for a $5 M_\odot$ equal-mass waveform starting at $15 \text{Hz}$. In both cases the error is largest at the low frequency end and decreases very rapidly as the frequency increases. 
This grid is actually not sufficiently fine to guarantee very small errors down to the BNS regime. There the mismatch due to interpolation error is about $0.1\%$. 
To push the phase error below $0.01 \text{rad}$ for a BNS starting at $10 \text{Hz}$ would require about $50000$ equispaced points.
The sparse grids defined below are much more compact and also more accurate at low frequencies.

To generate a sparse frequency grid we demand that the gridspacing as a function of frequency is chosen such that the \emph{local} spline interpolation error (i.e. we apply~\Eref{eq:cubic_spline_error} for every subset of 4 grid points) stays constant over the desired frequency range.  Going back to \Eref{eq:cubic_spline_error} we see that the gridspacing as a function of frequency has to be proportional to the reciprocal of the fourth root of the fourth derivative of the function to be approximated so that the error stays constant (if the equality were to hold).
Given an appropriately chosen gridspacing function $\Delta(f)$ a grid can then be generated in a very straightforward manner where the spline interpolation error stays approximately constant. This is outlined in Algorithm~\ref{alg:const-spline-error-points}.

Now I give examples on how to choose the gridspacing function:
Assuming a simplified amplitude $A \simeq f^{-7/6}$ we find $\Delta_A(f) \propto f^{31/24} \approx f^{1.3}$. In practice a linear dependence on frequency serves better to suppress amplitude error near the merger.
For the phase we look at the dominant terms in the phase evolution of the TaylorF2 approximant for an equal-mass non-spinning binary:
$\phi_\text{F2} \simeq -\frac{\pi}{4} + 0.2 (Mf)^{-1} + 0.014 (Mf)^{-5/3} + \bigO((Mf)^{-2/3})$. The mass-ratio changes only the prefactors and the spins enter at higher PN order. With the help of a Taylor expansion we then find $\Delta_\phi(f) \propto f^{17/12} \approx f^{1.4}$. 
In practice $\Delta_\phi(f) \propto f^{4/3} $ works well to keep the phase error balanced. The desired error threshold and, at the same time, the number of grid points can be set by adjusting the proportionality factor in $\Delta(f)$.

For the single spin model I choose $\Delta_A(f) = 0.1f$ and $\Delta_\phi(f) = 0.304 f^{4/3}$.
Starting at $f_\text{min}=0.0001/M$ application of algorithm~\ref{alg:const-spline-error-points} yields 78 amplitude points and 200 phase frequency points.
For the double spin model the waveforms are generated at a total mass of $12 M_\odot$ with an initial frequency of starting frequency of $8 \text{Hz}$ specified in LAL. It turns out that the lowest useable frequency is $f_\text{min} \eqsim 0.00062/M$ which is sufficient to fill the detector band down to $11Hz$.\footnote{Since this is a proof of principle paper, I choose to work with these double-spin waveforms; for practical models I intend to push this down to the $10 \text{Hz}$ cutoff demanded by aLIGO.}
There I choose $\Delta_A(f) = 0.06f$ and $\Delta_\phi(f) = 0.25 f^{4/3}$. This results in 95 amplitude points and 123 phase points. The number of points is less than for the single-spin model due to the higher total mass and starting frequency of the grid. These choices correspond to a maximum frequency domain amplitude error $\Delta A/A \sim 1\%$ and phase error of $\Delta\phi \sim 0.008 \, \text{rad}$.

\begin{figure}[htbp]
	\centering
		\includegraphics[width=0.6\textwidth]{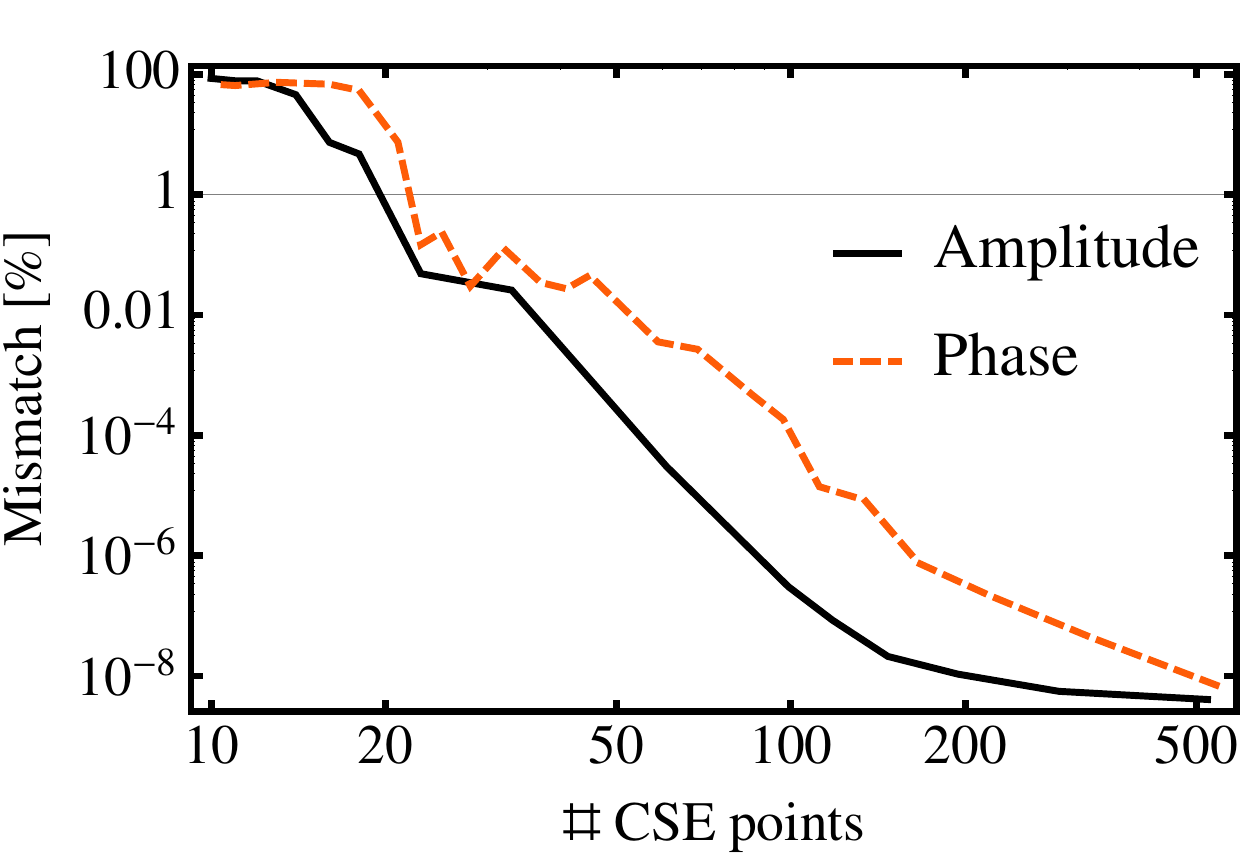}
	\caption{Mismatch error from frequency interpolation against frequency points for amplitude and phase for a $q=10$ non-spinning waveform at $400M_\odot$. The error for the CSE points falls off fairly smoothly and reaches very small mismatches for a few hundred points. As discussed in the text we choose $\Delta_\phi(f) \propto f$ and $\Delta_\phi(f) \propto f^{4/3}$ while varying the proportionality constant.
	}
	\label{fig:Const-spline-error-points-matches-DS}
\end{figure}

With this method it is very simple to adjust the resolution desired at low and high frequencies by changing the exponent of $f$ in the gridspacing function, so that one can guarantee good accuracy for a wide range of total masses.
CSE points are by definition fairly evenly distributed in frequency and it is very fast and simple to generate them. 
\Fref{fig:Const-spline-error-points-matches-DS} shows how the mismatch decreases with the number of CSE frequency points. On the order of hundreds of points the interpolation error is already negligible for CSE points. The 100 to 200 points  chosen above lead to both leading to a mismatch of about $10^{-4} \%$ or less.

\begin{algorithm}
	\caption{Constant spline interpolation error frequency points.}
	\label{alg:const-spline-error-points}
	\begin{algorithmic}
		\State $\mathcal{G}_1 \gets f_\text{min}$
		\State $i \gets 1$
		\While {$\mathcal{G}_i \leq f_\text{max}$}
			\State $P \gets \mathcal{G}_i + \Delta(\mathcal{G}_i)$ \Comment{Compute next point from gridspacing function.}
			\If{$P \leq f_\text{max}$} 
				\State $\mathcal{G}_{i+1} \gets P$ \Comment{Add point.}
			\Else
				\State $\mathcal{G}_{i+1} \gets f_\text{max}$ \Comment{Complete grid with endpoint.}
			\EndIf
			\State $i \gets i+1$
		\EndWhile
	\end{algorithmic}	
\end{algorithm}

\section{Building SVD-based reduced order models}
\label{sec:build_model}

This section deals with the singular value decomposition (SVD) and discusses some of its known properties as given in standard texts~\cite{GolubVanLoan,Demmel} that are especially useful for reduced order modeling. We proceed in three steps: First, the SVD is used to obtain a compressed reduced basis representation of a collection of input waveforms. Second, projection coefficients of the input waveforms are calculated in terms of the reduced basis. Finally, a reduced order waveform model is assembled from the reduced basis and projection coefficients interpolated over the parameter space.
Such a strategy has been pursued by Cannon et al for building time-domain reduced order models and working directly with the waveform strain~\cite{Cannon:2010qh,Cannon:2011xk,Cannon:2011rj}.

The preprocessed amplitudes and phases are collected in the columns of separate waveform (or template) matrices $\mathcal{T}_\mathcal{A}, \mathcal{T}_\Phi \in \mathbb{R}^{m \times n}$, 
\begin{equation}
	\mathcal{T} = \left[ \tau_1 | \cdots | \tau_n \right] \in \mathbb{R}^{m \times n}
\end{equation}
where $n$ is the total number of input waveforms, each waveform $\tau_i$ is given on a common grid of length $m$ and $\mathbb{R}^{m \times n}$ denotes the set of $m \times n$ matrices with entries in $\mathbb{R}$.
We may choose to represent the waveforms at a large number of frequency points so that $m \gtrsim n$. On the other hand, if the number of grid points has been reduced to a sparse set (see section \ref{sec:sparse_frequency_points}) then we usually have $m \ll n$.

Since we are dealing with a two- (or later when including both spins with a three-) dimensional parameter space we need to define a mapping that packs the waveforms on the rectangular parameter space grid $\mathcal{Q} \times \mathcal{X}$ into a single vector. First, we label a starting waveform at a specific mass-ratio and spin by the two-dimensional index pair $(i_q, i_\chi)$, where $i_q = 1,\dots, n_q$, and $i_\chi = 1,\dots, n_\chi$. We then define a flat index $i := (i_q - 1) n_\chi + i_\chi$ which ranges from $ i = 1,\dots,n_q n_\chi$. We place the waveforms into $\mathcal{T}$ successively as enumerated by this index. The two dimensional indices can be recovered as $i_\chi = i \mod n_\chi$ and $i_q = 1 + \lfloor i/n_\chi \rfloor$.

We perform a singular value decomposition (SVD)~\cite{GolubVanLoan,Demmel} of the waveform matrices
\begin{equation}
	\label{eq:SVD}
	\mathcal{T} = V \Sigma U^T,
\end{equation}
where $V = \left[ v_1 | \cdots | v_m \right] \in \mathbb{R}^{m \times m}$ and 
$U = \left[ u_1 | \cdots | u_n \right] \in \mathbb{R}^{n \times n}$ are orthogonal matrices.
$\Sigma = \diag(\sigma_1,\dots,\sigma_p) \in \mathbb{R}^{m \times n}$ is a diagonal matrix made up from the \emph{singular values} $\sigma_1 \geq \sigma_2 \geq \dots \geq \sigma_p \geq 0$, and $p = \min(m,n)$.
The $\sigma_i^2$ are the eigenvalues of $\mathcal{T}^T \mathcal{T}$.
The $v_i$ and $u_i$ are the \emph{left} and \emph{right singular vectors} of $\mathcal{T}$, respectively.
If $\mathcal{T}$ has $r$ positive singular values then $\rank(\mathcal{T}) = r$.

To obtain a reduced basis we can calculate a \emph{truncated SVD} of the waveform matrices where we produce a rank-reduced approximation for a $k < \rank(\mathcal{T})$ with the k-th partial sum of the \emph{outer-product expansion} of the input matrix $\mathcal{T}$
\begin{equation}
	\label{eq:SVD-truncation}
	\mathcal{T}_k = \sum_{i=1}^k \sigma_i v_i u_i^T
\end{equation}
and obtain a reduced set of basis vectors. 
We then have
\begin{equation}
	\min_{\rank(B)=k} \lVert \mathcal{T} - B \rVert_F = \lVert \mathcal{T} - \mathcal{T}_k \rVert_F = \sqrt{\sum_{i=k+1}^r (\sigma_i)^2},
\end{equation}
i.e. $\mathcal{T}_k$ is the best rank-$k$ approximation to $\mathcal{T}$ in the Frobenius norm $\lVert A \rVert_F = \sqrt{\sum_{i=1}^m\sum_{j=1}^n \lvert a_{ij}\rvert^2}$.
The Frobenius norm can also be expressed by the vector 2-norm as $\lVert A \rVert_F^2 = \sum_{i=1}^n \lVert a_i \rVert_2^2$, where $a_i$ denotes the $i$th column of $A$.

Note that the square of the Frobenius norm of the truncation error at rank $k$
\begin{equation}
	\label{eq:waveform_Frobenius_truncation_error}
	\lVert \mathcal{T} - \mathcal{T}_k \rVert_F^2 = \sum_{i=1}^n \lVert \delta \tau_i \rVert_2^2 
	= \lVert \sum_{i=k+1}^n \sigma_i v_i u_i^T \rVert_F^2 = \sum_{i=k+1}^r (\sigma_i)^2	
\end{equation}
is equal to the sum of the squares of the 2-norms of the errors $\delta\tau_i$ in each waveform amplitude or phase $\tau_i$ which make up the columns of $\mathcal{T}$. A useful estimate of the mean amplitude or phase error introduced by the approximation is the RMS norm of the truncation error $\frac{1}{\sqrt{n}} \lVert \mathcal{T} - \mathcal{T}_k \rVert_F $.

The basis for the amplitude / phase space is given in the columns $\mathcal{B}_i$ of the matrix $\mathcal{B} := V_n \in \mathbb{R}^{m \times n}$ if $m>n$, $\mathcal{B} := V \in \mathbb{R}^{m \times m}$ if $m \leq n$ and a full rank basis is desired. If $m<n$, then the information from the $n$ waveforms at $m$ grid points is contained in a basis of dimension $m$.
To compress the model one can define a \emph{reduced basis} of rank $k$
\begin{equation}
	\mathcal{B}_k = V_k = \left[ v_1 | \cdots | v_k \right] \in \mathbb{R}^{m \times k} \quad \text{for} \quad k < r \leq m.
\end{equation}
For any $k$ the columns of $V_k$ are an optimal orthonormal basis for the starting waveforms.
We usually drop the label $k$ on the rank-$k$ reduced basis.

\begin{figure}[htbp]
	\centering
		\includegraphics[width=0.6\textwidth]{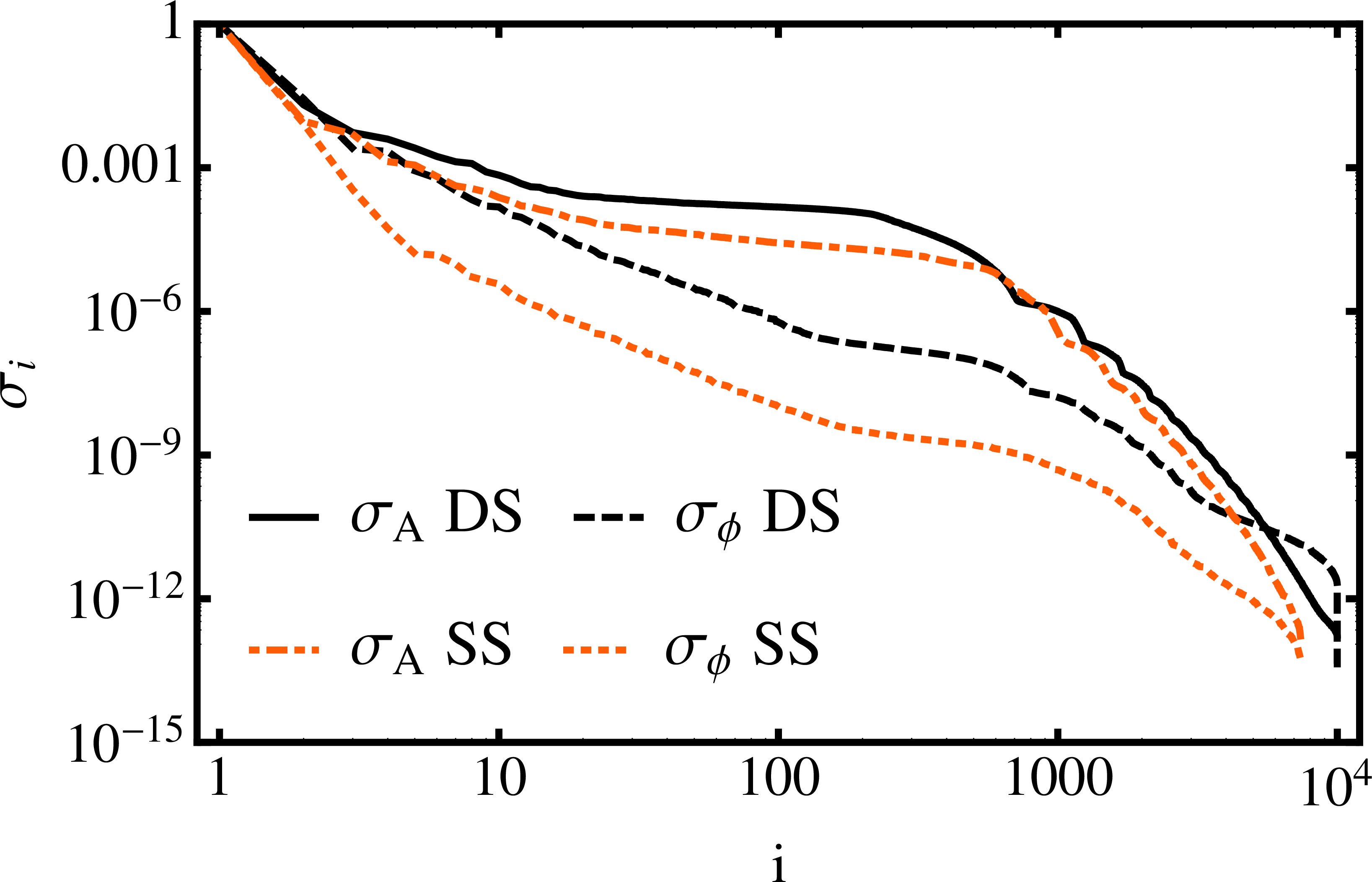}
	\caption{
	Falloff of the (normalized) singular values $\sigma_A$ and $\sigma_\phi$ for the amplitude and phase matrices, respectively, for the double spin (black) and single spin starting waveforms (orange). 
	The singular values fall off slower for the double spin model, especially for the waveform phase.
	The set of waveforms is larger for the double spin model and there is more variation in the waveform (especially the phase) due to the added dimensionality. 
	}
	\label{fig:singular-values}
\end{figure}

If the singular values decay sufficiently fast, then the truncation error is small, i.e. $\mathcal{T}_k$ is a good compressed and fast approximation of $\mathcal{T}$.
Fig.~\ref{fig:singular-values} illustrates that for the first few modes the singular values $\sigma_i$ of the amplitude and phase waveform matrices fall off very rapidly, then settle down to a slower decay before an accelerated plunge beyond $i \sim 1000$. The fact that the values are already reasonably small for $i$ on the order of a hundred motivates truncating the SVD after this many terms. Whether one thereby obtains an accurate compressed representation of the waveforms needs further investigation. In~\Sref{sec:errors} and~\Fref{fig:SVD-truncation-errors-vs-rank-SS} I will study the meaning of the truncation error in the Frobenius norm~\Eref{eq:waveform_Frobenius_truncation_error} and compare it with the ``mismatch'' between waveforms introduced by this approximation.

Given the reduced bases $\mathcal{B}_\mathcal{A}$ and $\mathcal{B}_\Phi$ we compute projection coefficient vectors $\mu$ for any given input waveform amplitude or phase $\tau \in \mathbb{R}^m$ as follows
\begin{equation}
	\mu(\tau) := \mathcal{B}^T \tau \in \mathbb{R}^k,
\end{equation}
where we have dropped the amplitude or phase labels for brevity.
We collect the projection coefficient vectors for all input waveforms in matrices $\mathcal{M}_\mathcal{A}$ and $\mathcal{M}_\Phi$ with entries
\begin{equation}
	\label{eq:projection_coefficients_M}
	\mathcal{M}_{ji} = \mu_j(\tau_i) = \left( \mathcal{B}^T \mathcal{T} \right)_{ji} \in \mathbb{R}^{k \times n}.
\end{equation}
Comparing with~\Eref{eq:SVD} we see that $\mathcal{M} = \mathcal{B}^T \mathcal{T} = \Sigma U^T$ for a full-rank basis $\mathcal{B} = V$.
It follows that the projection coefficient matrices are ordered in the same way as the individual waveforms in $\mathcal{T}$. To undo the packing of the waveforms in the matrices $\mathcal{M}$ we just partition the linear index $i$ that enumerates the waveforms in $\mathcal{T}$ and obtain a tensor
\begin{equation}
	\label{eq:projection_coefficients_M_partitioned}
	\mathcal{M}_{j,i_q, i\chi} = \mu_j(\tau_{(i_q, i_\chi)}) \in \mathbb{R}^{k \times n_q \times n_\chi}.	
\end{equation}

To complete the model we define the projection coefficient vectors at any location in the chosen parameter space by suitable interpolants $I_\otimes[\mathcal{M}](q,\chi) \in \mathbb{R}^k$ for the amplitude and phase coefficient tensors $\mathcal{M}_\mathcal{A}, \mathcal{M}_\Phi$. For each input waveform we have two corresponding $k$-vectors of projection coefficients (for amplitude and phase) that are interpolated over the parameter space. This is discussed in~\Sref{sec:int_par_space}.
The frequency domain reduced order model is then given by
\begin{empheq}[box=\mybluebox]{equation}
	\label{eq:model_function}
	\mkern-12mu \tilde h_\text{m}(q,\chi; M; Mf) := A_0(q,\chi,M) \;
		I_f\left[ \mathcal{B}_\mathcal{A} \cdot I_\otimes[\mathcal{M}_\mathcal{A}](q,\chi) \right](Mf)\;
		\exp\left\{ i \, I_f\left[ \mathcal{B}_\Phi \cdot I_\otimes[\mathcal{M}_\Phi](q,\chi) \right](Mf) \right\},		
\end{empheq}
where $\cdot$ denotes matrix multiplication, $I_f[\cdot]$ interpolates vectors in frequency on a suitable grid, and $A_0(q,\chi,M_\text{tot})$ is an amplitude prefactor.

The amplitude normalization is restored as follows. We save the normalization factors that are applied to each amplitude before taking the SVD and compute an interpolant $N_A(q,\chi)$ over the parameter space (using techniques discussed in~\Sref{sec:int_par_space}). 
We restore the LAL amplitude scaling for a binary at arbitrary distance $D$ and optimal orientation by computing
\begin{equation}
	\label{eq:amplitude_normalization_prefactor}
	A_0(q,\chi,M) = N_A(q,\chi) \frac{1 \text{Mpc}}{D(M)},
	\quad \text{where} \quad  
	D(M) := \frac{1 \text{Mpc}[\text{m}]}{M M_\odot[\text{m}]} \text{Mpc},
\end{equation}
where $M_\odot[\text{m}] = c M_\odot[\text{s}]$.

For optimal orientation the two polarisations of the dominant $l=m=2$ mode in the frequency domain are well approximated by
\begin{align}
	\tilde h_+ (f) 			&=   \widetilde{\text{Re} \, h} = \frac{\tilde h(f) + \tilde h^*(-f)}{2} \approx \frac{1}{2} \tilde h(f), \\
	\tilde h_\times (f) &= - \widetilde{\text{Im} \, h} = -\frac{\tilde h(f) - \tilde h^*(-f)}{2i} \approx \frac{i}{2} \tilde h(f)
\end{align}
since $\lvert \tilde h^*(-f) \rvert \ll  \lvert \tilde h(f) \rvert$. We have conjugated the input $h_{2,2}$ waveforms so that the power is in the positive frequencies $f>0$.
The inclination $\iota$ of the binary only changes an overall amplitude factor for aligned-spin waveforms and the final expression for the polarizations is
\begin{align}
	\tilde h_+ (f) 			&\approx \frac{1}{2}(1 + \cos^2\iota) \left(\frac{1}{2} \tilde h_m(f)\right), \\
	\tilde h_\times (f) &\approx \cos\iota 										\left(\frac{i}{2} \tilde h_m(f)\right).
\end{align}


\section{SVD and greedy reduced basis methods} 
\label{sec:comparison_between_svd_and_greedy_reduced_basis_methods}

I give a brief comparison of the SVD and \emph{greedy basis}~\cite{Field:2013cfa,Field:2011mf} methods. I contrast the falloff of singular values against the greedy error and show an example of how the greedy basis method picks configurations in the parameter space.

The greedy reduced basis algorithm (see e.g. Algorithm 1 in~\cite{Field:2013cfa}) starts with a seed basis element and iteratively picks waveforms with the largest error in the orthogonal projection onto the basis from a set of input waveforms (also called the ``training space'') and adds them to the basis. The basis elements are orthonormalized with a Gram-Schmidt algorithm.

Either the SVD or the greedy basis algorithm can be used to find a reduced orthonormal basis from a set of starting waveforms. Let $n$ be the number of waveforms given on $m$ gridpoints, assume $n > m$, and let $k$ be the number of reduced basis functions desired. The greedy basis method has complexity $\bigO(m \, n \, k)$ (see e.g.~\cite{Canizares:2013ywa}) as opposed to the SVD's more costly $\bigO(n \, m^2)$ and it can be trivially parallelized.  
For a full basis $k=m$ and the cost of the two methods is comparable.
In contrast to the greedy basis method, parallel implementations of the SVD are readily available and easy to use. For a sparse frequency grid of about $m=200$ points, computing the SVD is very cheap, and takes about one second.

\begin{figure}[htbp]
  \centering
    \includegraphics[width=0.47\textwidth]{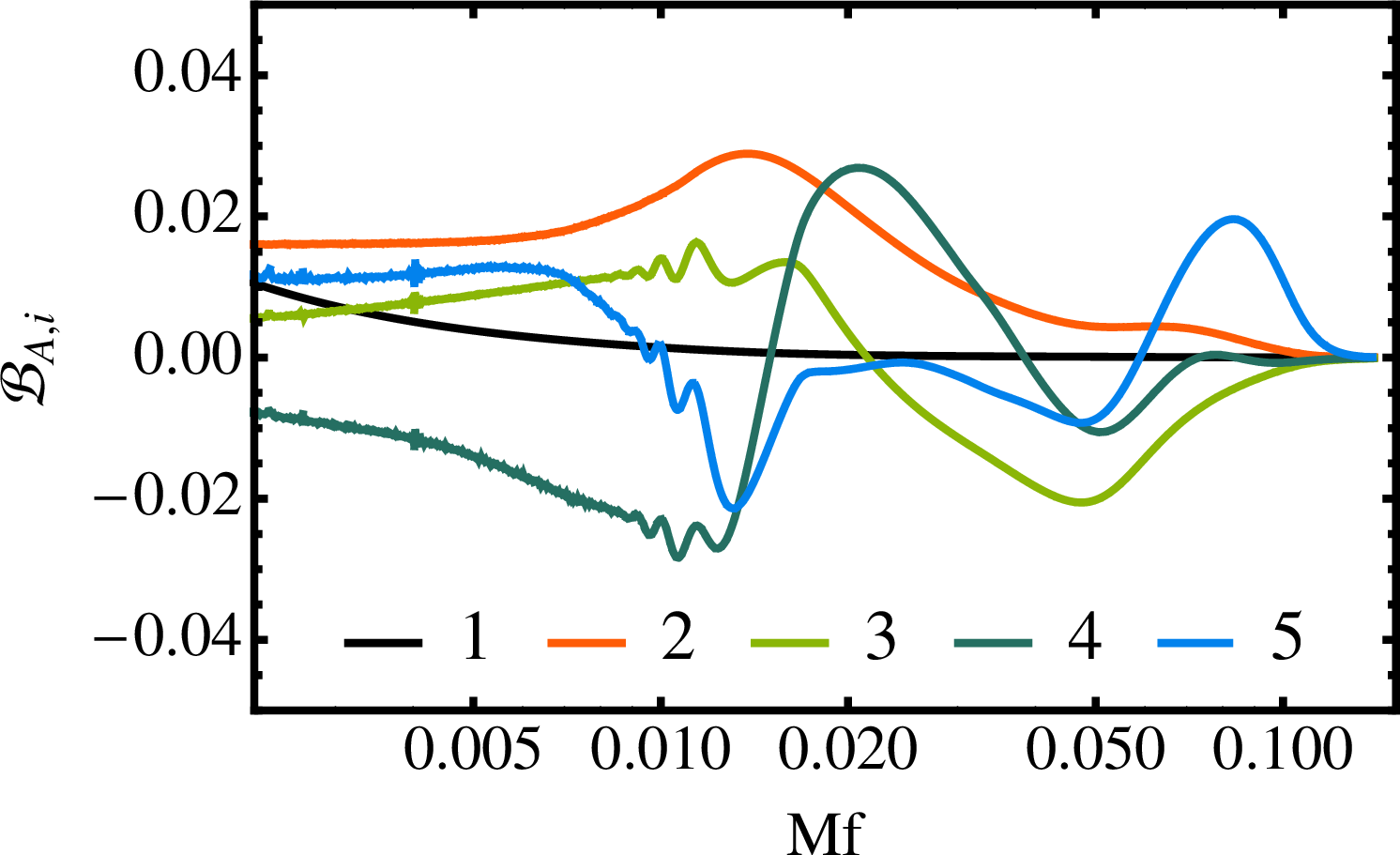}
		\includegraphics[width=0.47\textwidth]{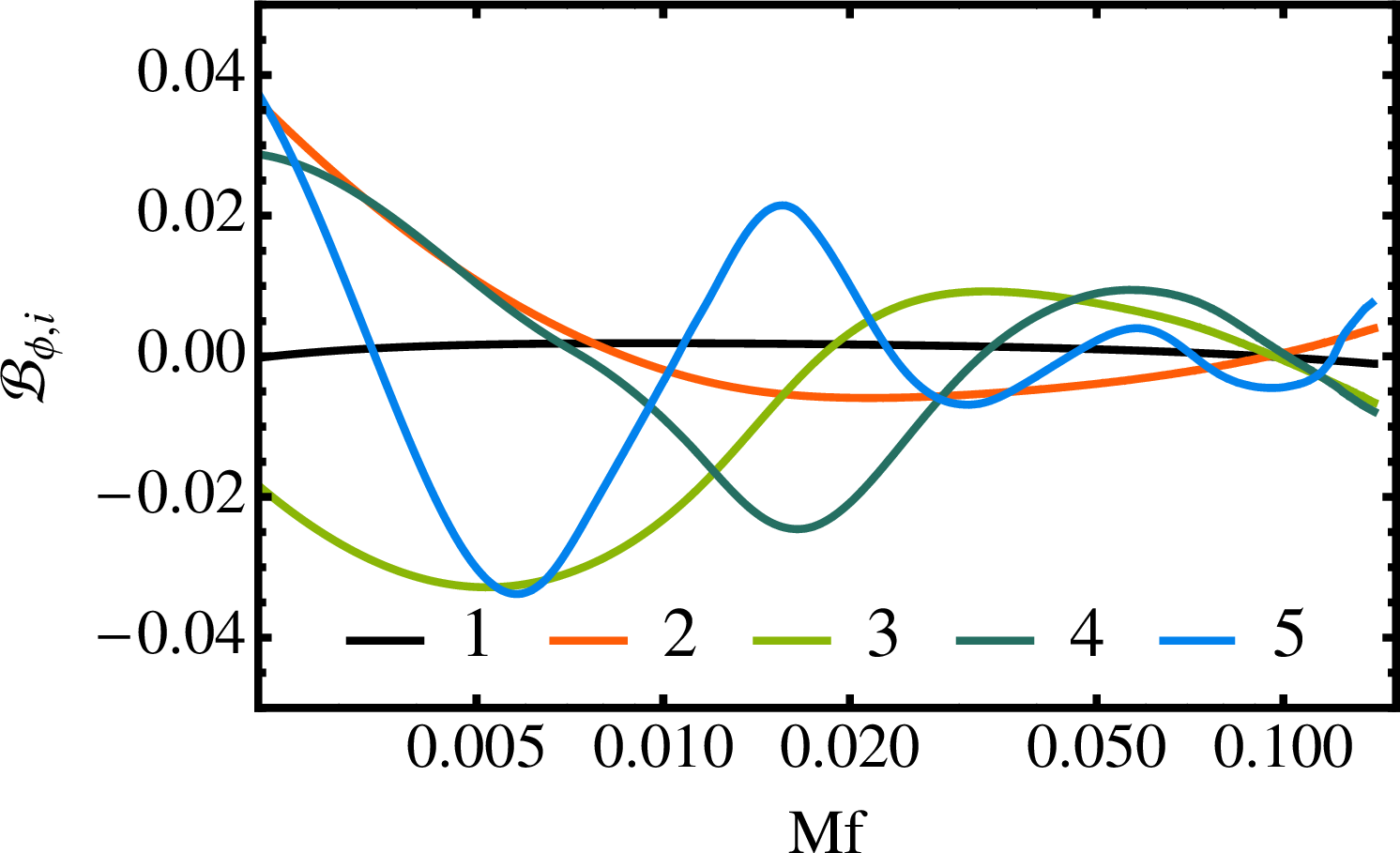}
  \caption{The first 5 amplitude and phase SVD modes for the single spin model.}
  \label{fig:SVD-SS-modes}
\end{figure}

\begin{figure}[htbp]
  \centering
	\includegraphics[width=0.47\textwidth]{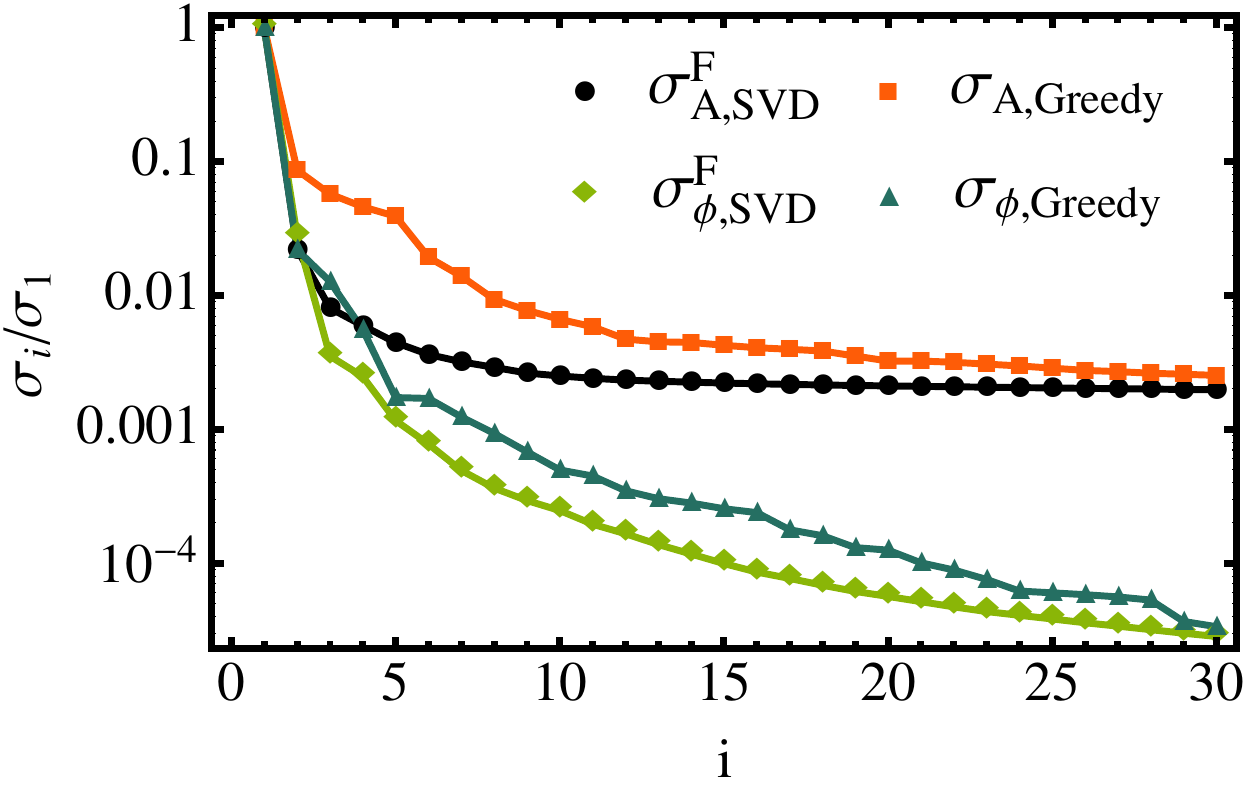}
  \caption{The Frobenius norm of the SVD truncation error $\sigma^F_i := \sqrt{\sum_{j=i+1}^r (\sigma_j)^2}$ (see~\Eref{eq:waveform_Frobenius_truncation_error}) is shown versus the greedy error for the first 30 waveforms chosen from the starting waveform space for the double spin model.
Note that while the greedy error bounds the \emph{worst} approximation error, the Frobenius norm of the SVD truncation error bounds the \emph{average} approximation error.
}
  \label{fig:sigma_i_DS_SVD_vd_Greedy}
\end{figure}

For both methods the reduced basis waveforms only resemble the physical behavior of frequency domain amplitudes and phases for the first basis function. Higher basis functions show corrections at finer scales and are oscillatory (see \Fref{fig:SVD-SS-modes}). The falloff of the SVD truncation error~\Eref{eq:waveform_Frobenius_truncation_error} and greedy errors is shown in~\Fref{fig:sigma_i_DS_SVD_vd_Greedy}. It is worth noting that when using a truncated basis the SVD truncation error is, by construction, an average projection error. In contrast, the projection error associated with reduced basis elements chosen by the greedy algorithm minimizes local projection errors (see e.g. Algorithm 1 in~\cite{Field:2013cfa}) and thus bounds the worst approximation error. This is consistent with the behavior shown in~\Fref{fig:sigma_i_DS_SVD_vd_Greedy}.

\begin{figure}[htbp]
  \centering
    \includegraphics[width=0.49\textwidth]{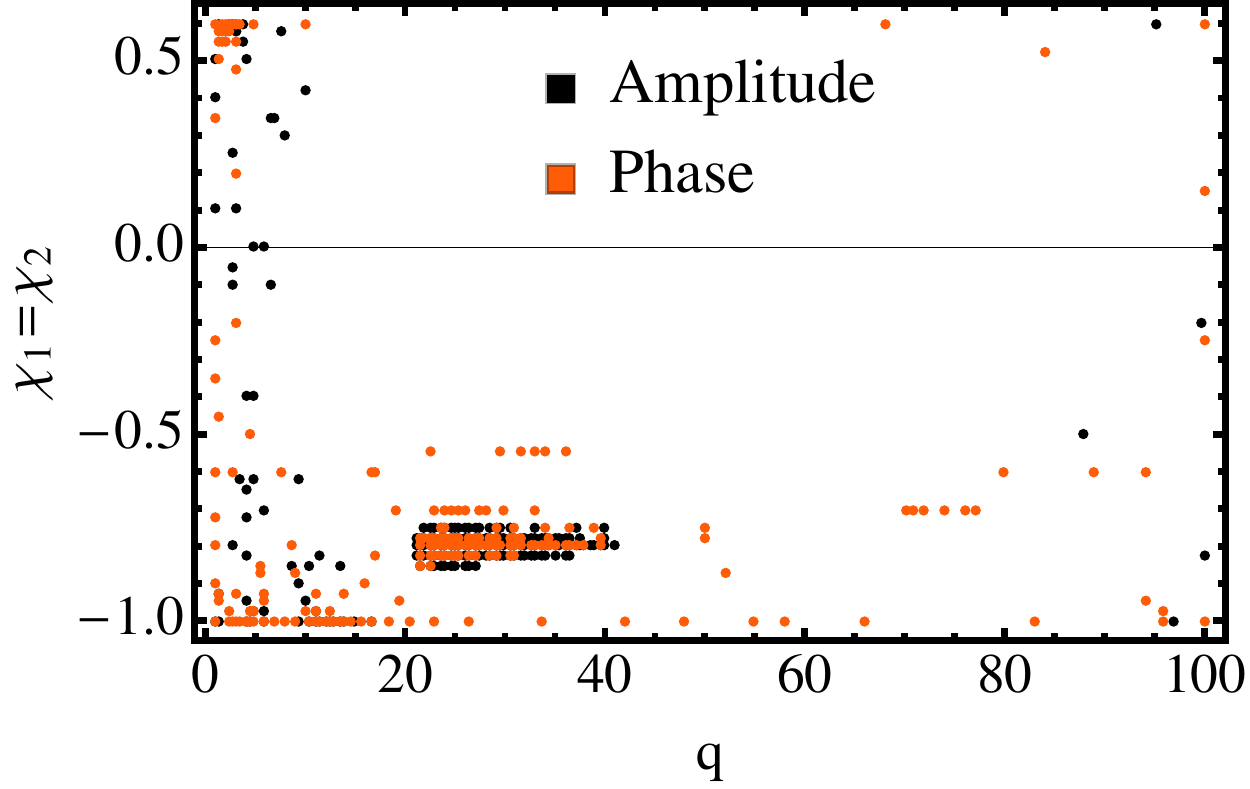}
		\includegraphics[width=0.49\textwidth]{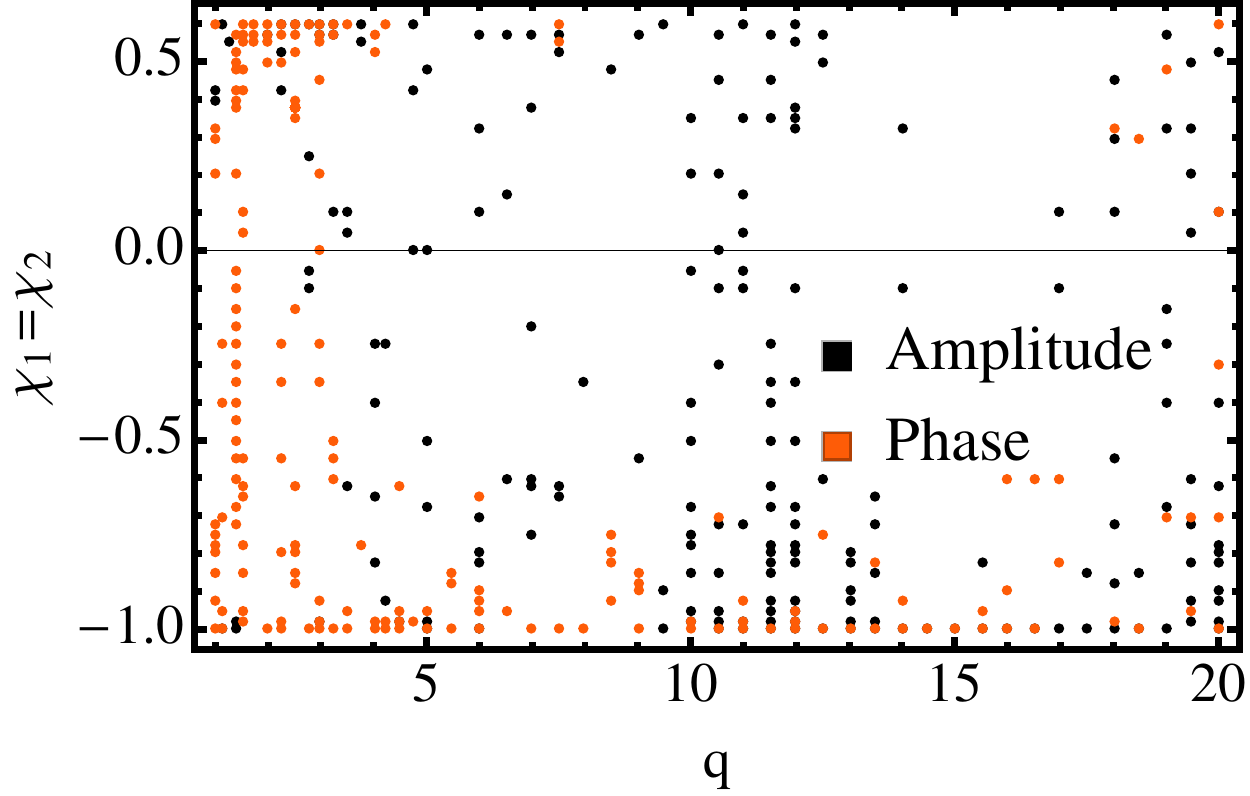}
  \caption{The first 200 waveforms chosen from the equal-spin starting waveforms by the greedy reduced basis algorithm. The left panel shows the whole range in mass-ratio, while the right panel zooms into the range $q \leq 20$. Note the clustering of points near the boundaries of the domain. The cluster at $\chi_1=\chi_2\sim 0.8$ and $q=20-40$ suggests that the SEOBNRv1 model waveforms vary very strongly in this region (see  also~\Fref{fig:Phase-error-sparse-ROM-model-histogram-CSE}). This is due to undersampling of the NQCs in this region in the SEOBNRv1 LAL code.}
  \label{fig:greedy_basis_picks}
\end{figure}

An advantage of the greedy algorithm is that it directly exposes which waveforms are picked from the space of input waveforms and then orthogonalized and promoted to basis vectors. Given a sufficiently dense set of waveforms in a subdomain of the physical parameter space this method provides valuable information on the variation of waveforms with the parameters. It can help identify problems in the underlying model when basis waveforms cluster in some region of the domain.
\Fref{fig:greedy_basis_picks} shows the first 200 basis waveforms chosen from the single spin training space. Apart from clustering near the boundaries of the parameter space domain (similar to what was observed in~\cite{Field:2012if}), basis waveforms are clumped together from $q \sim 20-40$ and $\chi \sim -0.8$. This clustering shows up both for the independent amplitude and phase basis.
This behavior is surprising as one would expect the waveforms to smoothly depend on mass-ratio and spin and may indicate that the source model contains unphysical artifacts, although here way beyond its calibration range $q \leq 6$.

The clustering in~\Fref{fig:greedy_basis_picks} and the peaks apparent in the amplitude coefficients in~\Fref{fig:projection_coefficients} (see also~\Fref{fig:Phase-error-sparse-ROM-model-histogram-CSE} for mismatches) turned out to be due to undersampling of the non-quasi-circular coefficients in the LAL implementation of SEOBNRv1~\cite{ABcomment}. Since this study primarily deals with a proof of concept of building surrogates and since the regeneration of all input waveforms is very expensive the surrogate will not be rebuilt here. It is also instructive to see that reduced order modeling techniques can help in testing waveform models. Ultimately, I plan to build a surrogate model for the successor to SEOBNRv1 (SEOBNRv2) which covers the full spin range $\chi_i \in [-1,1]$. This is the model that will be implemented in the LAL code.

If the source model is restricted to a range where it is deemed trustworthy (e.g. because it has been calibrated there) the method can show what the most physically important waveforms are and can provide clues where to perform NR simulations which can subsequently be used to improve current EOB and phenomenological waveform models. Given $n$ NR waveforms it is not possible to predict what the next most important NR waveform is that should be simulated without recourse to a trusted model.


\section{Interpolation over the parameter space}
\label{sec:int_par_space}

There are many choices for creating interpolants $I[\mathcal{M}](q,\chi)$ of the projection coefficients over the parameter space. In the following I consider primarily \emph{tensor product} spline interpolation, and briefly mention alternative methods. The discussion of the different interpolation methods assumes a two-dimensional parameter space; extension to a three-dimensional space $(q,\chi_1,\chi_2)$ is straightforward.

A natural way to interpolate the projection coefficients over the parameter space is to use a \emph{tensor product} expansion. The tensor product interpolant $\mathcal{I}_\otimes[f]$ of a scalar function $f(q,\chi)$ on the Cartesian product of intervals $\Omega = [q_\text{min},q_\text{max}] \times [\chi_\text{min}, \chi_\text{max}]$ is given simply in terms of the product of 1-dimensional basis functions $\Psi_q$, $\Psi_\chi$
\begin{equation}
	\label{eq:TP_interpolation}
	\mathcal{I}_\otimes[f](q,\chi) = 
	\sum_{i=1}^{n_q} \sum_{j=1}^{n_\chi} \bar f_{ij} \left(\Psi_q \otimes \Psi_\chi\right)_{ij} (q, \chi) =
	\sum_{i=1}^{n_q} \sum_{j=1}^{n_\chi} \bar f_{ij} \Psi_{q,i}(q) \Psi_{\chi,j}(\chi).
\end{equation}
The points $(q_i, \chi_j)$ are taken from a rectangular parameter space grid $\mathcal{Q} \times \mathcal{X} \in \Omega$ with
\begin{equation}
	\mathcal{Q} \times \mathcal{X} = \{ (q_i, \chi_j) : i=1,\dots,n_q ; j=1,\dots,n_\chi \}
\end{equation}
and $\bar f_{ij}$ are the discrete expansion coefficients of $f$ in terms of the tensor product basis 
$\Psi_q \otimes \Psi_\chi$. 
We consider either cubic \emph{B-splines}~\cite{deBoor} or \emph{Chebyshev polynomials}~\cite{CanutoVolI} as basis functions $\Psi_\bullet$. Tensor product interpolants of vector valued functions are simply defined by component and generalizations to higher dimensions are straightforward.

One-dimensional cubic splines are piecewise polynomials of degree 3 with continuous first and second derivatives. Different choices are possible at the endpoints. A convenient condition that only requires the knowledge of the function values and no derivatives is the ``not-a-knot'' spline, where the first and last two spline pieces are taken to be the same polynomial. This is equivalent to the continuity of the third derivative at the second and at the penultimate grid point~\cite{Gautschi}.
While the spline interpolant can be constructed directly by solving a system of linear equations from interpolatory and continuity conditions, it turns out to be convenient to introduce localized basis functions, B-splines, to facilitate generalizations.

Per default I use tensor product cubic splines as conveniently available in Mathematica~\footnote{Mathematica also provides tensor product Hermite interpolants where derivatives of the function need to be specified in addition to the function values. They are of comparable accuracy and speed to spline interpolants, but need a prescription for approximating derivatives if they are not available.}. This method provides good accuracy, efficiency and does not require further adjustments if the variation of the coefficients is resolved by the grid. If the expansion coefficients $\bar f_{ij}$ and the B-splines are precomputed the evaluation cost of a $d$-dimensional tensor product cubic spline interpolant goes as $4^d$ multiplications plus the cost of evaluating the non-vanishing B-splines at the desired physical parameter values.

\begin{figure}[htbp]
  \centering
		\includegraphics[width=0.905\textwidth]{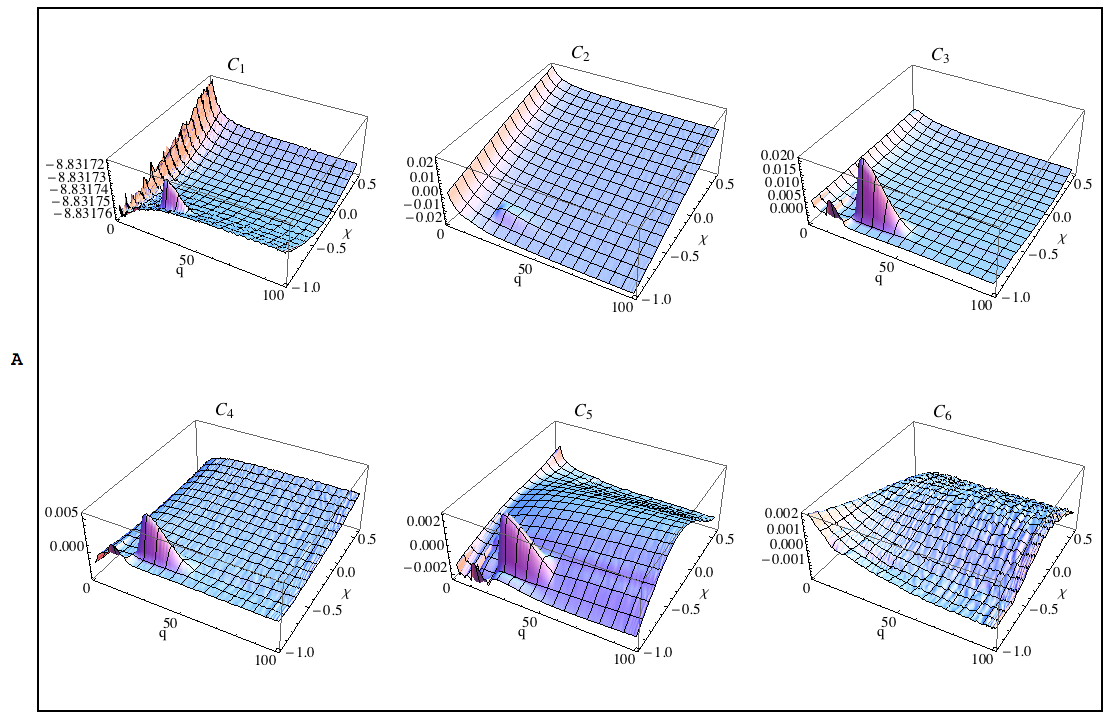}
    \includegraphics[width=0.9\textwidth]{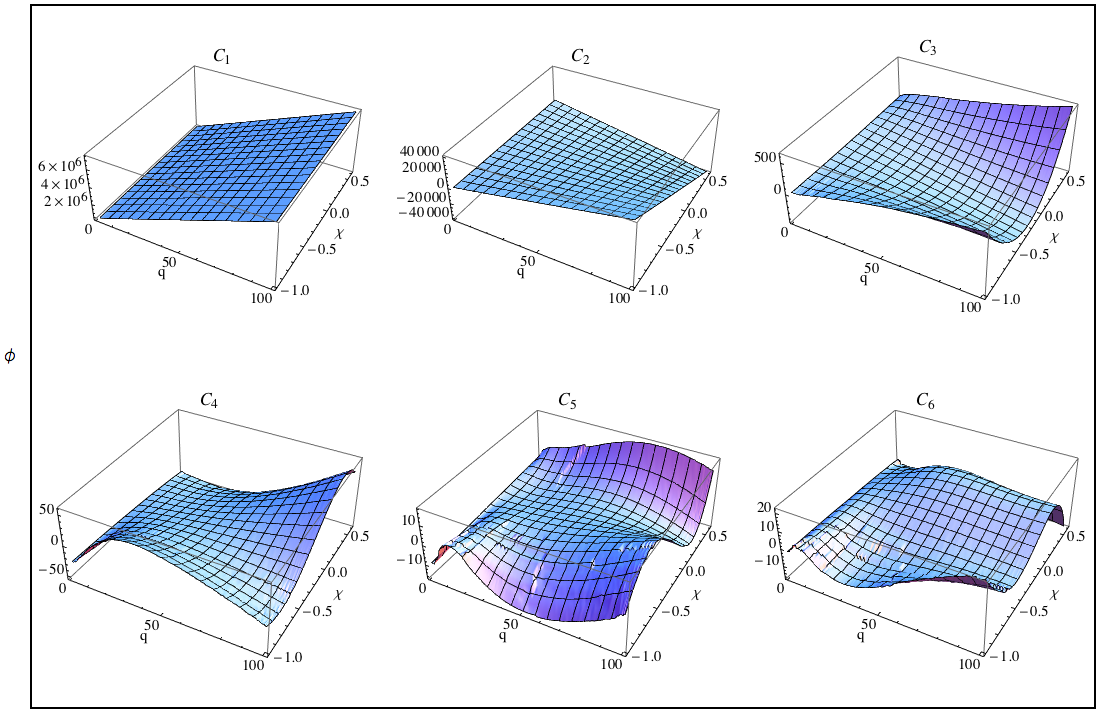}
  \caption{The dominant projection coefficients for amplitude (top panel) and phase (bottom panel) for the single-spin model over the parameter space.
The matrix-valued coefficients $C_j(i_q, i\chi) = \mathcal{M}_{j,i_q, i\chi}$
as defined in~\Eref{eq:projection_coefficients_M_partitioned}
correspond to the highest few singular values $\sigma_j$. Here we show them for $j = 1, \dots, 6$.
The first few coefficients of the phase appear fairly smooth up to $j=6$. Non-smoothness appears for higher coefficients along $\chi \sim -1$ and, to a lesser degree, along $\chi \sim 0.6$.
The inspiral part of the amplitudes has been fit to PN to suppress noise.
Then the amplitude coefficients are fairly smooth as well. More interestingly, there are pronounced peaks or dips for $q\sim 30$ and $\chi \sim -0.8$. This happens in the same region where greedy reduced basis functions cluster as seen in~\Fref{fig:greedy_basis_picks} and is due to undersampling of the NQCs in this region in the SEOBNRv1 LAL code.
}
  \label{fig:projection_coefficients}
\end{figure}

As can be seen in~\Fref{fig:projection_coefficients} the first few projection coefficients for the phase are smooth functions. If the amplitudes are computed from the untreated input amplitudes, the coefficients turn out to be very rough near equal mass. Most likely this is due to noise from the long ODE integration and subsequent FFT that is apparent upon visual inspection. Smooth amplitude coefficients (as shown in~\Fref{fig:projection_coefficients}) can be obtained by fitting the inspiral part of the input amplitudes against a PN amplitude and combining the smooth PN fit with the unchanged data for higher frequencies. Here, the fit extended up to one quarter of the frequency of the innermost stable circular orbit (ISCO) of a test particle around a Schwarzschild black hole, $f_\text{ISCO} = v_\text{ISCO}^3/(\pi M)$, where $v_\text{ISCO} = 1/\sqrt{6}$.
This change of the amplitudes has only a very small effect on the model, leading to mismatches of less than $10^{-4}$.

While it would be interesting to study the behavior of the B-spline interpolation error over the parameter space as a function of the density of input waveforms in order to optimize the choice of input configurations, such a study is beyond the scope of this paper.
Instead, we can compare B-spline interpolants of the model $\mu_j^\text{model} = I_\otimes[\mathcal{M}]_j (\bar q,\bar\chi)$ against projection coefficients $\mu_j^\text{test} = B_{ji} \tau_i$ computed for a test amplitude or phase $\tau$ with randomly chosen mass-ratio and spin $(\bar q, \bar\chi)$ for each SVD mode $j$.
The difference $\Delta\mu_j = \mu_j^\text{model} - \mu_j^\text{test}$ is then a measure of the parameter space interpolation error in the model at parameters $(\bar q, \bar\chi)$ and can be computed for the amplitude and phase coefficients.
A physically more meaningful error measure is the amplitude or phase error caused by the error $\Delta\mu_j$ in the projection coefficients in SVD mode $j$,
\begin{align}
	\label{eq:projection-coefficient-error}
	\Delta\phi_j &= \max_{i=1}^m \mathcal{B}^\Phi_{ij} \Delta\mu^\Phi_j\\
	\Delta A_j 	 &= \max_{i=1}^m \mathcal{B}^\mathcal{A}_{ij} \Delta\mu^\mathcal{A}_j,
\end{align}
where we take the maximum of the error over the discrete frequency points.

\begin{figure}[htbp]
	\centering
		\includegraphics[width=0.47\textwidth]{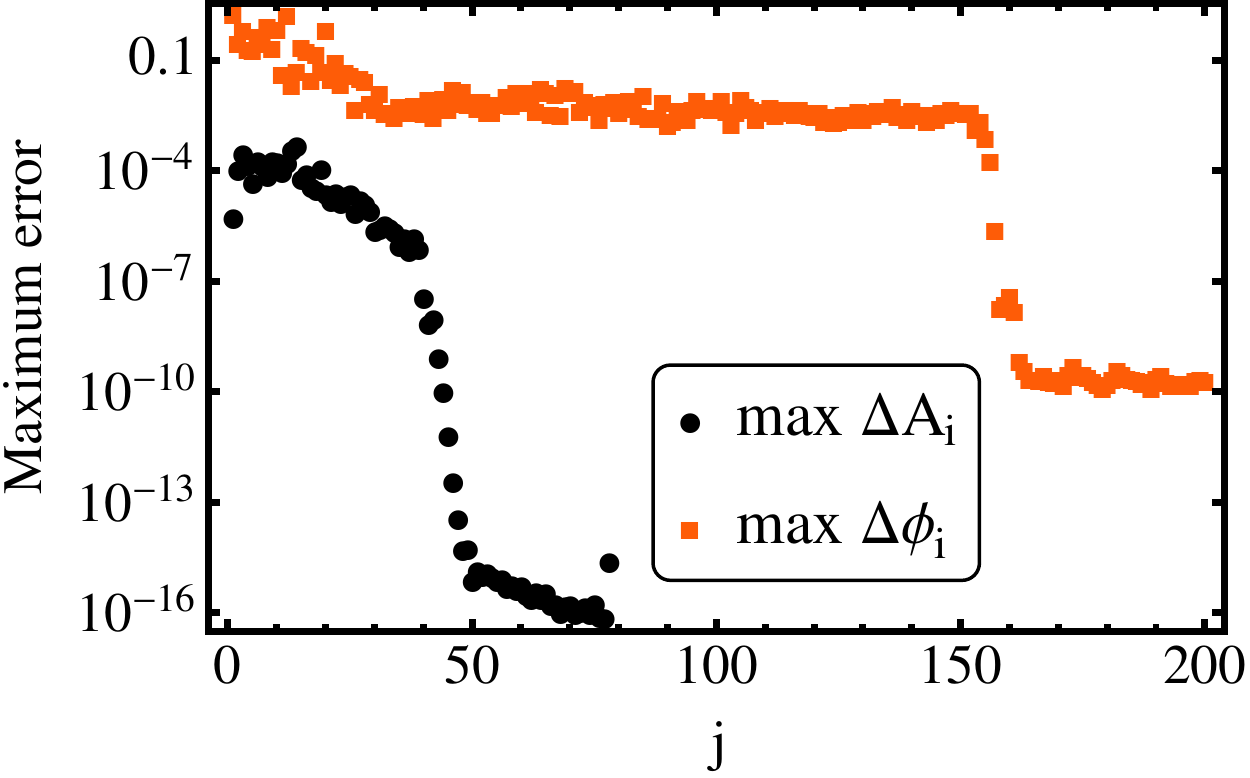}
	\caption{The amplitude and phase error (see~\Eref{eq:projection-coefficient-error}) due to errors in the parameter-space interpolation of the projection coefficients for each SVD mode $j$ maximized over a set of 1000 test waveforms. The average error is about one order of magnitude better than the maximum error.
	}
	\label{fig:Max_error_proj_coefficients}
\end{figure}

\Fref{fig:Max_error_proj_coefficients} shows the maximum of this amplitude and phase error over a set of 1000 random test waveforms that are also used in~\Sref{sec:results-SS}. The errors are largest for the first few SVD modes and decrease gradually. Beyond $j \sim 50$ for the amplitude and $j \sim 160$ for the phase the errors fall off rapidly. This is consistent with the singular values reaching zero at these modes. This behavior is different from the more gradual falloff of the singular values seen in~\Fref{fig:singular-values} and is due to the representation of the amplitude and phase on a set of 78 and 200 sparse frequency points, respectively and allows for early truncation of the SVD expansion in the model.

In addition to tensor product spline interpolation I also tried tensor product Chebyshev~\cite{CanutoVolI} interpolation and \emph{radial basis functions} (RBF)~\cite{Buhmann} which allow for general grids and scattered data. I found spline interpolation to be more efficient and since the evaluation speed of the model is more important than storage for most applications, I chose to use spline interpolation for the models build in this study.
Similar to techniques for solving PDEs the division of the parameter space domain into a number of smaller subdomains may allow Chebyshev interpolation and radial basis functions help perform better. This technique would also allow the use of local refinement for tensor product expansions.

Even though the most important configurations in the parameter space (e.g.. as given by the greedy reduced basis method method~\cite{Field:2013cfa,Field:2011mf}) are distributed in an irregular way over the physical domain, a structured tensor product grid of waveforms is advantageous, because interpolation methods on tensor product grids are more accurate than methods designed for unstructured grids.

\section{Other sources of errors}
\label{sec:errors}

We have already come across the main sources of errors that arise in building surrogate models in the previous sections. Here I summarize these findings, study several further errors and draw some conclusions.
The error due to a rank reduced SVD was pointed out in~\Sref{sec:build_model} and will be studied in more detail in this section.
In~\Sref{sec:int_par_space} I mentioned the error due to interpolation of the projection coefficients over the parameter space, while in~\Sref{sec:sparse_frequency_points} I investigated the interpolation error in frequency due to the choice of a reduced set of grid points for the waveforms. 
Some of these error sources can be related to norms of amplitude and phase errors. Ultimately we seek a more physical interpretation and wish to study the impact of the individual errors in terms of the waveform mismatch.

\subsection{SVD truncation error} 
\label{sub:svd_truncation_error}

One of the main ideas behind the approach pursued in this paper is a natural splitting of the frequency domain waveform into amplitude and phase and their representation by separate bases. These functions are nonoscillatory and ``simpler'' than the real and imaginary part of the strain and thus we expect to be able to find a more efficient compressed representation. Moreover, we can exploit additional freedom in finding optimal approximations if we treat amplitudes and phases completely separately. Unsurprisingly, we then found in~\Sref{sec:sparse_frequency_points} that the amplitude requires less frequency points than the phase in terms of a desired mismatch error due to interpolation. A disadvantage of performing this separation is that there is no direct relation between the singular values and the match. Such relations have been pointed out in the literature when the amplitude and phase were treated jointly and I address them briefly here.

Cannon et al~\cite{Cannon:2010qh} derive a relation between the singular values and the fractional SNR loss.
They work directly with the complex (whitened) time-domain waveform. 
Field et al~\cite{Field:2013cfa} give a relation between the greedy error of the reduced basis and the white-noise time-domain inner product between a waveform and its orthogonal projection onto the span of the basis.
Such relations do not directly allow the computation of the detector noise weighted mismatch for a range of system masses. Since these relations are not applicable in the approach pursued in this paper I compute the mismatch numerically.

\begin{figure}[htbp]
	\centering
		\includegraphics[width=0.47\textwidth]{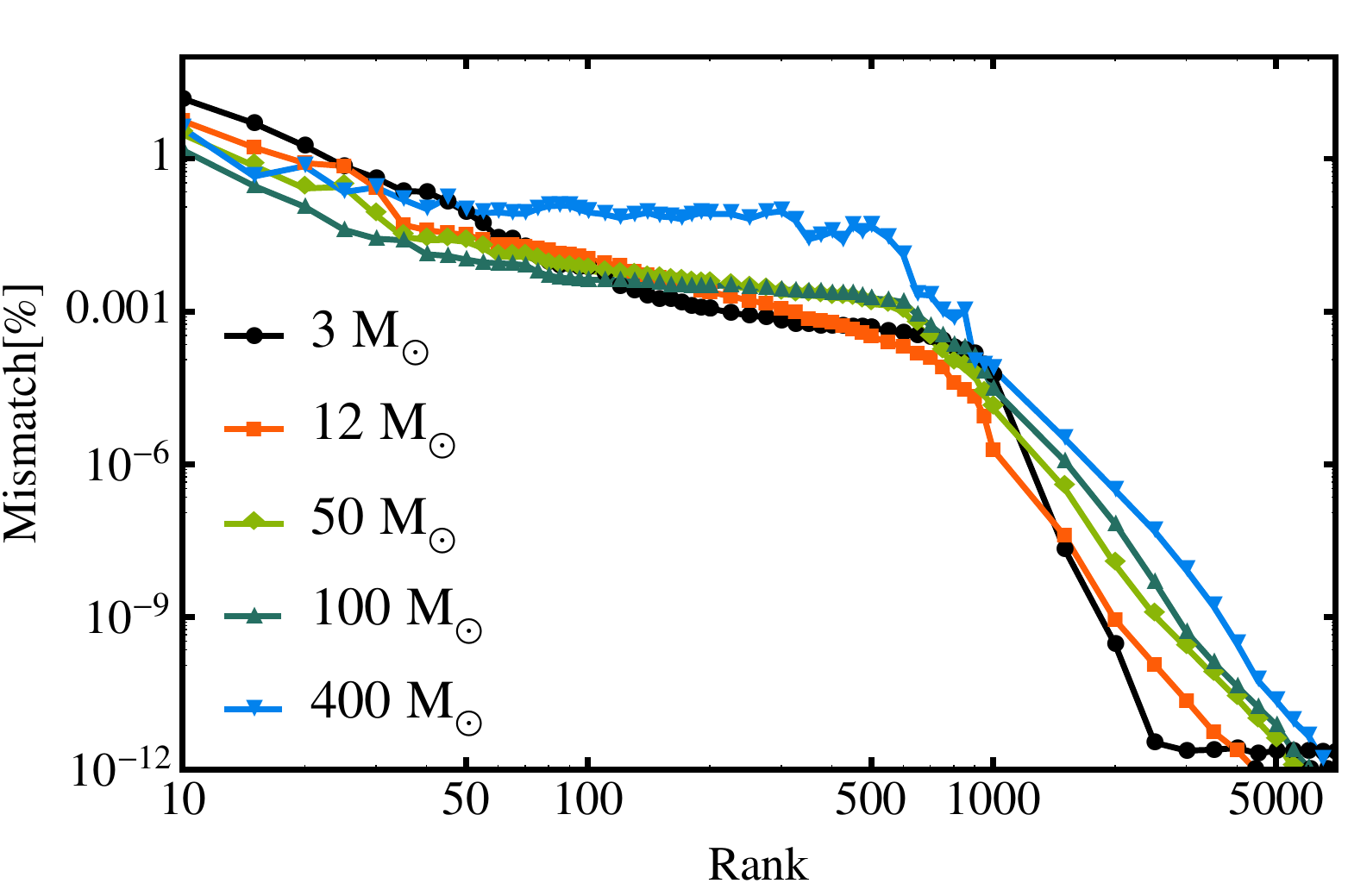}
		\includegraphics[width=0.48\textwidth]{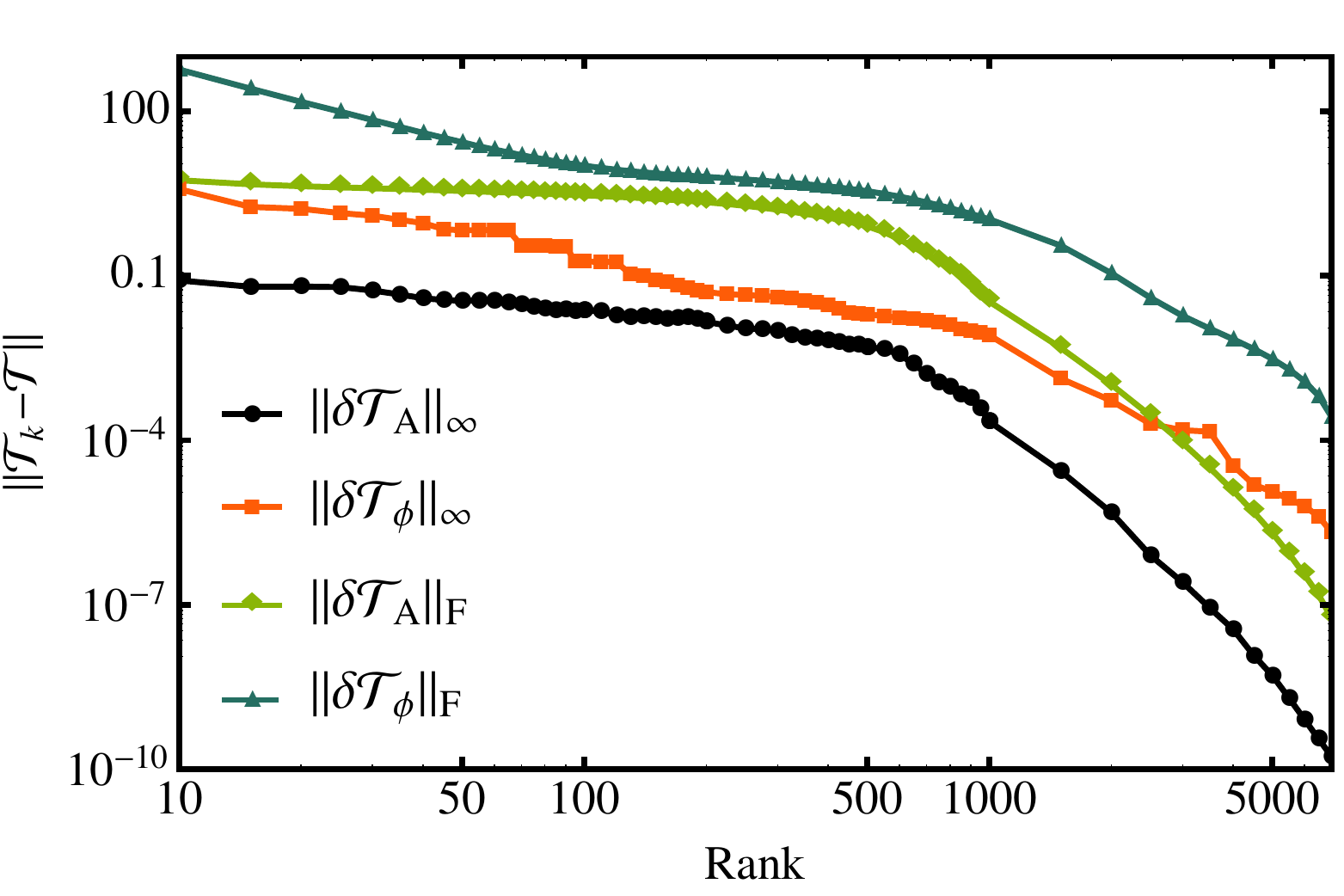}
		\includegraphics[width=0.47\textwidth]{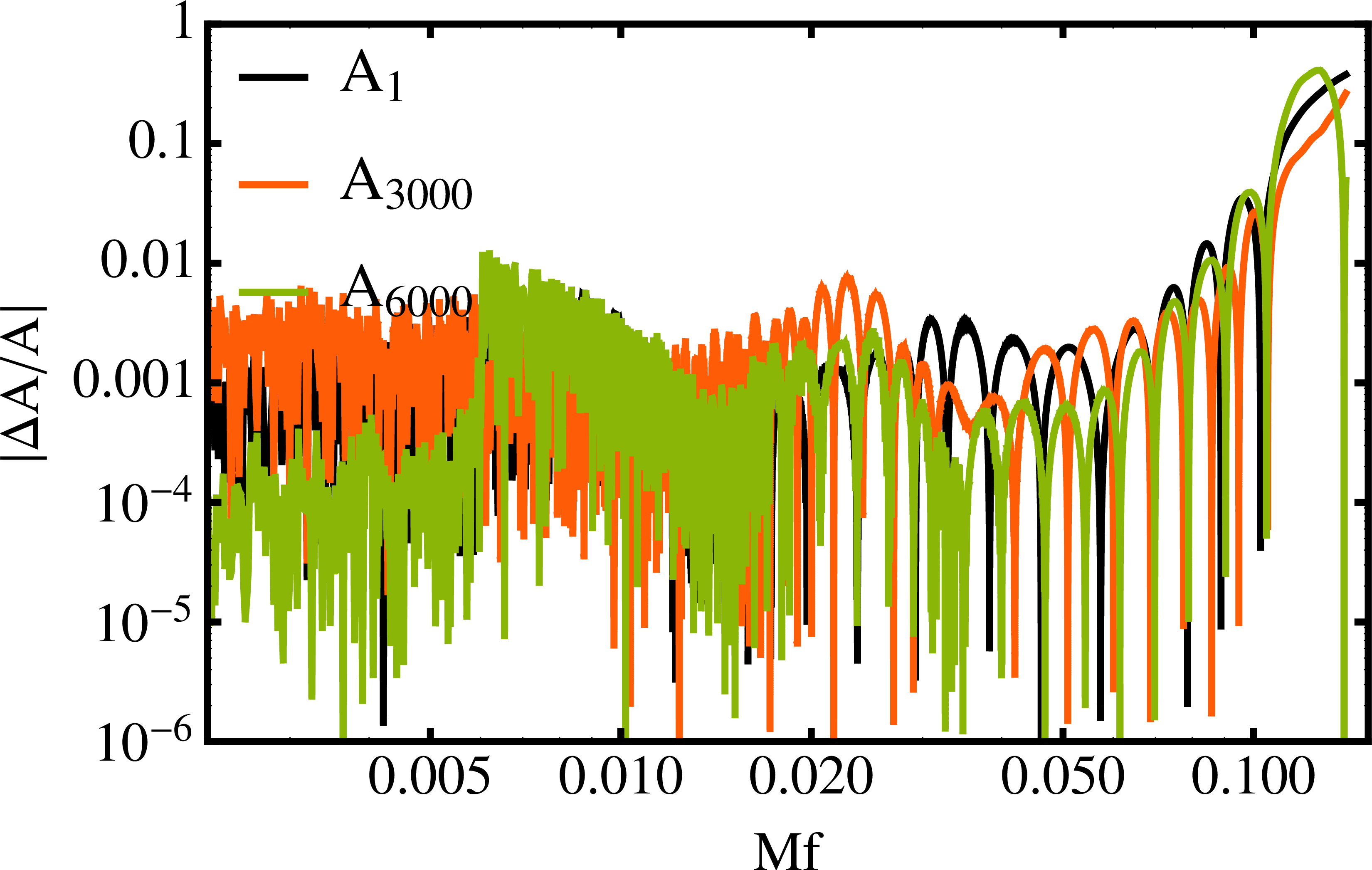}
		\includegraphics[width=0.48\textwidth]{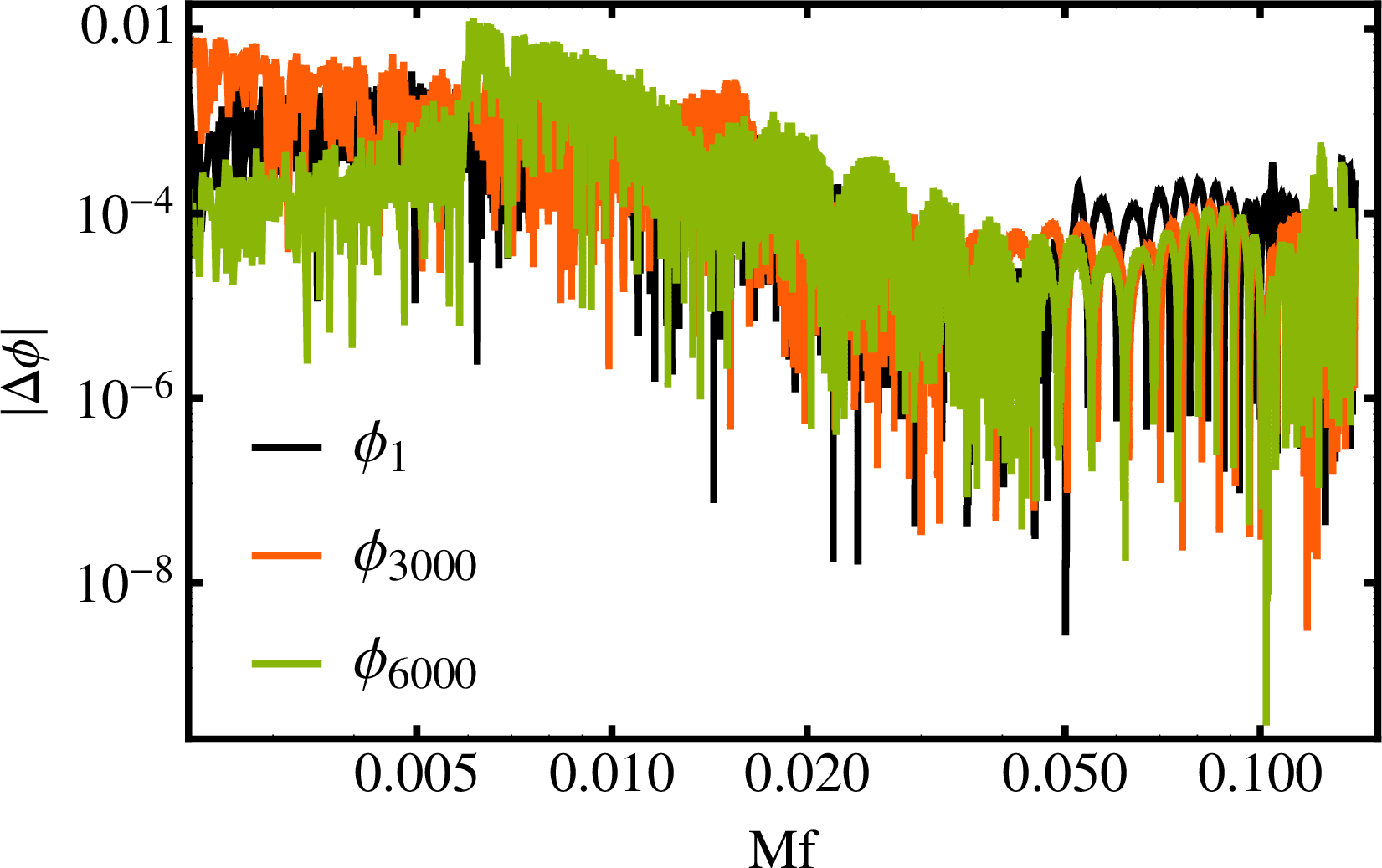}
	\caption{SVD truncation error against rank for the single spin waveforms on the 10000 point equispaced grid.
	We show mismatches of rank $k$ waveforms against the full waveforms for a range of total masses (top left) and show the $\ell_\infty$ and Frobenius norms of the error in the rank $k$ waveform matrices (top right).
	The bottom panel exemplifies how the SVD truncation error manifests itself in sample amplitudes and phases for a rank 200 truncation; the indices refer to the flat index $i:= (i_q-1)n_\chi + i_\chi$ numbering the input waveforms.
	}
	\label{fig:SVD-truncation-errors-vs-rank-SS}
\end{figure}

In~\Fref{fig:SVD-truncation-errors-vs-rank-SS} I show the decay of the SVD truncation error with rank for the single spin waveforms. The error is quantified both by the mismatch of the full model against a rank $k$ model (top left) and the $\ell_\infty$ or Frobenius norm\footnote{
Note that since the Frobenius norm yields an average $\ell_2$-norm of the waveform error, it can grossly overestimate the maximum phase or amplitude error. The $\ell_2$ and $\ell_\infty$ vector norms are related via the inequality
$\lVert x \rVert_\infty \leq \lVert x \rVert_2 \leq \sqrt{n}\lVert x \rVert_\infty$, where $x \in \mathbb{R}^n$.
}
of the error $\delta\mathcal{T} = \mathcal{T} - \mathcal{T}_k$ in the waveform matrix (top right).
In practise it is best to adjust $k$ so that the mismatch is smaller than desired.
Except for very high total mass the mismatch due to SVD truncation is near $\sim 0.001\%$ for $k$ of the order of a few hundred. There, the Frobenius norm of the starting phases is on the order of ten while the $\ell_\infty$ norm is near $0.1$.
The bottom panel of~\Fref{fig:SVD-truncation-errors-vs-rank-SS} gives an example of what the SVD truncation error looks like in terms of amplitude and phase error for selected waveforms. Both errors are highly oscillatory but their overall size is useful to remember and compare against, for instance, the amplitude and phase errors resulting from interpolation in frequency. 

A final remark on a peculiar feature of SVD-based models with separated bases for amplitude and phase is in order.
The SVD employs the Euclidean inner product to orthogonalize the basis functions. 
While this makes sense for the amplitude this is somewhat ``unnatural'' for the phase, as it enters the overlap in a complex exponential. Thus, the Euclidean inner product does not take into account that only the phase difference $\phi_1 - \phi_2$ appears in the overlap. 


\subsection{Further sources of error} 
\label{sub:further_sources_of_error}

The \emph{interpolation error over the parameter space} was mentioned in~\Sref{sec:int_par_space}. It arises due to the interpolation of the projection coefficients $\mathcal{M}$. This error depends on smoothness of the projection coefficients (see~\Fref{fig:projection_coefficients}) and the inherent accuracy of the chosen interpolation method. I compared different types of interpolants and settled for the robust choice of tensor product cubic splines. Since the smoothness of the coefficients is a priori unknown the interpolation error can only be computed by investigating the faithfulness of a given ROM model against waveforms that were not used in the construction of the model. Such a study with a set of waveforms at random locations in the model domain is carried out in~\Sref{sec:results-SS} and~\Sref{sec:double-spin-model}. The mismatch error for the single spin model is shown in~\Fref{fig:Faithfulness-sparse-ROM-model-histogram-CSE} and~\Fref{fig:Faithfulness-sparse-ROM-model-contours}. The fidelity of the double spin model is tested in~\Fref{fig:Faithfulness-full-DS-histogram}. The mismatch is below $0.1\%$ for the double-spin model, but up to $1\%$ for the single-spin model which spans a vast range in the mass-ratio and encounters artifacts from the SEOBNRv1 source model.

An additional error source for the single-spin model is \emph{hybridization error}. For the single-spin model I intended to extend the model into the binary neutron star domain. Due to the very high cost of generating SEOBNRv1 waveforms at these total masses I resorted to hybridization with the TaylorF2 approximant (see~\ref{sec:hybrids} for details).
The TaylorF2-hybrids have a mismatch less than $0.001 \%$ with SEOBNRv1 waveforms that cover the whole detector band at mass $1.35(1+q) M_\odot$ and $15\,$ Hz lower frequency cutoff.
This error can be reduced further by producing longer waveforms and hybridizing at lower frequencies or can be avoided altogether if one is willing to pay the price for significantly more expensive waveform generation at higher mass-ratios.

\emph{Inaccuracies in the computation of the overlap}: While not a problem of the model I caution the reader about a technical problem that appears in the computation of the overlap.
In the construction of the models I subtracted a linear frequency fit from the input phases. This operation does not alter the physical content in the phase and is in principle neutralized by the maximization of time and phase shifts that is performed in the computation of the overlap.
However, proper maximization over time-shifts requires a very fine frequency resolution in the presence of large linear dephasing between two waveforms. The models tend to have a very different phase evolution than SEOBNRv1 waveforms that were generated afterwards to test the model and the models may \emph{appear} more unfaithful than they are.
The remedy for this technical difficulty is either to subtract a linear fit from the phase difference of the waveforms in the computation of the overlap or to use substantial zero-padding. Zero-padding factors of up to 100 control this error sufficiently by providing a fine \emph{sinc interpolation} (see e.g.~\cite{SmithDFTbook}) of the complex SNR which allows accurate maximization over time shifts.

Initially I built ROM models from input waveforms on a fine equally spaced grid and employed SVD rank truncation to make the model more compact. There, I started with waveform matrices $\mathcal{T} \in \mathbb{R}^{m \times n}$ with a comparable number of grid points $m$ and number of waveforms $n$. If the waveforms are linearly independent, $\rank \mathcal{T} = \min(m,n) \sim m \sim n$. Later on I constructed ROM models in a different way: I first interpolated the input amplitudes and phases onto a set of $m$ sparse frequency points and then performed the SVD. In that case $m \ll n$ and the rank of the waveform matrix is now $m$. This leads to a loss of information and an associated error. I will refer to it as the ``rank-reduction'' error that stems from using fewer frequency points than input waveforms in the SVD.
In the absence of SVD truncation the square basis matrix $\mathcal{B} = V \in \mathbb{R}^{m \times m}$ has the same rank as the waveform matrix $\mathcal{T}$ that I feed into the SVD. The error is due to the approximation of the $n$ input waveforms in the waveform matrix by $m$ SVD modes. If I kept the full matrix $U \in \mathbb{R}^{n \times n}$ times the singular values I could reconstruct the full waveform matrix (except for numerical errors in the computation of the SVD). However I only keep a smaller set of projection coefficients $\mathcal{M} \in \mathbb{R}^{m \times n}$.
With this setup further SVD truncation at a lower rank $k < m$ is less useful and does not lead to significant further compression. 
An example of this ``rank-reduction'' error is discussed in~\Sref{sec:double-spin-model} and~\Fref{fig:Faithfulness-full-DS-histogram}.

To summarize, in practice I build ROM models by first generating a set of sparse frequency points for the amplitude and phase, interpolating the starting waveforms onto those sets, performing a full SVD of the waveform matrices and finally interpolating over the parameter space. The errors introduced by the individual approximations combine in complicated and nonlinear ways. Even though it does not seem to be possible to predict the fidelity of the final ROM model, one can check its faithfulness explicitly by computing matches against an independent set of randomly generated waveforms (see~\Sref{sec:results-SS} and ~\Sref{sec:double-spin-model}).

In the models constructed for this paper the interpolation of projection coefficients over the parameter space turned out to be the dominant source of error (see~\Sref{sec:results-SS} and~\Sref{sec:double-spin-model}). To reduce this error a finer spacing of input waveforms in regions where the coefficients are not sufficiently smooth is required. How much refinement is needed can only be explored iteratively and is evidently a very costly endeavor. Furthermore one has to be aware that while the fidelity of a model can be increased by adding more input waveforms, its evaluation speed will be reduced and storage requirements will be increased.



\section{Results for single-spin models}
\label{sec:results-SS}

Here I summarize results for models using a single-spin approximation and are built from equal-spin $\chi_1 = \chi_2$ waveforms similar to the approach used in phenomenological models for aligned-spin binaries~\cite{Purrer:2013ojf,Ajith:2011,Santamaria:2010yb,Ajith:2011ec}.
Individual error sources are discussed at length in~\Sref{sec:errors}. In this section I focus on \emph{faithfulness} results for single-spin models, that is, the match between a given SEOBNRv1 waveform and the surrogate model using the same parameters. 
Matches were computed with a frequency spacing of $0.5 \text{Hz}$.

\begin{figure}[htbp]
  \centering
	\includegraphics[width=0.49\textwidth]{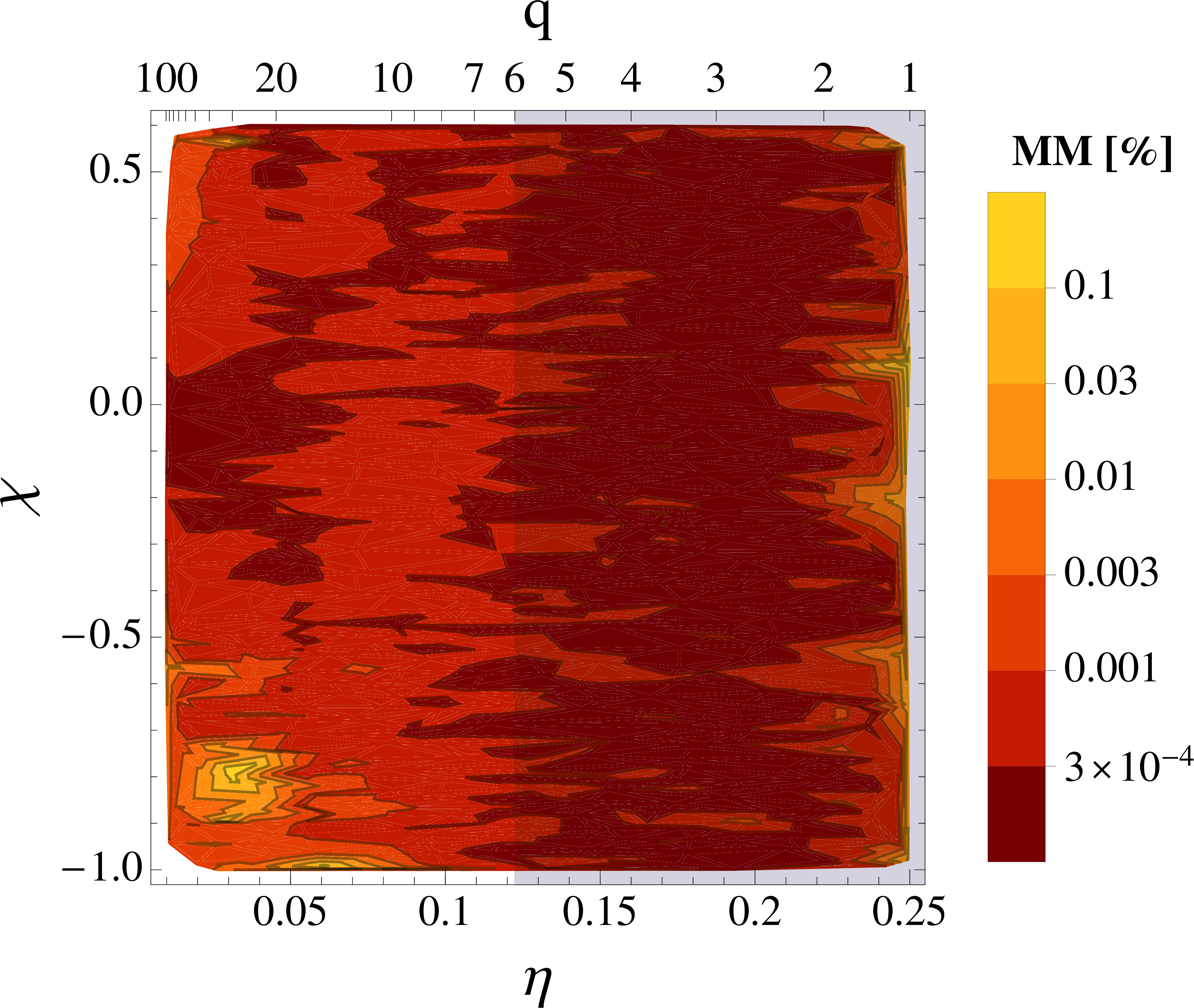}
	\includegraphics[width=0.46\textwidth]{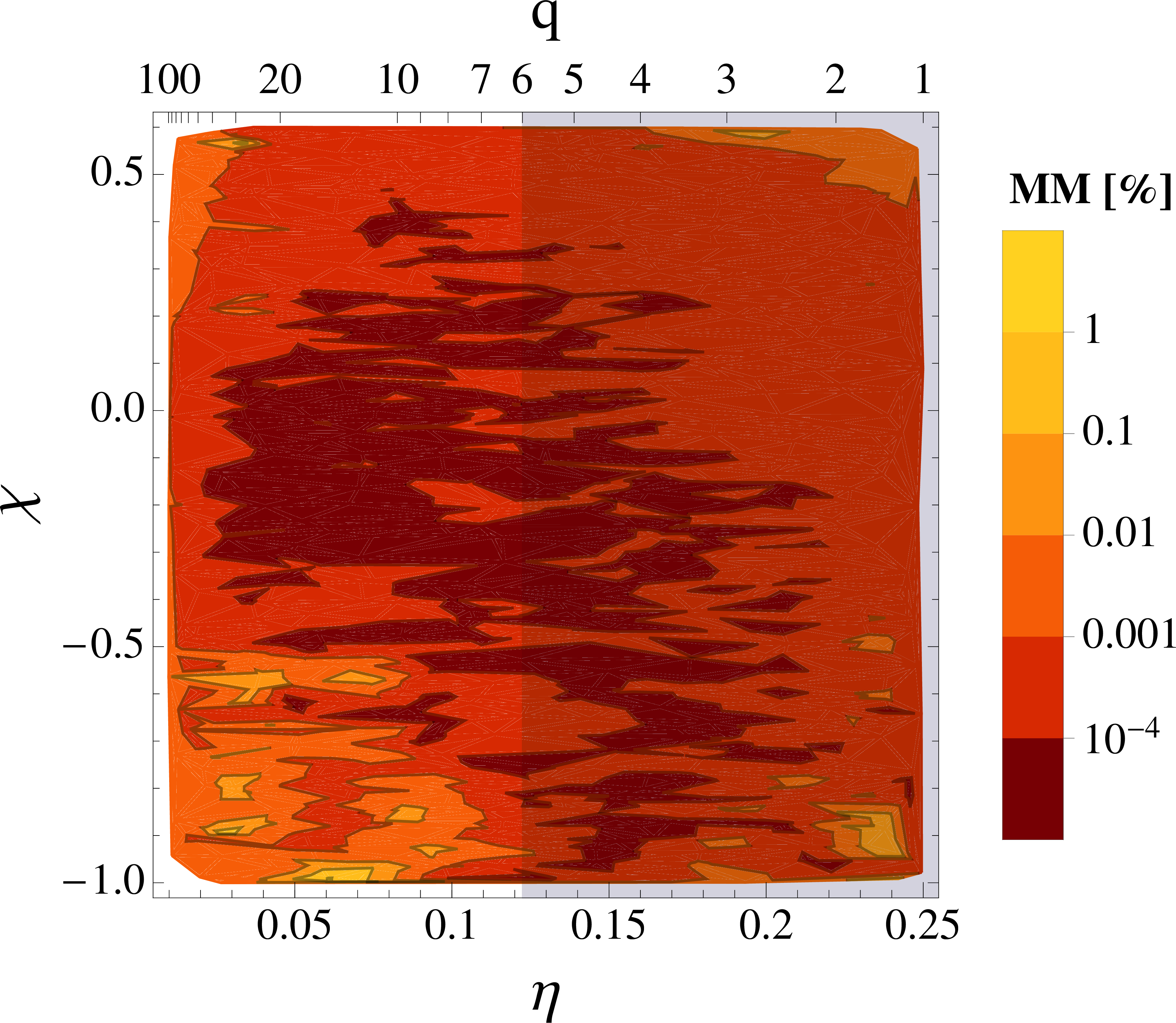}
	\caption{Faithfulness of a full CSE equal-spin $\chi=\chi_1=\chi_2$ tensor-product spline SVD model against SEOBNR waveforms at $m_1=1.4M_\odot$ (left) and $M=100M_\odot$ (right). The calibration region $q \leq 6$ of SEOBNRv1 is slightly shaded.
	In the left panel, the mismatch gets worse in the NS-NS region. This is due to phase error introduced at low frequencies and disappears at slightly higher masses. The mismatch at $m_1=1.4M_\odot$ is below $0.2\%$.
	In both panels isolated regions have mismatches that are significantly higher than the ambient value. These are clustered in the high mass-ratio and high anti-aligned spin region. At $M=100M_\odot$ the mismatch in the isolated regions can be as high as $1\%$ and is due to undersampling of NQCs in the SEOBNRv1 LAL code.
	\Fref{fig:Faithfulness-sparse-ROM-model-histogram-CSE} shows histograms for the same data displayed in this plot.
	}	
  \label{fig:Faithfulness-sparse-ROM-model-contours}
\end{figure}

In~\Fref{fig:Faithfulness-sparse-ROM-model-contours} the mismatch between a full CSE SVD model and 1000 equal-spin waveforms randomly distributed in the symmetric mass-ratio $\eta$ and $\chi=\chi_1=\chi_2$ is displayed. The left panel shows a contour plot of the mismatch at fixed mass $m_1 = 1.4 M_\odot$. The mismatch rises slightly towards equal mass. This is due to phase error at low frequencies.
The mismatch at $m_1 = 1.4 M_\odot$ is below $0.2\%$ for all values of $q$ and $\chi$.
More interesting are localized ``islands'' where the mismatch is higher than the ambient value. The dominant island in the left panel is near $q\sim 30$ and $\chi \sim -0.8$. These do not look as serious as one would expect from~\Fref{fig:projection_coefficients}. The matches are worse there because the model does not capture the behavior of SEOBNRv1 very accurately in this region; but one would expect the matches to be much worse if the model's amplitude and phase projection coefficients were very smooth there, contrary to the peaks visible in~\Fref{fig:projection_coefficients}.

\begin{figure}[htbp]
	\centering
	\includegraphics[width=0.9\textwidth]{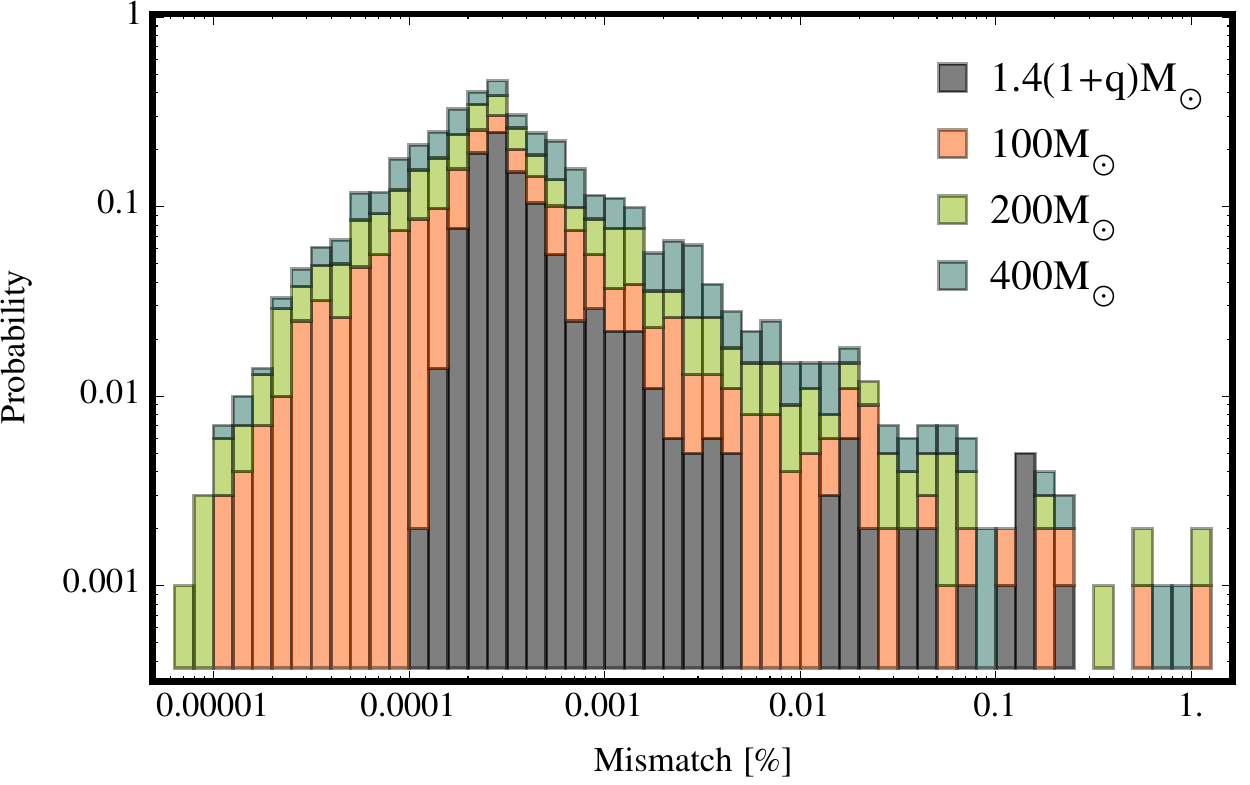}
	\caption{Faithfulness of an single-spin ROM model against 1000 random equal-spin SEOBNRv1 waveforms. The histogram shows mismatches for $m_1=1.4M_\odot$, $M=100M_\odot$, $M=200M_\odot$, $M=400M_\odot$. 
	Compare with~\Fref{fig:Faithfulness-sparse-ROM-model-contours} for contour plots of the same data for $m_1=1.4M_\odot$, $M=100M_\odot$.
	Two configurations at high mass-ratio and high anti-aligned spin lead to mismatches as high as $1\%$ and are further discussed in the text.
	}
	\label{fig:Faithfulness-sparse-ROM-model-histogram-CSE}
\end{figure}

\begin{figure}[htbp]
	\centering
	\includegraphics[width=0.47\textwidth]{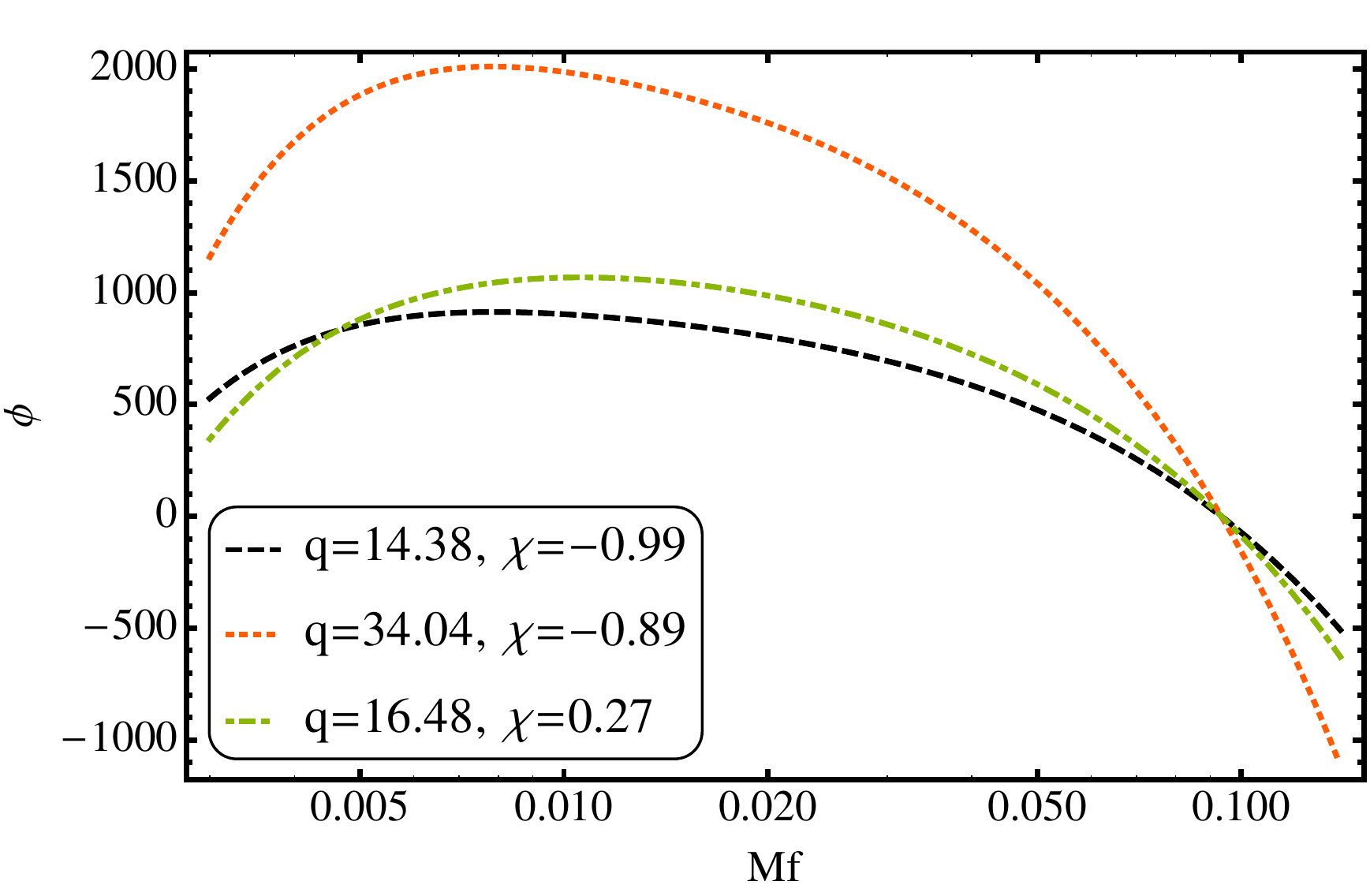}
	\includegraphics[width=0.47\textwidth]{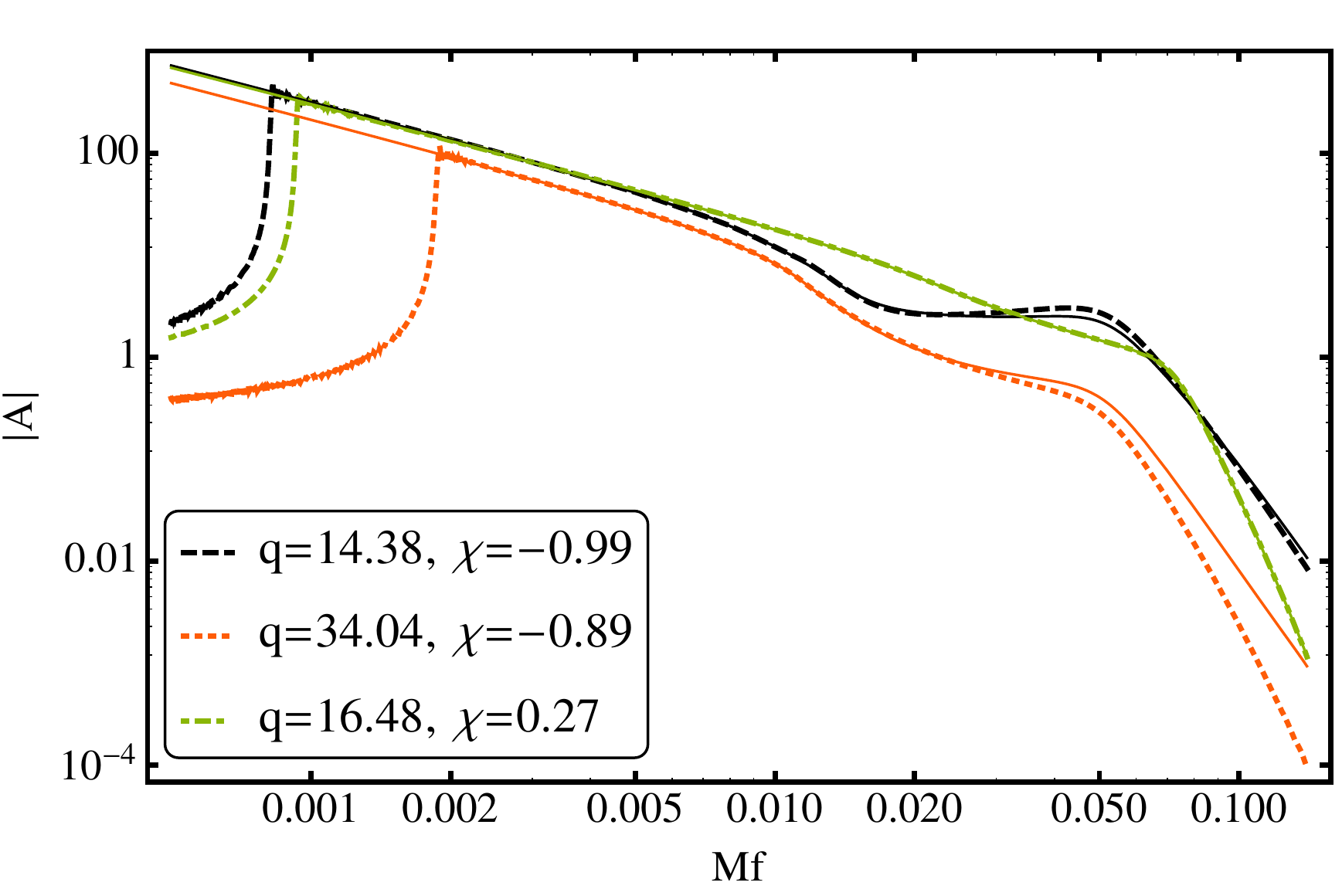}
	\includegraphics[width=0.47\textwidth]{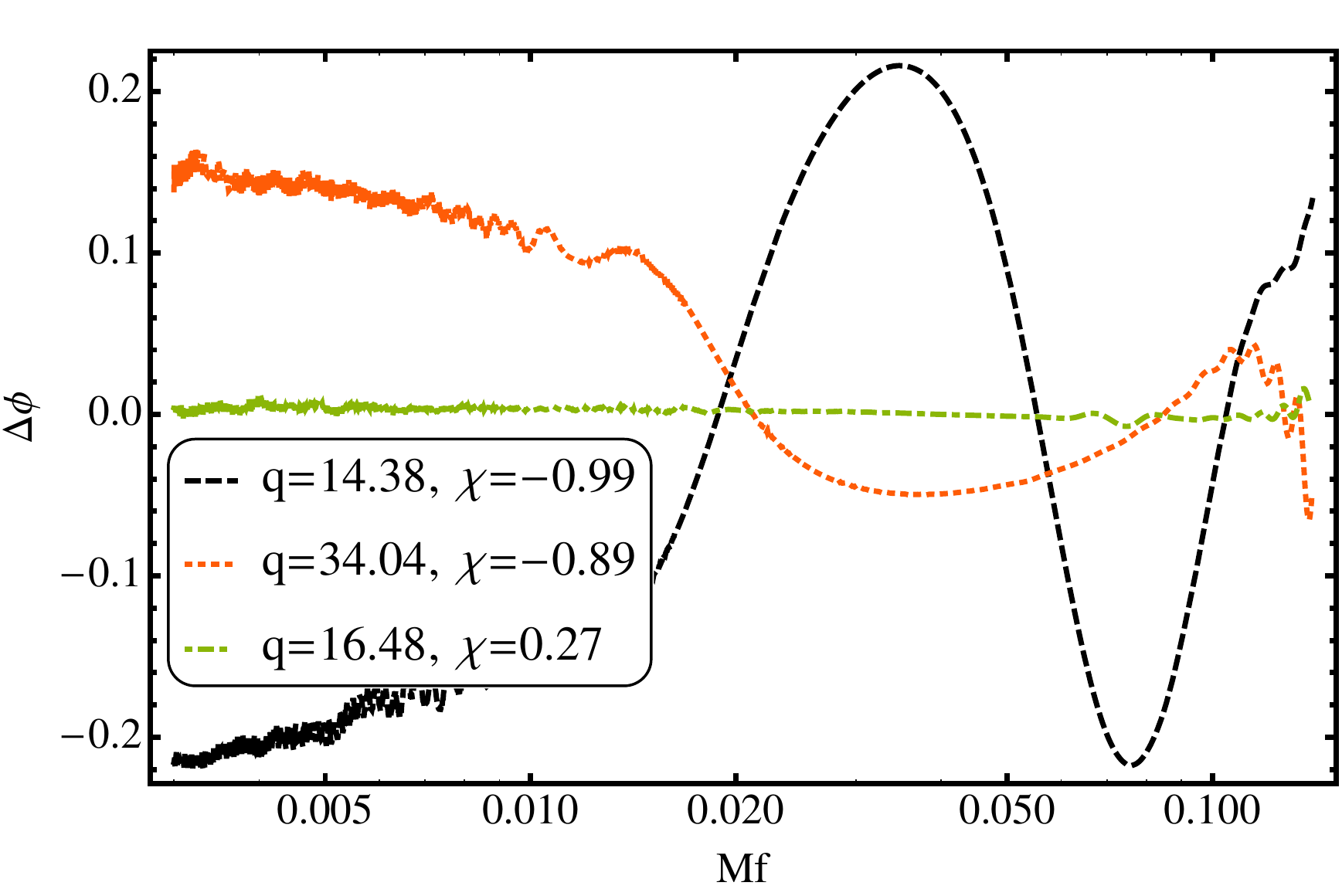}
	\includegraphics[width=0.47\textwidth]{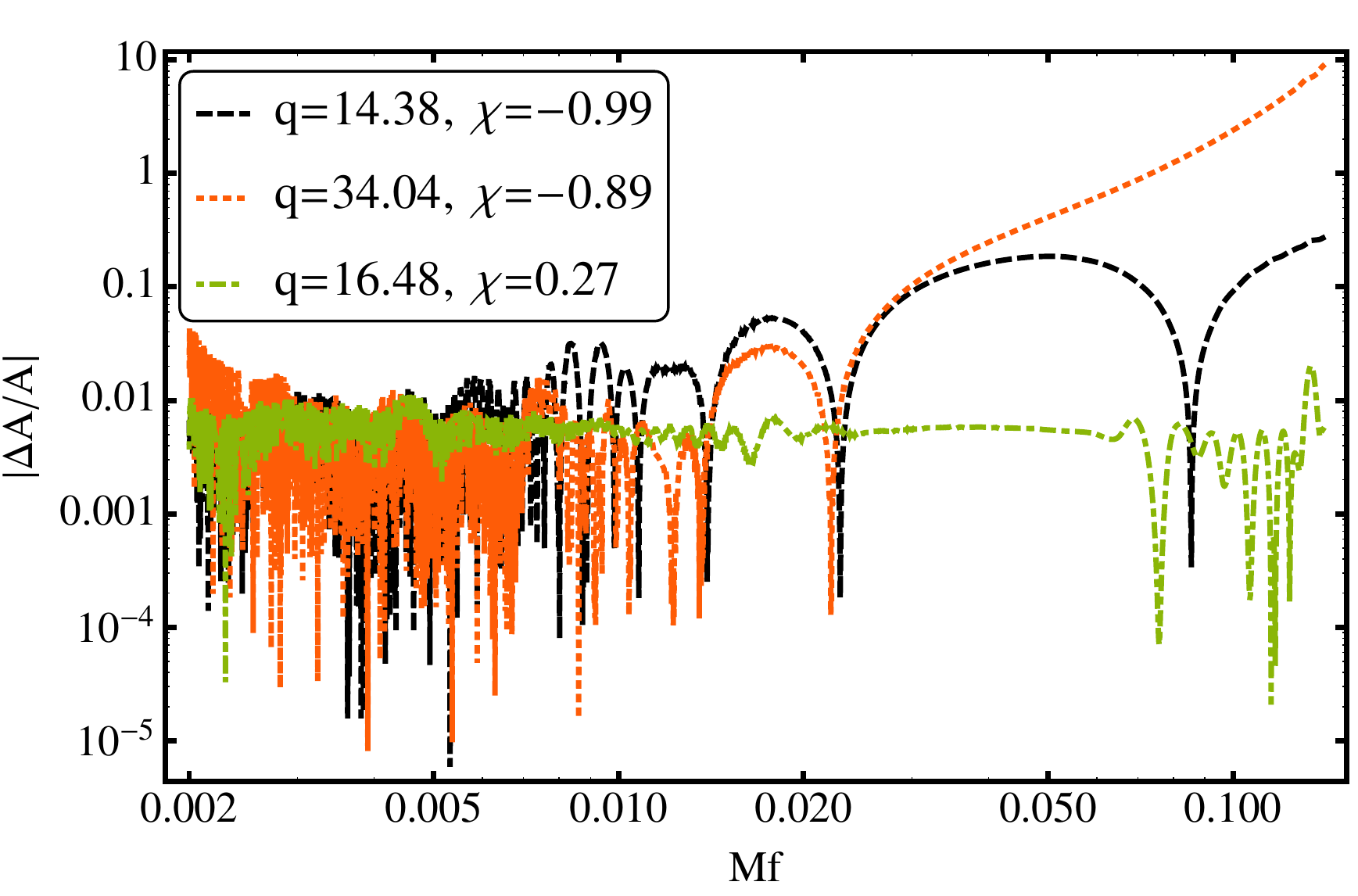}
	\caption{
		The left panels shows the phase (top) and phase error (bottom) between the SEOBNRv1 waveform and the CSE ROM model for the two configurations with highest mismatch at $100 M_\odot$ as shown in~\Fref{fig:Faithfulness-sparse-ROM-model-contours} and~\Fref{fig:Faithfulness-sparse-ROM-model-histogram-CSE}. For the sake of comparison a ``typical'' configuration is shown as well. A linear fit in frequency has been subtracted. Only frequencies required for computing the overlap at $100 M_\odot$ and $15 \,$ Hz or higher are shown. 
		In the right panels the amplitudes (top) and amplitude errors (bottom) of the SEOBNRv1 test waveforms and the model amplitudes (thin lines) are shown. 
	}
	\label{fig:Phase-error-sparse-ROM-model-histogram-CSE}
\end{figure}

The right panel of~\Fref{fig:Faithfulness-sparse-ROM-model-contours} plots the mismatch at a total mass of $100 M_\odot$. There we see multiple islands at high negative spin.
Isolated configurations (compare with the histogram in~\Fref{fig:Faithfulness-sparse-ROM-model-histogram-CSE}) have mismatches above $0.5\%$.
These are located at $q\sim 14.38$, $\chi_i \sim -0.99 $
and $q \sim 34.04$, $\chi_i \sim -0.89$.
Except for these two configurations the mismatch is below $0.3\%$.
These configurations are close to ``clusters'' in the greedy reduced basis (see \Fref{fig:greedy_basis_picks}) and are due to undersampling of NQCs in the SEOBNRv1 LAL code in these parameter space regions.
The phase error for these configurations is shown in~\Fref{fig:Phase-error-sparse-ROM-model-histogram-CSE}.
As can be seen from~\Fref{fig:Phase-error-sparse-ROM-model-histogram-CSE} the phase and amplitude error for these configurations is nonlinear at fairly high frequencies.
\Fref{fig:Faithfulness-sparse-ROM-model-histogram-CSE} shows histograms of mismatches against the CSE single-spin ROM model for the same random configurations as in~\Fref{fig:Faithfulness-sparse-ROM-model-contours}. In addition histograms are also shown for total masses $200 M_\odot$ and $400 M_\odot$.

\begin{figure}[htbp]
	\centering
	\includegraphics[width=0.9\textwidth]{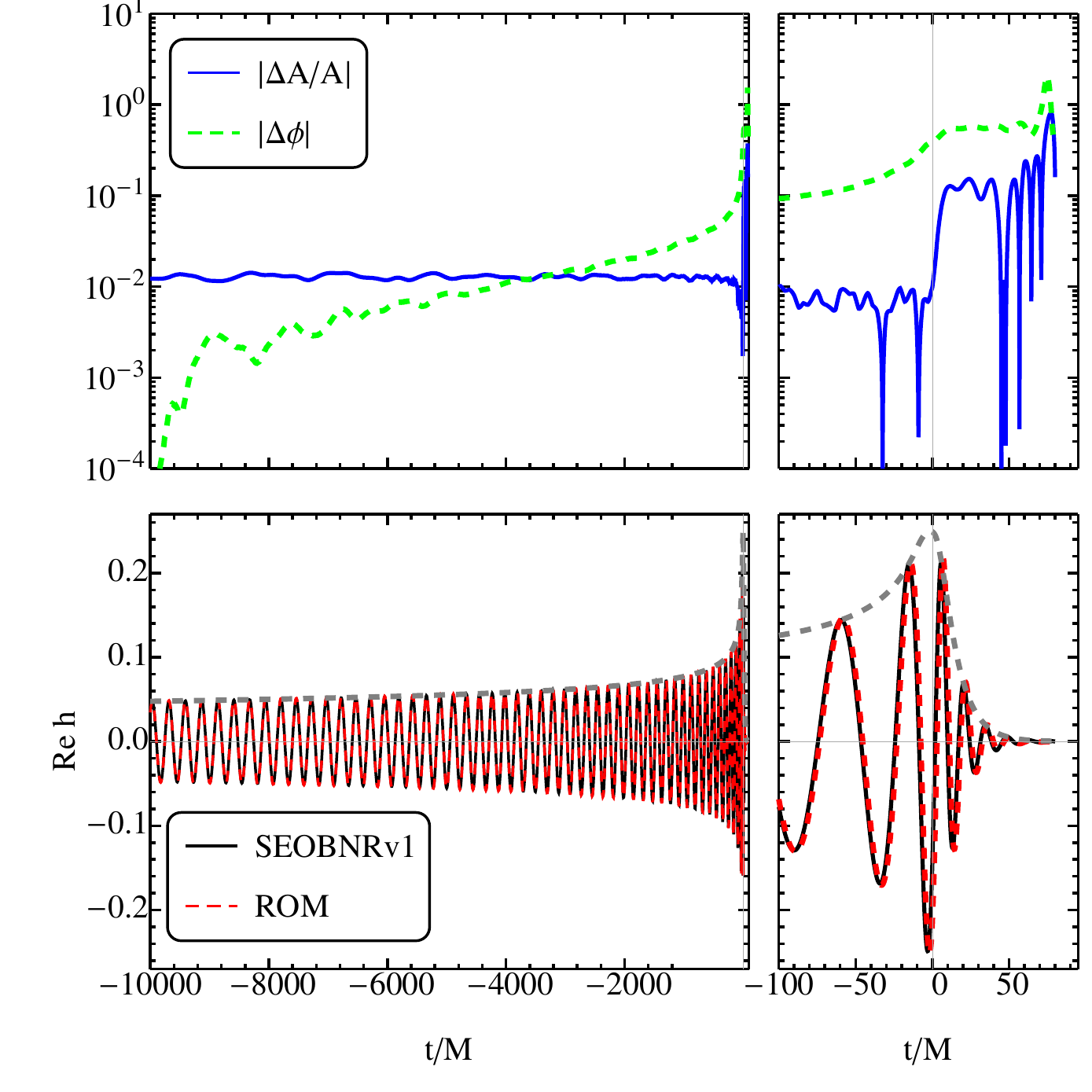}
	\caption{Comparison of iFFT of ROM model waveform (red, dashed) with SEOBNR time-domain waveform (black) for $q=1.11$, $\chi_1=\chi_2=-0.963$. The waveforms have been aligned in time at the amplitude maximum and aligned in phase
	so that the phases agree well before the merger. The entire waveform is $\sim 5 \times 10^7 M$ long.
	}
	\label{fig:TD-comparison}
\end{figure}

\begin{figure}[htbp]
	\centering
	\includegraphics[width=0.47\textwidth]{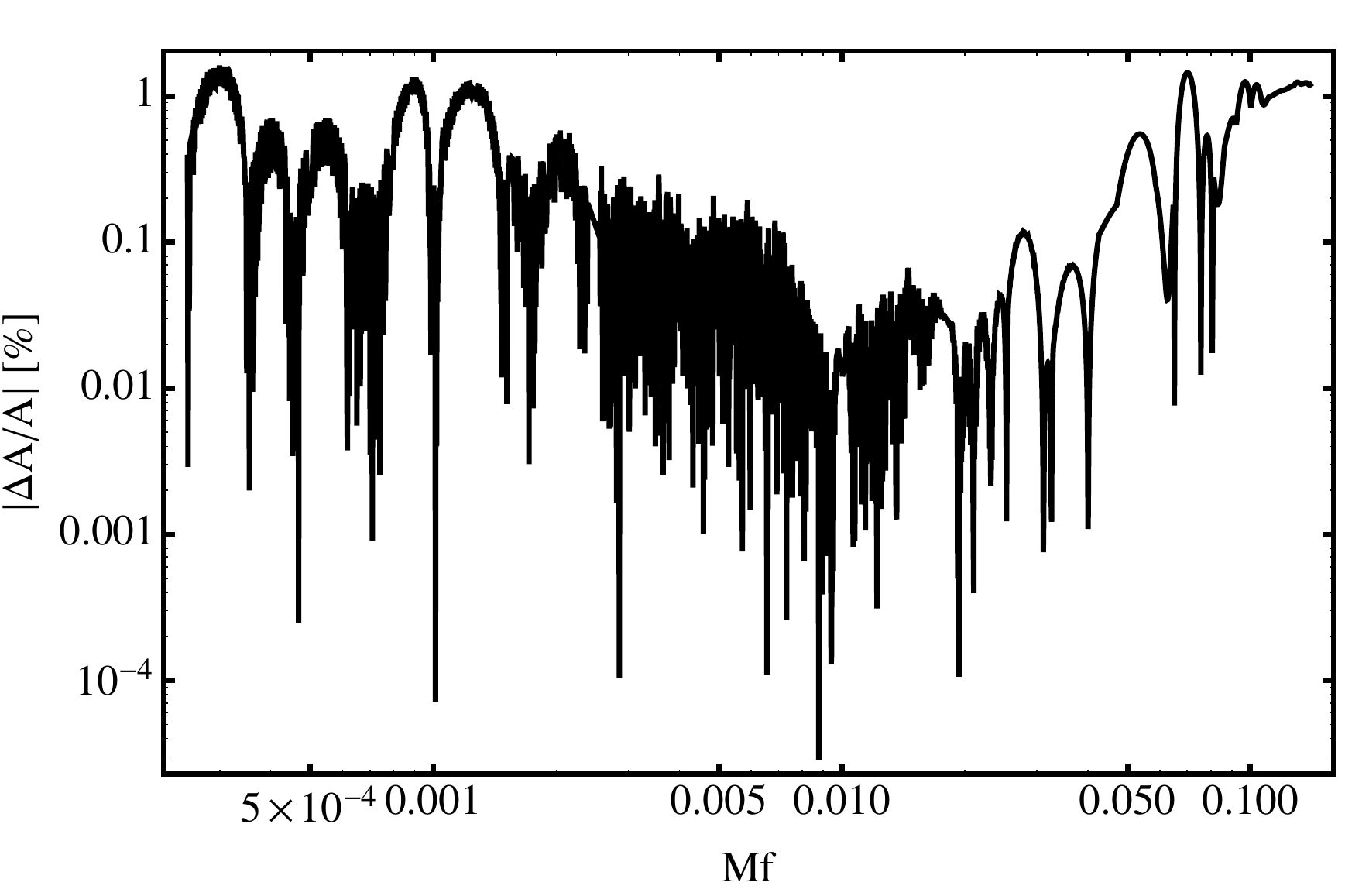}
	\includegraphics[width=0.47\textwidth]{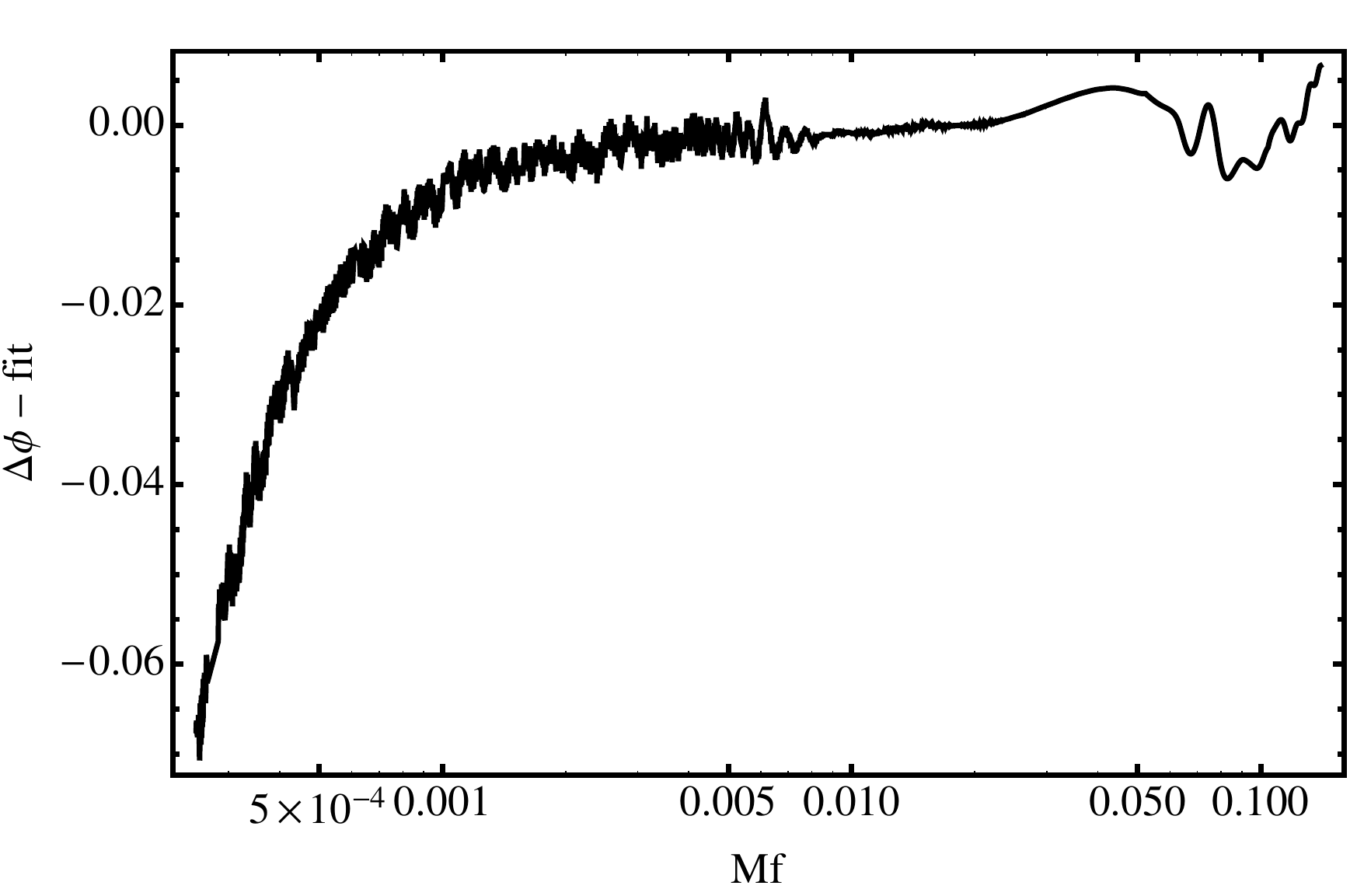}
	\caption{Frequency domain amplitude and phase errors for the configuration $q=1.11$, $\chi_1=\chi_2=-0.963$ shown in~\Fref{fig:TD-comparison}.
	}
	\label{fig:TD-comparison-FD-errors}
\end{figure}

To corroborate the fact that the ROM model is of high fidelity beyond high faithfulness I show a time-domain comparison
in~\Fref{fig:TD-comparison}. The inverse FFT of the ROM model is compared against an SEOBNRv1 waveform. The upper panel shows the amplitude and phase errors. The amplitude error is about $1\%$ right until the amplitude maximum, then rises to $10\%$ and more in the final stages of the ringdown. Similarly, the phase error (with alignment at $t=- 10000 M$) is below $\sim 0.1 \,$ rad until shortly before the merger, rises to $\sim 0.6 \,$ rad through merger and peaks at $2\,$ rad well into the ringdown. The bottom panel plots the real part of the strain waveform. The disagreement in the waveform is barely visible near the merger. The accuracy of the ROM model is comparable to amplitude and phase errors quoted for the SEOBNRv1 model in~\Sref{sec:SEOBNRv1}.
The mismatch at $M > 12M_\odot$ is smaller than $0.005\%$. The frequency domain phase error (see~\Fref{fig:TD-comparison-FD-errors}) is below $0.015 \,$ rad from $Mf \sim 0.0009$ ($15\,$ Hz lower cutoff at $12 M_\odot$) to $Mf =0.14$. The frequency domain amplitude error is below $1\%$.
The phase error in the Fourier domain should not be confused with the phase error in the time-domain, which is a more familiar quantity.

\section{Results for double spin models}
\label{sec:double-spin-model}

So far the discussion has mainly focussed on the single spin model. Since the source SEOBNRv1 model allows for two independent spin components I have also constructed a double spin model in a more restricted parameter space region. Here I discuss some of the choices made for that model and assess its fidelity.

Due to the increase in dimensionality and the indications that the source model may not be correct at higher mass-ratios (see the clustering of waveforms in~\Fref{fig:greedy_basis_picks}) I choose to restrict the extent of the parameter space covered to $q \leq 10$. While this goes beyond the $q=6$ calibration domain of SEOBNRv1, this choice is motivated by the model's $0.4\%$ mismatch against a non-spinning $q=10$ NR waveform in the NRAR catalog.

\begin{figure}[htbp]
	\centering
		\includegraphics[width=0.7\textwidth]{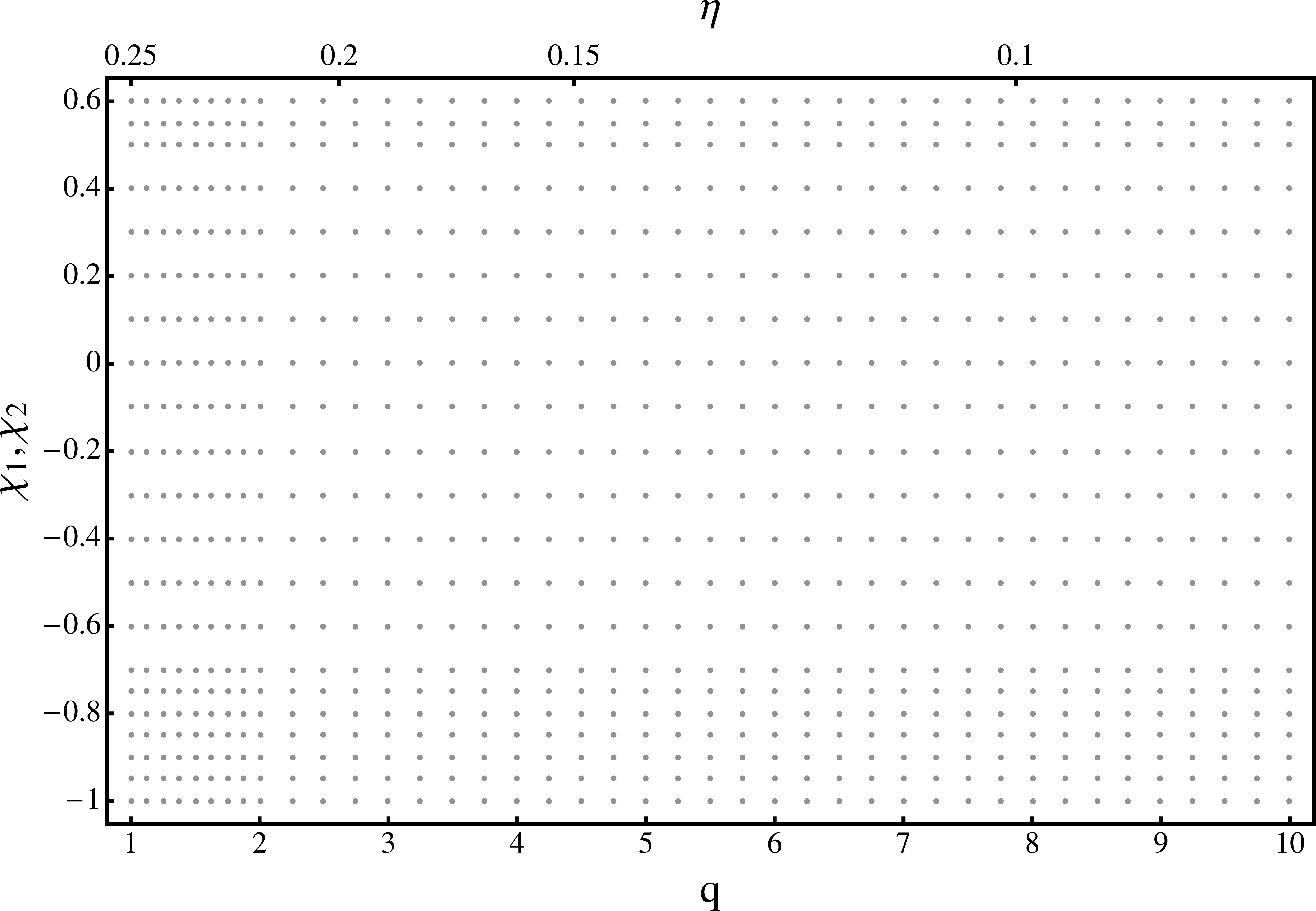}
	\caption{The $41 \times 21 \times 21 = 18081$ SEOBNRv1 starting waveforms used to construct the double-spin model.
	A projection onto a plane spanned by the mass-ratio and one of the spin directions is shown. The second spin direction uses the same selection of points.}
	\label{fig:plots_Templates_aligned_DS}
\end{figure}

I pick a spacing of $\Delta q = 0.25$ and $\Delta\chi_1=\Delta\chi_2=0.1$ with additional waveforms near equal-mass and near the spin boundaries $\chi_i=-1$ and $\chi_i=0.6$. In total I use $41\times 21\times 21 = 18081$ waveforms on a tensor product (rectangular) grid as shown in~\Fref{fig:plots_Templates_aligned_DS}. The waveforms are sampled at a rate of $f_s = 16384 Hz$ and a specified initial frequency of $8 \text{Hz}$ for the LAL code. Due to the high computational cost I generate these waveforms at a higher total mass of $12 M_\odot$ for all mass-ratios considered here in order to avoid hybridization. This eliminates the hybridization error which would be significantly larger at $12M_\odot$ than for BNS waveforms. With a generation time of several minutes this is feasible at this higher total mass. 
The lowest usable frequency in frequency domain waveforms is $Mf \simeq 0.00062$. This is sufficient to fill the detector band down to about $11 \text{Hz}$. For practical models I will generate waveforms that reach the $10 \text{Hz}$ lower cutoff required for aLIGO, but for the proof of principle models discussed here the results displayed use a lower cutoff of $11 \,$ Hz.

\begin{figure}[htbp]
	\centering
	\includegraphics[width=0.9\textwidth]{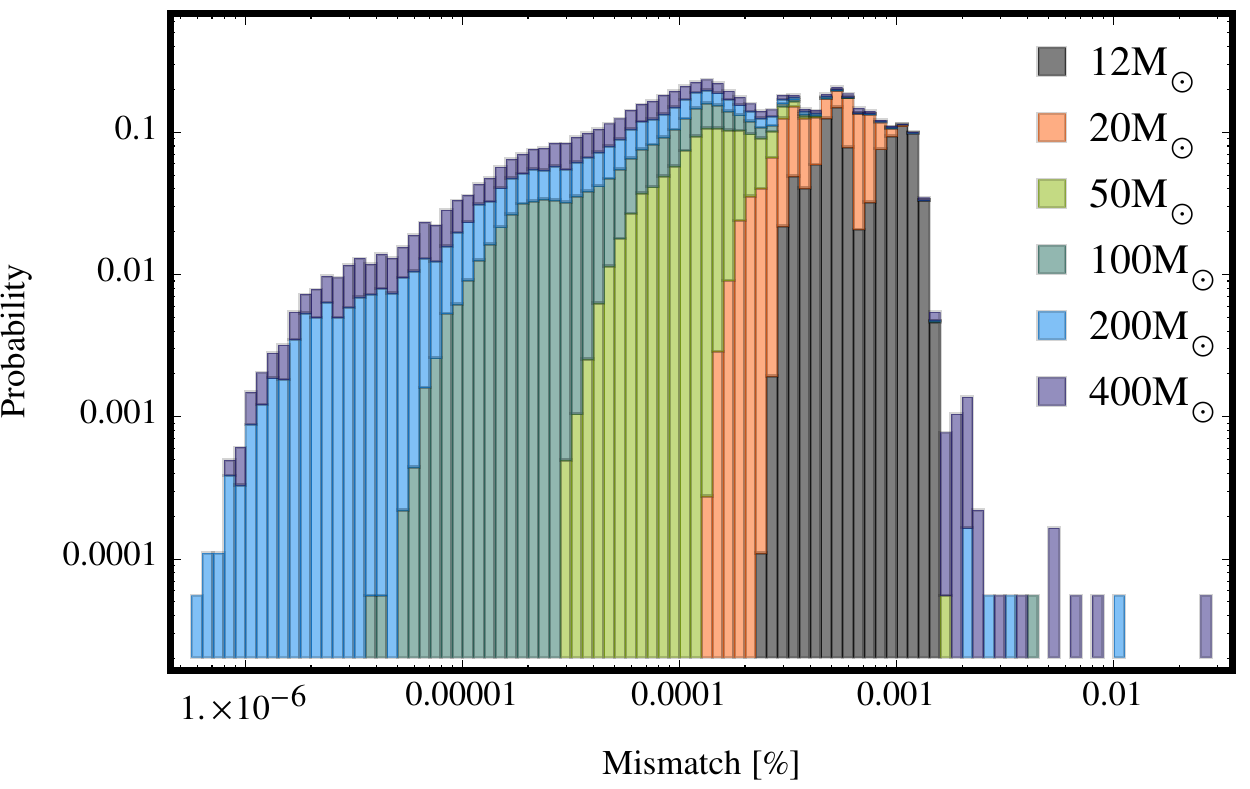}
	\caption{Faithfulness of full double-spin model with CSE points against the set of SEOBNRv1 waveforms the model was built from at a range of total masses. The tail of the mismatch is below $\sim 0.01\%$. This is a measure of the error due to interpolating the waveforms onto the CSE grid combined with the error due to rank reduction from the SVD. The individual errors are show in~\Fref{fig:Faithfulness-modelwfs-CSE-DS-histogram}.
	The mismatch is even smaller on a 10000 point equispaced grid.
	Substantial zero-padding (here a factor 100) or subtraction of a fit from the phase difference is required to get accurate results. 
	}
	\label{fig:Faithfulness-modelwfs-DS-histogram}
\end{figure}

\begin{figure}[htbp]
	\centering
	\includegraphics[width=0.47\textwidth]{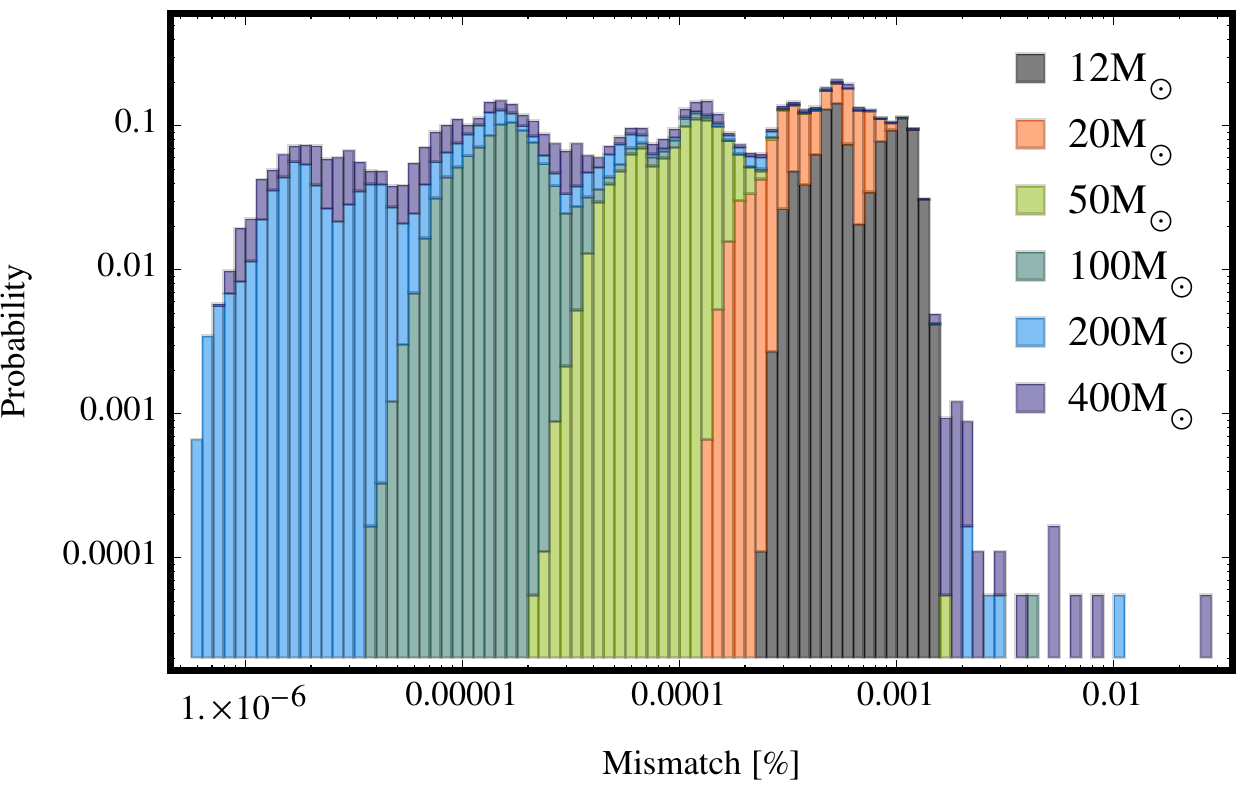}
	\includegraphics[width=0.47\textwidth]{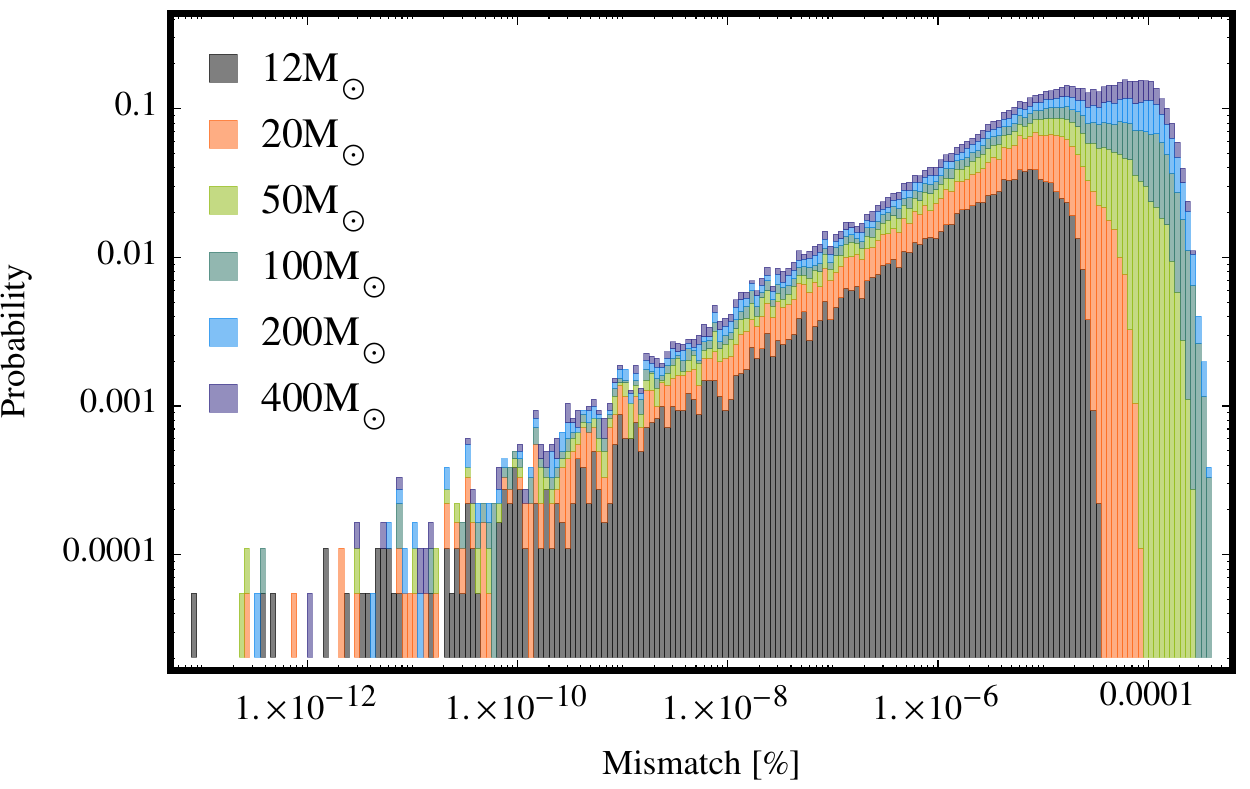}
	\caption{
	Frequency interpolation error (left panel) and rank-reduction error (right panel) for the double spin model. The combination of these errors is shown in~\Fref{fig:Faithfulness-modelwfs-DS-histogram}.
	The left panel shows the mismatch of CSE waveforms against full SEOBNRv1 waveforms for the set of double spin model waveforms for a range of total masses. This mismatch is purely due to frequency interpolation error introduced by interpolating a very dense grid obtained from the FFT of the generated SEOBNRv1 waveforms onto the CSE grids for amplitude and phase.
	The right panel shows the faithfulness of full CSE double-spin model against the set of CSE SEOBNRv1 waveforms the model was built from for a range of total masses.  The mismatch shown here is purely due to ``rank reduction'' in building the model as the number of frequency points is smaller than the number of starting waveforms.
	This mismatch is smaller than $4 \times 10^{-4} \%$.
	}
	\label{fig:Faithfulness-modelwfs-CSE-DS-histogram}
\end{figure}

I interpolate the waveforms obtained from the FFT of time-domain SEOBNRv1 waveforms onto a constant spline interpolation error (CSE) grid as discussed in~\Sref{sec:const_spline_int_error_points}. I have already investigated the error introduced by this interpolation in frequency in~\Fref{fig:Faithfulness-modelwfs-DS-histogram}, but I study it here again for all input waveforms. First I show the faithfulness of the CSE model against the set of input waveforms in~\Fref{fig:Faithfulness-modelwfs-DS-histogram}. The mismatch is below $0.01\%$ for almost all waveforms and total masses considered. It arises from two individual error sources: (a) the mismatch induced by the interpolation of each input waveform onto the CSE amplitude and phase points in frequency and (b) the ``rank-reduction'' error that stems from using fewer frequency points than input waveforms in the SVD.
These errors are investigated individually in~\Fref{fig:Faithfulness-modelwfs-CSE-DS-histogram}. We see that the frequency interpolation error (left panel) can be as high as $\sim 0.01 \%$, while the rank-reduction error (right panel) is below $4 \times 10^{-4} \%$ for all total masses considered. This is consistent with the total mismatch shown in~\Fref{fig:Faithfulness-modelwfs-DS-histogram} and the bound on the combination of mismatches given in~\Eref{eq:mismatch_triangle_inequality}. This shows that the mismatch due to frequency interpolation is the larger of these two error sources.

\begin{figure}[htbp]
	\centering
	\includegraphics[width=0.9\textwidth]{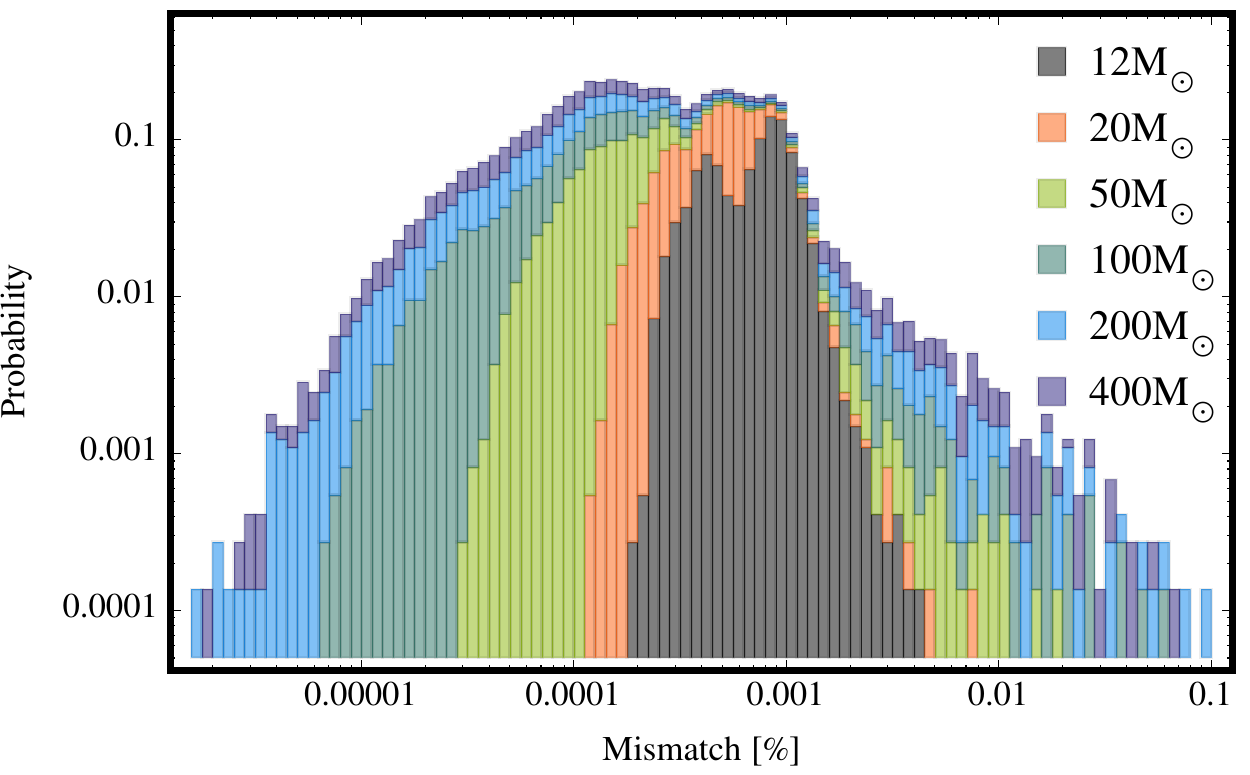}
	\caption{Faithfulness of full double-spin model with CSE points against SEOBNRv1 waveforms at a range of total masses  at about 7300 randomly chosen points in the domain of the model for the ``zero-detuned high-power'' aLIGO noisecurve. The tail of the mismatch is always below $0.1\%$ and below $0.01\%$ for systems with total mass below $20 M_\odot$. Substantial zero-padding (here a factor 100) or subtraction of a fit from the phase difference is required to get accurate results. The increase in mismatch compared to~\Fref{fig:Faithfulness-modelwfs-DS-histogram} is due to interpolation over the parameter space. 
	}
	\label{fig:Faithfulness-full-DS-histogram}
\end{figure}

Finally, I test how faithful the CSE double-spin model is against a set of about $7200$ SEOBNRv1 waveforms which are generated at random points in the parameter domain of the model. As shown in~\Fref{fig:Faithfulness-full-DS-histogram} the mismatch for the ``zero-detuned high-power'' aLIGO noisecurve is better than $0.01\%$ up to $20 M_\odot$, better than $0.02\%$ up to $50 M_\odot$ and below $0.1\%$ beyond $100 M_\odot$. The mismatch is comparable for the ``early
aLIGO''~\cite{G1000176} noisecurve, better at low total masses and slightly worse at high total masses.
Comparing with~\Fref{fig:Faithfulness-modelwfs-DS-histogram} we see that interpolation error over the parameter space is the dominant error source in the final model.
The highest mismatches arise predominantly for configurations near mass-ratio $q=10$ and near extremal negative spin on the larger BH, i.e. near two of the boundaries of the parameter domain.

\section{Discussion}
\label{sec:discussion}

I give a brief comparison of SVD and greedy reduced basis methods, summarize the accuracy, performance, speed and storage requirements of the SEOBNRv1 reduced order models discussed in this paper. Finally I address how the techniques presented here can be extended to build models for GWs from generic binaries.

A matter of practical importance are the storage requirements of reduced order models.
As a point of reference we define as the starting size the set of input waveforms on a $10000$ point frequency grid, which is about $3.6 \times 10^8$ double precision floating point numbers for the double-spin input set considered here. The final size of the reduced basis for the double-spin CSE model is $123^2 + 95^2 = 24154$ floating point numbers. However the model also includes projection matrices for all input waveforms. This is the dominant contribution to storage size. The overall size of the double-spin CSE model is about $30 \text{MB}$.
Compared to the size in bytes of the starting waveforms the information has been compressed by roughly a factor of $70$.
The total storage size is determined by the number of sparse frequency points (this determines the size of the basis matrices) and the number of input waveforms covering the parameter space (this sets the size of the projection coefficients). To reduce the data required for the model, the number of input waveforms could be reduced. This would lead to an increase in the interpolation error over the parameter space which is already the dominant ROM error. An alternative to achieve higher compression is to use fits to the projection coefficients instead of interpolation. Fits require in general a bit of tweaking and may introduce further errors, while interpolation is automatic and robust.

Looking back at~\Eref{eq:model_function} the evaluation cost of the ROM consists of evaluating the tensor product interpolants of the projection coefficients $I_\otimes[\mathcal{M}](q,\chi)$, matrix multiplication between the reduced bases $\mathcal{B}$ and the $I_\otimes[\mathcal{M}]$, and the construction and evaluation of the one-dimensional cubic splines $I_f$ for the desired number of frequency points.
For full reduced bases $k=m$, the total cost goes as $\bigO(m) + \bigO(m^2 n) + \bigO(L \log m)$, where $m$ is the number of sparse frequency points, $n$ the number of input waveforms and $L$ the number of model evaluations in frequency.
The solution of the linear systems for constructing the tensor product interpolants over the parameter space can be cached; therefore the associated cost is not included here.

Let us mention that we could have used a rank reduced SVD to make models based on a sparse set of frequency points (such as CSE points) still more compact. The size of the projection coefficients $\mathcal{M} \in \mathbb{R}^{k \times n}$ and the reduced basis matrix $\mathcal{B}_k \in \mathbb{R}^{m \times k}$ decreases when truncating the SVD at rank $k < r \leq m$. However, since the number of frequency points $m$ was already chosen to be small, there may not be much further leeway for compression without significant loss of fidelity. Therefore the SVD truncation technique is more useful when working with a large set of frequency points.

As discussed in~\Sref{sec:SEOBNRv1} the SEOBNRv1 model was found to be faithful to $0.5\%$ for total masses between $20 M_\odot$ and $200 M_\odot$ using the aLIGO ``zero-detuned high-power'' noisecurve. I showed in~\Sref{sec:double-spin-model} that the double-spin surrogate model is faithful to SEOBNRv1 to about $0.1\%$ for total masses above $100 M_\odot$ and $0.01\%$ for $\sim 50 M_\odot$ or lower. Clearly the ROM error is much smaller than the uncertainty in the original model. We can then estimate that the mismatch between ``true'' NR waveforms (against which SEOBNRv1 has been compared) and the surrogate model presented here is better than $1\%$ at high total masses and better than $0.6 \%$ below $50 M_\odot$ by using the triangle inequality~\cite{Ohme:2011rm}
\begin{equation}
	\label{eq:mismatch_triangle_inequality}
  \mathcal{M}(h_\text{ROM}, h_\text{true}) \leq \left( \sqrt{\mathcal{M}(h_\text{ROM}, h_\text{SEOBNRv1})} + \sqrt{\mathcal{M}(h_\text{SEOBNRv1}, h_\text{true})} \right)^2.
\end{equation}
Note that the actual total mismatch may be much lower than the bound obtained from this inequality.
Since ROM error is much smaller than the uncertainty in the original model it is not necessary to improve the surrogate model even more.

An important practical question for waveform models is how low the mismatch must be so that PE calculations are not compromised. This can be answered by requiring that the error of the model against a set noise level be \emph{indistinguishable}~\cite{Lindblom:2008cm,Ohme:2011rm} for a detector.
For the waveform error $\lVert \delta h \rVert$ to be smaller than unity the mismatch needs to satisfy $\mathcal{M} \lesssim 1/(2 \rho^2)$. Assuming that the SEOBNRv1 is ``true'' we see then that the double-spin model with its modeling mismatch of $\mathcal{M} \sim 0.1\%$ is sufficiently accurate for PE up to SNR $20$ for total masses above $50 M_\odot$ and up to SNR $50$ for lower total masses.
This is comparable or better than the $1\%$ mismatch spacing of templates used in the SVD-based model tested in~\cite{Smith:2012du} for PE applications.
It should be noted that the above criterion on the mismatch is a sufficient but not a necessary criterion and it can be too stringent as shown in~\cite{Littenberg:2012uj} for the nonspinning EOBNRv2 model~\cite{Pan:2011gk}.

This paper presents a general prescription for building SVD-based frequency domain models for GWs from aligned spin binaries and the first practical reduced order model of SEOBNRv1 that can be used for GW data analysis. The ROM model can be used for searches, but also for parameter estimation where such a model is urgently needed to enable in-depth studies that have so far not been possible due to the prohibitive cost of generating SEOBNRv1 waveforms on the fly. A reduced order model for SEOBNRv2 will be made available in the publicly accessible LAL software package~\cite{LAL-web}.

The modeling techniques discussed here can be applied to \emph{any} aligned spin model. Producing a model for the followup SEOBNR model~\cite{Taracchini:2013rva} would be useful in order to assess the improvements over the parameter space, lift the restriction on aligned spins $\chi_i < 0.6$ that is present in the current model and to provide a model with even higher fidelity. The techniques can be extended in a straightforward way to create reduced order models including higher modes. 

Models of GWs from precessing binaries can be built either on a mode-by-mode basis, or, following the way recent models have been constructed, by twisting up (using Wigner rotations) aligned spin waveforms with time- or frequency-dependent angle functions that describe the motion of the orbital plane. Thus, in addition to the aligned spin waveform, the angle functions will need to be modeled in an efficient manner~\cite{Pan:2013rra,Hannam:2013oca,Lundgren:2013jla}.
Modeling the full precessing parameter space without an (approximate) mapping between aligned spin and precessing waveforms is a daunting task and due to the lack of a sufficient number of NR waveforms or IMR models that take this approach it is only applicable in the PN regime. There, a reduced order model could shed light on degeneracies that are suppressed by the decomposition of the space using Wigner rotations.


\section{Acknowledgements} 

It is a pleasure to thank M.~Hannam, F.~Ohme, A.~Buonanno, M.~Tiglio, S.~Field, C.~Galley, R.~Smith, C.~Hanna, S.~Husa, P.~Ca\~nizares-Martinez, B.~Sathyaprakash, S.~Fairhurst, P.~Sutton for useful discussions and comments.
I thank the anonymous referees for their detailed comments and suggestions which helped to improve the presentation of the paper.
This work has made use of Mathematica packages written by S.~Husa, M.~Hannam, F.~Ohme and myself.
I acknowledge support by Science and Technology Facilities Council grants ST/I001085/1 and ST/L000962/1.


\appendix


\section{Hybridization}
\label{sec:hybrids}

SEOBNRv1 waveforms are very expensive to generate, in particular, when the total mass becomes close to that of a neutron star. As one goes to higher mass-ratios the lowest meaningful system mass is going to be higher than $q+1$ times the mass of a neutron star. To ease the computational burden it makes sense to generate waveforms at mass-ratio $q$ at these total masses. However, in order to build a model and used reduced basis techniques such as the SVD, all waveforms have to be defined on the \emph{same} frequency grid. To achieve this we create frequency domain TaylorF2 hybrids and follow a procedure outlined in~\cite{Purrer:2013ojf, Santamaria:2010yb}. 

The calculation of SEOBNRv1 frequency domain waveforms was already described in~\Sref{sec:templates}. The matching procedure aligns $\tilde h_\text{PN}(f;t_0,\phi_0) = \tilde h_\text{F2}(f) + 2\pi f t_0 + \phi_0$ and $\tilde h_\text{EOB}(f)$ by a least squares fit over a frequency fitting interval $[f_1,f_2]$. We then determine the matching frequency $f_m \in [f_1,f_2]$ at which the NR and EOB phases coincide by a root-finding algorithm. The PN and EOB amplitudes are aligned separately without any freedom to adjust parameters.
The hybridization interval for amplitude and phase is chosen as follows
\begin{align}
	M f_{m,\phi} 	&= M f_0 [1.03, 1.15]\\
	M f_{m, A} 		&= M f_0 [1.05, 1.2],
\end{align}
where $f_0$ is the lowest usable frequency for each waveform. This quantity rises with the mass-ratio.

The impact of hybridization on the ROMs is very small since the added low frequency content is only needed to perform the SVD of the waveform matrix and is not going to be in the detector frequency band when the system mass is chosen to be higher than $q+1$ times the mass of a neutron star. In that sense any smooth low frequency extension could be used, even if it were unphysical.

\section{Greedy frequency points}
\label{sec:greedy_frequency_points}

This section discusses an alternative algorithm for producing sparse frequency points. The basic idea is to successively add grid points at which the relative or absolute errors are highest to a skeleton grid.

Separate greedy algorithms (see~\ref{alg:greedy-amplitude-points} and \ref{alg:greedy-phase-points}) are used to find a set of amplitude or phase points $\mathcal{G}$. Let $\mathcal{A}$ / $\Phi$ be a subset of input waveform amplitudes / phases that are spline interpolated over the frequency grid $G$. 
The subsets $\mathcal{A}$ / $\Phi$ can be chosen as the first, say 150, greedy basis configurations (as obtained from Algorithm 1 in~\cite{Field:2013cfa}). This will cover the major variations in the functions and so ensure that all waveforms will be interpolated to high accuracy.

We use different error measures for amplitudes and phases. We discuss the algorithm for the amplitude first and then point out the differences in the algorithm used for the phases. We start with a skeleton grid consisting of just the endpoints. In each iteration a new amplitude point is found in~\ref{alg:greedy-amplitude-points} by first computing relative error vectors $e_i$ and their $\ell_\infty$ norms $N_i$ for each amplitude in the set $\mathcal{A}$. Then we find the index $k$ of the error vector with the largest error norm and the grid point at which the error is largest in $e_k$ that is not yet part of our current grid. We iterate until a desired error tolerance has been obtained.
To generate the phase points in~\ref{alg:greedy-phase-points} we use a combination of relative and absolute errors, subtract a linear fit from each phase difference to remove the influence of time and phase shifts which are of no interest.
These algorithms are inspired by the \emph{empirical interpolation method}~\cite{Barrault:2004,Field:2013cfa,Canizares:2013ywa} in the fact that the point with the largest pointwise error is promoted to be a new grid point. 

One can use e.g. cubic spline interpolation for its simplicity and robustness, although high order polynomial interpolation can be used as well. In applying this method to subsets of SEOBNRv1 input waveforms considered in this paper I found that the error falls off steeply for the first about 50 points, but then tends to threshold, especially for the amplitude. This surprising flattening would benefit from future study. The error becomes smoother as the size of the subsets $\mathcal{A}$ / $\Phi$ is increased. Lower errors can be obtained for smaller subsets.

\begin{algorithm}
	\caption{Greedy frequency points for amplitude.}
	\label{alg:greedy-amplitude-points}
	\begin{algorithmic}
		\State $\mathcal{G} = \{ f_\text{min}, f_\text{max} \}$
		\State $err \gets 1$
		\While {$err \geq tol$}
			\State $\mathcal{A}_\mathcal{G} \gets \mathcal{I}[\mathcal{A}[\mathcal{G}]]$ \Comment{Interpolate amplitudes onto current greedy grid.}
			\For{$i \gets 1, \len(\mathcal{A})$} \Comment{Loop over selected waveform amplitudes}
				\State $e_i \gets \left(\lvert \mathcal{A}_i[G] - (\mathcal{A}_\mathcal{G})_i[G] \rvert \right) / \mathcal{A}_i[G]$	\Comment{Compute relative error for each amplitude.}
				\State $N_i \gets \lVert e_i \rVert_\infty$
			\EndFor
			\State $err \gets \sum_i N_i / \len(\mathcal{A})$
			\State $k \gets \argmax(N)$ \Comment{Find the error vector with the largest error norm.}
			\State $S \gets \text{sort}[\{G, e_k\}^T, 2]^T_1$ \Comment{Sort grid points in descending order of the largest errors.}
			\State $p \gets (S \setminus \mathcal{G})_1$ \Comment{Find grid point with the largest error that is not yet part of the grid $\mathcal{G}$.}
			\State $\mathcal{G} \gets \mathcal{G} \cup \{ p \}$ \Comment{Place a new point there.}
		\EndWhile
		\State $\mathcal{G} \gets \text{sort}(\mathcal{G})$
	\end{algorithmic}	
\end{algorithm}

\begin{algorithm}
	\caption{Greedy frequency points for phase.}
	\label{alg:greedy-phase-points}
	\begin{algorithmic}
		\State $\mathcal{G} \gets \{ f_\text{min}, f_\text{max} \}$
		\State $err \gets 1, iter \gets 1, \gamma \gets 10^{-3}$
		\While {$err \geq tol$ and $iter \leq maxiter$}
			\State $\Phi_\mathcal{G} \gets \mathcal{I}[\Phi[\mathcal{G}]]$ \Comment{Interpolate phases onto current greedy grid.}
			\State $\Delta\Phi[G] \gets \Phi[G] - \Phi_\mathcal{G}[G]$
			\For{$i \gets 1, \len(\Phi)$} \Comment{Loop over selected waveform phases}
				\State $e_{\text{rel},i} \gets \left(\lvert \Delta\Phi_i[G] \rvert \right) / \left( 1 + \lvert \Phi_i[G] \rvert \right)$	\Comment{Compute relative error for each phase.}
				\State $e_{\text{abs},i} \gets \lvert| \Delta\Phi_i[G] -\text{Fit}_i \rvert|$ \Comment{Subtract linear fit from phase difference.}
				\State $e_i \gets e_{\text{rel},i} + \gamma e_{\text{abs},i}$
				\State $N_i \gets \lVert e_i \rVert_\infty$
			\EndFor
			\State $err \gets \sum_i N_i / \len(\Phi)$
			\State $k \gets \argmax(N)$ \Comment{Find the error vector with the largest error norm.}
			\State $S \gets \text{sort}[\{G, e_k\}^T, 2]^T_1$ \Comment{Sort grid points in descending order of the largest errors.}
			\State $p \gets (S \setminus \mathcal{G})_1$ \Comment{Find the grid point with the largest error that is not yet part of the grid $\mathcal{G}$.}
			\State $\mathcal{G} \gets \mathcal{G} \cup \{ p \}$ \Comment{Place a new point there.}
		\EndWhile
		\State $\mathcal{G} \gets \text{sort}(\mathcal{G})$
	\end{algorithmic}	
\end{algorithm}


\section*{References}

\bibliography{ROM}

\end{document}